\documentclass[11pt]{article}
\usepackage[utf8]{inputenc}
\usepackage{fullpage}
\usepackage[letterpaper, portrait, margin=1.in]{geometry}
\usepackage{amsmath}
\usepackage{amssymb}
\usepackage{textgreek}
\usepackage{xcolor}
\usepackage{inconsolata}
\usepackage[protrusion=true,expansion=true]{microtype}
\usepackage[noblocks]{authblk}
\usepackage[sorting=none,style=numeric-comp]{biblatex}
\usepackage[hyperfootnotes=false]{hyperref}
\usepackage[multiple]{footmisc}
\usepackage{graphicx}
\usepackage{csquotes}    
\usepackage{enumitem}
\usepackage{ragged2e}
\usepackage{epigraph}
\usepackage{etoolbox}
\usepackage{lipsum}

\newcommand{\secauthors}[1]{{\hfill\begin{minipage}[t]{0.7\textwidth}
{\raggedleft \textit{contributed by #1}\par}\end{minipage}\vspace{1.em}}}

\newcommand{\recommendation}[1]{{\hfill\centerline{\begin{minipage}[t]{0.93\textwidth}
{\justifying \textbf{Recommendation:} \textit{#1}\par}\end{minipage}}\vspace{1.em}}}

\newcommand\snowmass{\begin{center}\rule[-0.2in]{\hsize}{0.01in}\\\rule{\hsize}{0.01in}\\
\vskip 0.1in Submitted to the  Proceedings of the US Community Study\\ 
on the Future of Particle Physics (Snowmass 2021)\\ 
\rule{\hsize}{0.01in}\\\rule[+0.2in]{\hsize}{0.01in} \end{center}}

\usepackage[firstpage=true]{background}
\backgroundsetup{contents={\parbox{6.5in}{\snowmass}}, scale=1,placement=top,opacity=1,color=black,position={3.25in,1.2in}}

\title{Snowmass21 Accelerator Modeling Community White Paper \\
\vspace{0.5cm}
{\large by the Beam and Accelerator Modeling Interest Group (BAMIG)\thanks{AccBeamModelSnowmass21@lbl.gov}}
}

\author[13]{Authors (alphabetical): S. Biedron} 
\author[1]{L. Brouwer}
\author[7]{D.L. Bruhwiler}
\author[7]{N. M. Cook}
\author[6]{A. L. Edelen}
\author[1]{D. Filippetto}
\author[9]{C.-K. Huang}
\author[1]{A. Huebl}
\author[15]{T. Katsouleas}
\author[4]{N. Kuklev}
\author[1]{R. Lehe}
\author[12]{S. Lund}
\author[1]{C. Messe}
\author[10]{W. Mori}
\author[6]{C.-K. Ng}
\author[9]{D. Perez}
\author[4,5]{P. Piot}
\author[1]{J. Qiang}
\author[6] {R. Roussel}
\author[2]{D. Sagan}
\author[11]{A. Sahai}
\author[9]{A. Scheinker}
\author[14]{E. Stern}
\author[16]{M. Th\'evenet}
\author[10]{F. Tsung}
\author[1]{J.-L. Vay}
\author[8]{D. Winklehner}
\author[3]{H. Zhang}

\affil[1]{Lawrence Berkeley National Laboratory, Berkeley, CA 94720, USA}
\affil[2]{Cornell University, Ithaca, NY 14853, USA}
\affil[3]{Thomas Jefferson National Accelerator Facility, Newport News, VA 23606, USA}
\affil[4]{Argonne National Laboratory, Lemont, IL 60439, USA}
\affil[5]{Northern Illinois University, DeKalb, IL 60115, USA}
\affil[6]{SLAC National Accelerator Laboratory, Menlo Park, CA 94025, USA}
\affil[7]{RadiaSoft LLC, Boulder, CO 80301, USA}
\affil[8]{Massachusetts Institute of Technology, Cambridge, MA, 02139, USA}
\affil[9]{Los Alamos National Laboratory, Los Alamos, NM 87545, USA}
\affil[10]{University of California at Los Angeles, Los Angeles, CA 90095, USA}
\affil[11]{University of Colorado Denver, Denver, CO 80204, USA}
\affil[12]{Michigan State University, East Lansing, MI 48824, USA}
\affil[13]{University of New Mexico, Albuquerque, NM 87106, USA}
\affil[14]{Fermi National Accelerator Laboratory, Batavia, IL 60563, USA}
\affil[15]{University of Connecticut, Storrs, CT 06269, USA}
\affil[16]{Deutsches Elektronen-Synchrotron DESY, 22607 Hamburg, Germany}

\date{\today}

\begin{document}

\maketitle

\newpage

\tableofcontents

\newpage

\section*{Executive summary}
\addcontentsline{toc}{section}{\protect\numberline{}Executive Summary}

Computer modeling is essential to beam and accelerator physics research, as well as to the design, commissioning and operation of particle accelerators. 
Somewhat surprisingly, despite accelerator physics being a field with extreme levels of coordination and long-range
planning for the research, design, construction, and operation of its largest accelerator complexes,
e.g., at CERN or at Fermilab, the development of beam and accelerator physics codes has often
been largely uncoordinated. This comes at a great cost, is not desirable and may not be tenable. 

Accelerator simulation is a large, complex topic, and much time and effort has been spent in
developing a large collection of simulation software that cover an expanding range of intertwined 
physics topics. The complexity of the overall endeavor has risen sharply in the last decade with the impetus 
to adapt the  algoritms  and  codes to rapidly changing computing hardware and software environments, 
and the additional task of having to reimagine the algorithms and recast them for quantum computing.
This is compounded by the need to infuse a rapidly evolving set of AI/ML technologies, which 
represent tremendous opportunities but can also be very disruptive.

After a summary of relevant comments and recommendations from various reports over the last ten years, 
this \emph{community} paper examines the modeling needs in accelerator physics, from the modeling of single beams and individual accelerator elements, to the realization of virtual twins that replicate all the complexity to model a particle accelerator complex as accurately as possible. 
A discussion follows on cutting-edge and emerging computing opportunities, such as advanced algorithms, AI/ML and quantum computing, computational needs in hardware, software performance, portability and scalability, and needs for scalable I/O and in-situ analysis.
Considerations of reliability, long-term sustainability, user support and training are covered next, followed by an overview of the benefits of ecosystems with integrated workflows based on standardized input and output, and with integrated frameworks and data repositories developed as a community. The last section highlights how the community can work more collaboratively and efficiently through the development of consortia and centers, and via collaboration with industry.

The following high-level recommendations are provided to synthesize the summary of recommendations for the topics discussed in the paper and listed afterward:

\begin{enumerate}
    \item Develop a comprehensive portfolio of particle accelerator and beam physics modeling tools in support of achieving Accelerator and Beam Physics Thrust Grand Challenges on intensity, quality, control, and prediction.
    \item Develop software infrastructure to enable end-to-end virtual accelerator modeling and corresponding virtual twins of particle accelerators. 
    \item Develop advanced algorithms and methods including AI/ML modalities and quantum computing technologies.
    \item Develop efficient and scalable software frameworks and associated tools to effectively leverage next generation high-performance and high-throughput computing hardware.
    \item Develop sustainable and reliable code maintenance practices, community benchmarking capabilities, and training opportunities to foster the cooperative application of accelerator software.
    \item Foster an open community that spans academia, national labs and industry to (a) develop software ecosystems, libraries, frameworks and standards, (b) curate data repositories, and (c) establish dedicated centers and distributed consortia with open governance models.
\end{enumerate}

\section*{Summary of recommendations (extended version)}
The Beam and Accelerator Modeling Interest Group (BAMIG) proposes the following recommendations to the Snowmass21 conveners:


\newcommand{\RecModelingNeeds}{Support the development of a comprehensive portfolio of particle accelerator and beam physics modeling tools for all types of particle accelerators (e.g., RF-based, plasma-based, structured-based wakefield, plasmonic), accelerator components (e.g., materials, superconducting magnets, structured plasmas), and which target the Accelerator and Beam Physics Thrust Grand Challenges  on intensity, quality, control, and prediction.}

\newcommand{\RecRF}{Support the development of modeling tools and methods that target the Accelerator and Beam Physics Thrust Grand Challenges (intensity, quality, control, and prediction), which will require modeling of collective effects with improved fidelity on long time scales, improving computational speed to allow for statistical ensembles and design optimization, and improved integration with realistic magnet and RF modeling.}

\newcommand{\RecPBA}{Support the development of modeling tools and methods that will enable start-to-end simulations that predict the full 6-D (+ spin) evolution of beams in a PBA-based linear collider, from their creation to their final focusing at the interaction point, and include all the intermediate phases of acceleration, transport, manipulations, collisions, etc. }

\newcommand{\RecSWFA}{Support the development of efficient and accurate algorithms capable of modeling the beam interaction with its wakefield over long interaction lengths in structures with arbitrary geometries and constitutive parameters.}

\newcommand{\RecPlasmonics}{Support the development of a new quantum-kinetic approach to model large-amplitude plasmons sustained by oscillations of ultra-dense conduction-band free electron Fermi gas. Modeling the dynamics of ionic lattice and the energy band structure under the influence of  PetaVolts per meter plasmonic fields of relativistic, nonlinear plasmons is critical to understand the effect on materials.}

\newcommand{\RecMaterials}{Support the development of automated scale-bridging methods that can autonomously parameterize higher-scale models from large numbers of lower-scale calculations, so as to enable predictive materials studies over a broad space of materials and conditions.}

\newcommand{\RecPlasmas}{Support the development and integration of fluid and kinetic codes to meet the modeling demands for a new class of structured plasma devices coupling macroscopic plasma properties with strict requirements on kinetic interactions.}

\newcommand{\RecMagnets}{Support the development of novel mixed finite element formulations and algorithms, and their implementation in open source software tailored for superconducting magnet design.}


\newcommand{\RecNextFrontier}{Support the development of accelerator modeling software that orchestrate interdisciplinary set of tools with standardized data representations to enable end-to-end virtual accelerator modeling and virtual twins of particle accelerators, which combine first-principle models together with machine learning-based surrogate models, for tunability from maximum precision for accurate and realistic accelerator design to maximum speed for online particle accelerator tuning.}

\newcommand{\RecInterdisciplinary}{Support interdisciplinary simulations especially efforts to establish standards that would ease the sharing of information and data across codes as well as standards for interfacing codes.}

\newcommand{\RecEVA}{Support the development of software that are capable of end-to-end virtual accelerator (EVA) modeling (Grand Challenge of Accelerator and Beam Physics) that incorporate all components (including both conventional and AAC sections) and all pertinent physical effects.}

\newcommand{\RecVirtualTwins}{Support the development of virtual twins of particle accelerators, which combine high-performance computing, first-principle models together with machine learning-based surrogate models, with tunability from maximum precision for accurate and realistic accelerator design to maximum speed for online particle accelerator tuning.}


\newcommand{\RecCuttingEdge}{Support the research and development on cutting-edge and emerging computing opportunities, including advanced algorithms, AI/ML methods,  quantum computing algorithms for beam and accelerator physics, as well as on the development of storage ring quantum computers.}

\newcommand{\RecAlgorithms}{Support research on algorithms, from refining the understanding of the properties and bottlenecks of existing algorithms to the elaboration of novel algorithms that exhibit better properties, remove the bottlenecks, and improve the speed and accuracy of accelerator modeling.}

\newcommand{\RecAIML}{Support the development of ML modeling techniques and their integration into accelerator simulation and control systems, with an emphasis on fast-executing (up to real-time) and differentiable models, continual learning and adaptive ML for time-varying systems and distribution shifts, uncertainty quantification to assess confidence of model predictions, and physics-informed methods to enable broader model generalization to new conditions and reduced reliance on large training data sets.}

\newcommand{\RecQIS}{Support quantum computing algorithm and code development for accelerator modeling, feasibility study on quantum computing implementation in accelerator modeling, and quantum computing education in accelerator community.}

\newcommand{\RecSRQC}{Support research and development of storage ring quantum computers.}


\newcommand{\RecComputationalNeeds}{
Support the development of increasingly powerful and specialized High-Performance Computing (HPC) and High-Throughput Computing (HTC) capabilities for accelerator modeling, 
and maintenance of  software to run efficiently on these  hardware (e.g., port of codes to GPUs) with efficient and scalable I/O, post-processing and in situ data analysis solutions, 
which will be needed to support ab initio modeling at increasing fidelity, training of surrogate models and AI/ML guided designs.
}

\newcommand{\RecHardware}{Support the development of increasingly powerful and specialized High-Performance Computing (HPC) and High-Throughput Computing (HTC) capabilities for accelerator modeling, 
which will be needed to support ab initio modeling at increasing fidelity, interactive and parallel data analysis, training of surrogate models and AI/ML guided designs.}

\newcommand{\RecSoftware}{
Foster the development and maintenance of codes that run efficiently on the latest hardware (e.g., add support for GPUs/FPGAs) by using maintainable single-source, portable solutions, and that are scalable on leadership-scale supercomputers with multiple levels of parallelization and support for effective dynamic load balancing.
}

\newcommand{\RecIO}{
Support the development and maintenance of efficient and scalable I/O, post-processing, in situ data analysis, and data sharing solutions in particle accelerator codes.
Coordinate on scientific data documentation, standardization, development of interactive and reproducible analysis workflows and foster data reuse.
}


\newcommand{\RecSustainability}{Provide sufficient resources for code maintenance, automated testing, benchmarking, documentation and code reviews. Convene a community effort to identify topics and teaching teams to deliver academic classes designed to foster sharing and cooperation, to be taught at the U.S. Particle accelerator school.}

\newcommand{\RecRobustness}{Establish and maintain open review, automated testing, validation and benchmark procedures for modeling tools and ensure reproducibility by tracking all changes in a documented and openly accessible manner.}

\newcommand{\RecUsability}{Consider maintenance of code reviews for improvements, documentation, installation and testing to be central to the mission of usable scientific software: establish open feedback and support channels, perform regular releases with change logs, use permissive open source licensing whenever possible and cover all scientific functionality with automated tests.}

\newcommand{\RecTraining}{Convene a community effort to identify topics and teaching teams to deliver academic classes designed to foster sharing and cooperation. The classes should be taught in the US Particle Accelerator School with course materials and linked tutorials/extensions regularly maintained and publicly posted.}


\newcommand{\RecEcosystem}{
Organize the beam and accelerator modeling tools and community through the development of (a) ecosystems of codes, libraries and frameworks that are interoperable via open community data standards, (b) open access data repositories for reuse and community surrogate model training, (c) dedicated Centers and distributed consortia with open community governance models and dedicated personnel to engage in cross-organization and -industry development, standardization, application and evaluation of accelerator and beam modeling software and data.
}

\newcommand{\RecWorkflows}{Foster the adoption and continued development of open community (meta-)data standards and their implementation in modeling tools and data acquisition for seamless integration of the community accelerator modeling tools into multiphysics workflows.}

\newcommand{\RecFrameworks}{Establish open, contributable, modular libraries and integrated frameworks for central computing, modeling and analysis tasks that foster the sharing  of common functionalities between applications, using open licenses and best practices/policies.}

\newcommand{\RecRepositories}{Establish open access data repositories and foster publishing of modeled and measured accelerator \& beam data to allow re-use (e.g. beam transport to applications), model training (e.g. AI/ML), preservation, recasting and reinterpretation.}

\newcommand{\RecCenters}{
Organize the beam and accelerator modeling community through the development of dedicated Centers and distributed consortia.
Dedicate resources to adopt open community governance models and dedicate personnel to engage in cross-organization and -industry cooperation.}


\begin{enumerate}
    \item {\bf Recommendation on Modeling needs:} \RecModelingNeeds
    \begin{enumerate}
        \item {\bf Subrecommendation on RF-based acceleration:} {\it \RecRF}
        \item {\bf Subrecommendation on plasma-based wakefield acceleration:} {\it \RecPBA}
        \item {\bf Subrecommendation on structure-based wakefield acceleration:} {\it \RecSWFA}
        \item {\bf Subrecommendation on PetaVolts per meter plasmonics:} {\it \RecPlasmonics}
        \item {\bf Subrecommendation on materials modeling for accelerator design:} {\it \RecMaterials}
        \item {\bf Subrecommendation on structured plasmas:} {\it \RecPlasmas}
        \item {\bf Subrecommendation on superconducting magnets:} {\it \RecMagnets}
    \end{enumerate}

    \item {\bf Recommendation on the next frontier: ultraprecise, ultrafast virtual twins of particle accelerators:} \RecNextFrontier
    \begin{enumerate}
        \item {\bf Subrecommendation on interdisciplinary simulations:} {\it \RecInterdisciplinary}
        \item {\bf Subrecommendation on end-to-end Virtual Accelerators (EVA):} {\it \RecEVA}
        \item {\bf Subrecommendation on virtual twins of particle accelerators:} {\it \RecVirtualTwins}
    \end{enumerate}

    \item {\bf Recommendation on cutting-edge and emerging computing opportunities:} \RecCuttingEdge
    \begin{enumerate}
        \item {\bf Subrecommendation on Advanced algorithms:} {\it \RecAlgorithms}
        \item {\bf Subrecommendation on Artificial intelligence, machine learning, and differentiable simulations:} {\it \RecAIML}
        \item {\bf Subrecommendation on Quantum computing:} {\it \RecQIS}
    \end{enumerate}

    \item {\bf Recommendation on computational needs:} \RecComputationalNeeds
    \begin{enumerate}
        \item {\bf Subrecommendation on hardware: CPU/GPU time, memory, archive:} {\it \RecHardware}
        \item {\bf Subrecommendation on software performance, portability and scalability:} {\it \RecSoftware}
        \item {\bf Subrecommendation on scalable I/O and in-situ analysis:} {\it \RecIO}
    \end{enumerate}

    \item {\bf Recommendation on sustainability, reliability, user support, training:} \RecSustainability
    \begin{enumerate}
        \item {\bf Subrecommendation on code robustness, validation \& verification, benchmarking, reproducibility:} {\it \RecRobustness}
        \item {\bf Subrecommendation on usability, user support and maintenance:} {\it \RecUsability}
        \item {\bf Subrecommendation on training and education:} {\it \RecTraining}
    \end{enumerate}

    \item {\bf Recommendation on community ecosystems \& data repositories:} \RecEcosystem
    \begin{enumerate}
        \item {\bf Subrecommendation on loose integration: Integrated workflows:} {\it \RecWorkflows}
        \item {\bf Subrecommendation on tighter integration: Integrated frameworks:} {\it \RecFrameworks}
        \item {\bf Subrecommendation on data repositories:} {\it \RecRepositories}
        \item {\bf Subrecommendation on centers \& consortia, collaborations with industry:} {\it \RecCenters}
    \end{enumerate}
\end{enumerate}

\newpage

\section{Introduction}
For particle accelerators---among the most complex and largest high-tech devices ever built---computational tools are indispensable. 
Computer simulations are critical to the design, commissioning, operation, and upgrading of accelerator facilities which cost many millions to billions of dollars.
It is thus widely recognized that the importance of accelerators to society and the high cost of new accelerator facilities demand that the most advanced and sophisticated high-performance computing (HPC) tools be brought to bear on modeling activities in accelerator science and technology~\cite{HEPAP,AAC_Roadmap_2016,ALEGRO,NagaitsevARXIV_ABP2021,ICFASagan2021,ICFAVay2021}.\cite{}

Accelerator simulation is a large, complex topic, and much time and effort has been spent in developing a large collection of simulation software, many developed by a single accelerator physicist and some by interdisciplinary collaborations of computational accelerator physicists, computer scientists, applied mathematicians and software engineers. 
Some address a single physics topic while others involve multiphysics, interdisciplinary frameworks, or workflows.
Some software has grown very large. Some include very sophisticated algorithms that are pushing the state-of-the-art in accelerator modeling, and sometimes in applications outside of accelerator physics (e.g., astrophysics). While some accelerator modeling research or design can be done on a laptop or workstation, others need the full power of the largest supercomputers. For a significant fraction of these activities, approximations, idealizations or others means to reduce the computational needs are necessary to fit the simulations within the available computer memory and runtime, eventually compromising accuracy and fidelity. In addition to pursue its longstanding tradition in research and development of novel algorithms, the accelerator modeling community is investigating emerging opportunities based on machine learning and quantum computing.

Despite accelerator physics being a field with extreme levels of coordination and long-range planning for the research, design, construction, and operation of its largest accelerator complexes, e.g., at CERN or at Fermilab, the development of beam and accelerator physics codes has often been largely uncoordinated. This comes at a great cost, is not desirable and may not be tenable. 
Due to developers retiring or moving on to other projects, numerous simulation programs have been completely abandoned or are seldom used.
This has resulted in a collection of codes that are not interoperable, use different I/O formats and quite often duplicate some physics functionalities using the exact same underlying algorithms. 
Frequently there is a huge impediment to maintaining these programs due to poorly-written code and lack of documentation. Additionally, many of the programs that are available tend to be ``rigid''. That is, it is generally difficult to modify a program to simulate something it is not designed to simulate \emph{a priori}. Adding a new type of lattice element that a particle can be tracked through is one such example~\cite{ICFASagan2021}.
Abandoned simulation programs represent a huge cost \cite{goble2014}, not only in terms of time and money spent in developing a program, but also in terms of researchers leveraging existing technology.  Indeed, a researcher who wants to simulate something that existing programs are unable to, will, due to time and monetary constraints, generally not be able to fully develop a comprehensive simulation program from scratch as compared to what could have been done if existing software could be leveraged. As simulation programs become more complex due to the ever-increasing demands placed upon machine performance, the situation will become worse if not addressed.

After a summary of relevant comments and recommendations from various reports over the last ten years, 
this paper examines the modeling needs in accelerator physics, from the modeling of single beams and individual accelerator elements, to the realization of virtual twins that replicate all the complexity to model a particle accelerator complex as accurately as possible. 
We then discuss cutting-edge and emerging computing opportunities, such as advanced algorithms, AI/ML and quantum computing, computational needs in hardware, software performance, portability and scalability, and needs for scalable I/O and in-situ analysis.
Considerations of reliability, long-term sustainability, user support and training are considered next, before discussing the benefits of ecosystems with integrated workflows based on standardized input and output, and with integrated frameworks and data repositories developed as a community. Last, we highlight how the community can work more collaboratively and efficiently through the development of consortia and centers, and via collaboration with industry.
A recommendation is proposed at the beginning of each section and subsection.

\section{\label{sec:reports}Previous Reports and Recommendations}
 
This section summarizes relevant comments and recommendations from various reports over the last ten years.
 
 \paragraph{2012 Office of HEP Accelerator R\&D Task Force report}

\epigraph{Software should help researchers optimize operating regimes and reduce the overall risk that underlies all modern accelerator design. Much of the current software has not taken advantage of the many computer improvements that have been developed in the last few decades. Such progress includes vastly increased processor speed, exploding memory capabilities, disk storage growth and cloud computing. There is not enough overall commercial demand for such high performance accelerator design software to have confidence this problem will be solved without US government intervention. 
\\... \\
Accelerators across the board also need advanced simulation studies, and long-term support for code development and maintenance is therefore needed.}{--- \textup{Office of HEP Accelerator R\&D Task Force report}, 2012}

\paragraph{2014 Particle Physics Project Prioritization Panel (P5) report}
\epigraph{
The present practice is to handle much
of the computing within individual projects. Rapidly evolving
computer architectures and increasing data volumes require
effective crosscutting solutions that are being developed in
other science disciplines and in industry. Mechanisms are
needed for the continued maintenance and development of
major software frameworks and tools for particle physics and
long-term data and software preservation, as well as investments
to exploit next-generation hardware and computing models.
Close collaboration of national laboratories and universities
across the research areas will be needed to take advantage of
industrial developments and to avoid duplication.\\
{\bf Recommendation 29:} Strengthen the global cooperation
among laboratories and universities to address computing
and scientific software needs, and provide efficient training
in next-generation hardware and data-science software
relevant to particle physics. Investigate models for the
development and maintenance of major software within
and across research areas, including long-term data and
software preservation.
\\... \\
Computing in particle physics continues to evolve based on
needs and opportunities. For example, the use of high-performance
computing, combined with new algorithms, is advancing
full 3-D simulations at realistic beam intensities of nearly all
types of accelerators. This will enable “virtual prototyping” of
accelerator components on a larger scale than is currently
possible.
}{--- \textup{
Report of the Particle Physics Project Prioritization Panel (P5)}, 2014}

 \paragraph{2015 HEPAP report}

\epigraph{
{\bf Recommendation 3.} Support a collaborative framework among laboratories and universities that assures sufficient support in beam simulations and in beam instrumentation to address beam and particle stability including strong space charge forces.
\\... \\
With Scenario B funding, an ambitious computational accelerator science program could be initiated to develop new algorithms, techniques, and generic simulation code with the goal of end-to-end simulations of complex accelerators that will guide the design, and improve the operations, of future accelerators of all types. Advancing the capabilities of accelerator simulation codes to capitalize on the drive toward exascale computing would have large benefits in improving accelerator design and performance. New computational algorithms coupled with the latest computer architectures are likely to reduce execution times for many classes of simulation code by several orders of magnitude, thereby making practical end-to-end simulations of complex accelerator systems. Such capabilities will enable cost-effective optimization of wakefield accelerators, as well as near-real-time simulations of large operational machines such as megawatt proton accelerators or a very high-energy proton-proton collider. In the near term, advanced simulation tools will maximize the productivity of R\&D for all future accelerators.
\\... \\
One area in which there has been some recent movement towards a nationally unified effort is in accelerator-related computation. Effort in the area has been boosted by funding from the SciDAC (Scientific Discovery through Advanced Computing) program jointly funded by ASCR (DOE Office of Advanced Scientific Computational Research) and HEP. One of the outgrowths of this effort is the CAMPA (Consortium for Advanced Modeling of Particle Accelerators) initiative from LBNL, SLAC, and Fermilab to establish a national program in advanced modeling of accelerators. There are, however, still many isolated simulation efforts within the program.
\\... \\
It is likely that there will be significant developments in accelerating technologies, both conventional (NCRF, SRF) and advanced (DWFA, PWFA, LWFA) technologies, in the coming decades. ... 
Accelerator physics and simulation support in these areas are crucial for making progress.
\\... \\
Computer simulations play an indispensable role in all accelerator areas. Currently, there are many simulation programs used for accelerator physics. There is, however, very little coordination and cooperation among the developers of these codes. Moreover there is very little effort currently being made to make these codes generally available to the accelerator community and to support the users of these codes. The CAMPA framework is an exception, and such activities should be encouraged.
\\
\vspace{3mm}
The direction of development in computer technologies makes it mandatory that the accelerator simulation codes (as well as all other HEP-related codes) adapt to modern computer architectures. High performance computers are another resource that HEP has not yet sufficiently exploited. The effort to coordinate such advanced computational activities for HEP is taking place within the Forum for Computing Excellence (FCE). Accelerator simulation effort in the direction of advanced computing should also be an integral part of the FCE, as are the other areas of HEP computation. An overall goal of this coordinated effort is to maintain and update mainline accelerator computer codes to take ad-vantage of the most modern computer architectures.
\\
\vspace{3mm}
Advances in simulations, as well as in computational capabilities, raise the exciting possibility of making a coherent set of comprehensive numerical tools available to enable virtual prototyping of accelerator components as well as virtual end-to-end accelerator modeling of beam dynamics. It should be possible to construct real-time simulations to support accelerator operations and experiments, allowing more rapid and detailed progress to be made in under-standing accelerator performance.
\\
\vspace{3mm}
Simulation efforts are vital for new accelerator development and supporting experimental accelerator R\&D studies. Such coherent efforts could be tailored after the successful LARP model that identified mutual study goals for assuring success of a given project (HL-LHC in the case of LARP) and supported collaboration among various university and laboratory partners.
}{--- \textup{HEPAP report}, 2015}%

\section{Modeling needs}
\recommendation{\RecModelingNeeds}

\subsection{RF-based acceleration}
\secauthors{J. Qiang, C. Mitchell, C-K. Ng}
\recommendation{\RecRF}

In an RF accelerator, a train of charged particles is accelerated to very high energy (TeVs) by RF cavities, and confined by magnetic elements for high energy physics applications. For such a high intensity charged particle beam, the collective effects from charged particle interactions among themselves and from the other beams play an important role in limiting the beam quality and accelerator performance. These collective effects include space-charge, intrabeam scattering, coherent synchrotron radiation, short-range and long-range wakefields, beam-beam, electron cloud and electron cooling effects.   

A number of computational methods have been developed to model these collective effects in an RF accelerator. The particle-in-cell method has been used to simulate space-charge, beam-beam, and electron cloud effects on massive parallel computers~\cite{warp,QIANG2000434,AMUNDSON2006229,Qiang2006,Furman2007,adelmann:opal}. The modeling of coherent synchrotron radiation effects was reviewed in a recent publication~\cite{Mayes2021}, where the recent advance has led to multi-dimensional and self-consistent simulation capability for the first time.
A Monte-Carlo method has been used to simulate the intrabeam scattering effect~\cite{Yu2009}. A Langevin method was developed to self-consistently simulate the intrabeam scattering effect by solving the Fokker-Planck equations with the Landau/Rosenbluth potential~\cite{Qiang2000b}.
The fast multipole method was developed to simulate both the space-charge and the electron cooling effects~\cite{ZHANG2011338,Marzouk2021}. 

Even though significant progress has been made in modeling RF accelerators, self-consistent modeling of the above collective effects remains challenging, especially for understanding dynamics on a long time scale. This includes long-term simulation of space-charge and beam-beam effects, self-consistent first-principles modeling of intrabeam scattering and electron cooling, electron cloud and coherent synchrotron radiation effects.  For example, one key goal involves the accurate prediction of halo formation and low-level beam loss at high intensity~\cite{LOI_halo_loss}.  
Fast advanced computational methods are needed to improve both the speed and the accuracy of modeling these effects. The computing time required must be sufficiently short to enable the large ensembles of runs needed for design optimization.  
Parallel programming paradigms that can make use of the
latest computer hardware such as multi-node GPUs are needed to further improve the speed of the simulation.
Studies are needed to understand numerical artifacts such as numerical noise associated with long-term simulations. A final area lies in improved magnet modeling and improved integration of RF and magnet models with existing tracking tools.  Methods such as surface methods that can include realistic external field effects are needed for improving the fidelity of RF accelerator designs.

The manufacture of accelerator structures and components is a major cost of an RF accelerator. Virtual prototyping of accelerator components has been a key process in the design and optimization of accelerators. Therefore, the ability to virtually prototype with HPC tools to create designs that work "out of the box" will substantially reduce the R\&D and operational costs of these accelerator components by eliminating the delicate, labor-intensive modifications employed in previous practices. Virtual prototyping requires multi-physics modeling capabilities for determining RF parameters such as shunt impedance, wakefield effects and higher-order-mode (HOM) damping, temperature distribution and thermal heat load, as well as mechanical stress and shape deformation~\cite{Xiao2019}.

To provide the required luminosity for a linear collider, it is critical to reduce emittance dilution in the main linac from machine tolerances such as cavity misalignments and imperfections~\cite{akcelik2008}. For example, in a superconducting linac, misalignments can arise from individual cavity misalignments and changes in the properties of coupled HOMs in a cryomodule with misaligned and deformed cavities. The RF parameters and fields of the HOMs can be evaluated for random distributions of cavity offsets (in a cryomodule) and cavity deformations along the full linac. The fields are then used for beam emittance dilution evaluation. A statistical analysis~\cite{lunin2018} using the constraints form realistic fabrication and component placement tolerances will be facilitated by the HPC capabilities of advanced simulation codes.


 

\subsection{Plasma-based wakefield acceleration}
\secauthors{J.-L. Vay, W. Mori, F. Tsung}
\recommendation{\RecPBA}

This section summarizes the modeling needs for plasma-based accelerators (PBAs). A more detailed description is given in \cite{ICFAVay2021}, Section 2.1.

Several aspects of plasma-based accelerators (PBA) are particularly challenging to model. 
These include  detailed kinetic modeling 
of the wake excitation;  trapping of background plasma electrons in the wakefields; ion motion for a nonlinear beam loading scenario;  the need to model the processes in three dimensions; the propagation of ultra-low emittance and energy spread beams in the plasma over meter  distances (many time steps), the long  (ps to ns) evolution of the perturbed plasma; the disparity in time and space scales of wake and driver evolution; and the large number of 3D simulations required to study the tolerances to nonideal effects.

The most widely used numerical tool to study plasma accelerators is the particle-in-cell (PIC) algorithm \cite{Birdsalllangdon}.
However, ab initio simulations of particle accelerators with PIC codes are limited by the number of time steps that are needed to resolve the beam propagation through the accelerator and by the number of grid cells that are sometimes needed to cover the wide disparity of spatial scales.
Several methods to reduce the computational cost, while maintaining physical fidelity, have been developed. 
{\it Reduced dimensionality:} When the beam and structure are nearly azimuthally symmetric, a two-dimensional ($r$-$z$) representation with a truncated series of azimuthal modes may be used \cite{godfrey1985iprop,LifschitzJCP2009, Benedetti_2010,davidson:15}.
{\it Quasistatic approximation:}
For PBA, one of the most successful methods to speed up simulations is the {\it quasistatic} approximation 
\cite{Sprangleprl90,mora:97, huang06, an:13,Mehrling_2014,li21a}, which relies on the separation of time scales to make approximations and  decouple the (long time scale) driver evolution and the (short time scale) plasma wave excitation, providing orders of magnitude speedup over standard PIC.  
{\it Boosted frame:}
An alternative to performing a quasi-static approximation to handle the disparity of scales is to shrink the range of scales by using the 
{\it Lorentz-boosted frame method}~\cite{Vayprl07}. With this method, the ultrahigh relativistic velocity of the driver is taken advantage of by using a frame of reference that is moving close to the speed of light, and  in which the range of space and time scales of the key physics parameters is reduced by orders of magnitude, lowering the number of time steps---thus speeding up the simulations---by the same factor.

While enormous progress has been made in the algorithms and codes to model PBAs, the modeling of a chain of tens to thousands of PBA stages for a multi-TeV collider is extremely challenging, needing further developments \cite{ALEGRO,ICFAVay2021}.
It is also essential to integrate physics models beyond single-stage plasma physics in the current numerical tools, including 
ionization and recombination, coupling with conventional beamlines, production of secondary particles, spin polarization, QED physics at the interaction point, collisions, and in some cases plasma hydrodynamics for long-term plasma/gas evolution and accurate plasma/gas profiles. Modeling PBA is currently challenging because  many beam-loading scenarios are still being considered. These choices may differ between accelerating electrons and positrons,  and the best numerical choices for each scenario are not the same.

\subsection{Structure-based wakefield acceleration}
\secauthors{P. Piot}
\recommendation{\RecSWFA}

Structure-based wakefield acceleration (SWFA) relies on high-charge “drive” bunches [{\cal O}(10–100 nC)] passing through slow-wave structures (SWSs) to excite electromagnetic wakefields. SWFAs can be configured in either two-beam acceleration (TBA) or collinear wakefield acceleration (CWA). The produced wakefields can be directly used to accelerate a delayed “main” bunch (CWA) or be out-coupled and guided to an optimized accelerating structure that accelerates the main bunch (TBA) in a parallel beamline. CWA offers a simpler configuration where both the drive and main bunches are transported along the same beamline; the TBA scheme decouples the drive and bunch beam dynamics at the expense of increased complexity (e.g., two parallel beamlines are required). The beam dynamics associated with the simultaneous transport of the accelerating main bunch and decelerating drive bunch is one of the major challenges in CWA.

Significant achievements in TBA research include accelerating gradients in excess of $\sim 300$~MV/m, wakefield-based power generation of $\sim 0.5$~GW,~\cite{1GWpower} and demonstration of staged acceleration~\cite{JING201872}. Similarly, recent progress along the CWA includes the beam-based demonstration of 300-MV/m gradient in a THz structure~\cite{OShea2016}, the generation of record transformer ratio of $> 5$ via improved longitudinal beam shaping~\cite{PhysRevLett.120.114801,PhysRevLett.124.044802}, and significant advances in the theoretical understanding of beam stability~\cite{PhysRevAccelBeams.21.031301}.

These developments have been possible due to consistent progress in software capable of modelling the beam-wave interactions in complex accelerating and power-extracting and transmission structures (PETSs). Modeling of the SWFA includes radiofrequency (RF) design of complex structures where the RF properties (e.g., $R/Q$, operating frequency, scattering parameters) are optimized using a frequency-domain solver for eigenmode or scattering parameters (e.g., such as {\sc omega3p} and {\sc s3p}~\cite{ACE3P}). Likewise, time-domain simulations are performed to understand the coupling (wakefield generation and acceleration) of an electron bunch with the mode supported by the structures. These investigations are generally performed with computer programs such as {\sc t3p}~\cite{ACE3P} and {\sc warpx}~\cite{WarpX}.

In unison with the roadmap elaborated in Ref.~\cite{usdoe-aac-2016} the demonstration of SWFA as a mature technology for a linear collider will require development to ($i$) perform integrated experiments in dedicated test facilities, ($ii$) explore high-efficiency schemes, and $(iii)$ push accelerating gradient.  Such developments can only be guided with high-fidelity simulations. Developing integrated experiments demands software capable of simulating the long-term beam dynamics within meter-scale structures. The target for high-efficiency acceleration relies on a fine control of the phase-space distribution of the drive and possibly main beams~\cite{abp-roadmap}. Understanding the precise evolution of these phase-space distributions over long interaction distances is critical to understanding the onset of the beam break-up (BBU) instability. Simulations over long integration time are also needed for TBA where the generation of GW-level power involves the deceleration of a train of bunches (with 100's ps separation) in a structure. Finally, pushing the accelerating gradient requires optimization of complex electromagnetic structures operated in a transient mode. Implementing boundaries with complex geometries  frequency-dependent electromagnetic and material properties is critical to the development of exotic structures and understanding of their dynamical response to intense fields~\cite{OShea2016}; see also Section~\ref{sec:material}.

\subsection{PetaVolts per meter plasmonics and Plasmonic acceleration}
\secauthors{Aakash Sahai along with T. Katsouleas, D. Filippetto and other contributors of the PV/m plasmonics Snowmass paper}

\recommendation{\RecPlasmonics}

PetaVolts per meter (PV/m) plasmonics effort \cite{PlasmonicsIEEE2021, PlasmonicsSPIE2021, CUDenverPCT, NanofocusingSPIE2022} focuses on access to extreme electromagnetic fields using plasmonics. Plasmonics is based upon the unique properties of condensed matter that stem from the structure of the ionic lattice which inevitably gives rise to electronic energy bands. The electrons that inherently occupy the conduction band in conductive materials are free to move across the entire lattice in response to external excitation. Plasmonic modes are collective oscillations of the Fermi gas that is constituted by the conduction band electrons \cite{Plasmon1953}.

Below we briefly summarize the PV/m initiative and its modeling needs. Further details are available in the above referred work as well as the independent PV/m Plasmonic Snowmass paper \cite{PVm-plasmonics-snowmass}.

The scope of PV/m plasmonics is limited to mechanisms where there exists a strongly correlated ionic lattice and corresponding electronic energy bands over relevant timescales. Under these conditions, quantum mechanical effects dominate. Specifically, our effort focuses on acceleration and focusing gradients as high as PetaVolts per meter \cite{PlasmonicsIEEE2021, PlasmonicsSPIE2021, CUDenverPCT, NanofocusingSPIE2022}.

PV/m plasmonics initiative \cite{PlasmonicsIEEE2021, PlasmonicsSPIE2021, CUDenverPCT, NanofocusingSPIE2022} introduces a novel class of non-perturbative plasmonic modes that are strongly electrostatic. These novel modes are excited by intense ultra-short charged particle beams. Optical photons cannot directly excite these strongly electrostatic modes as their energy is significantly smaller than metallic density plasmons. Moreover, high amplitude optical plasmons are technologically infeasible due to the large pre-pulse or pedestal energy in high-intensity optical pulses which ablate and disrupt the ionic lattice. This turns any sample into solid-state plasma which are out of the scope of the PV/m plasmonic effort.

The electrostatic nature of plasmons makes it possible to access unprecedented PetaVolts per meter electromagnetic fields. Strong excitation of the Fermi electron gas drives the oscillations to the wavebreaking limit. Because metals have an equilibrium free electron density $n_0$, as high as $\rm 10^{24}cm^{-3}$, the coherence or ``wavebreaking'' limit \cite{Wavebreaking1959} of EM fields is,

$$\rm E_{plasmon} = 0.1 \sqrt{ n_0(10^{24}cm^{-3}) } PVm^{-1}.$$

A comprehensive approach to modeling of relativistic and highly nonlinear plasmonic modes in conductive materials is required to appropriately incorporate the multi-scale and multi-physics nature of the underlying processes. Being strongly electrostatic, nonlinear plasmonic modes  significantly differ from the conventional optical plasmons and do not lend themselves to being modeled using purely electromagnetic codes. Optical plasmons are conventionally modeled using the Finite-Difference-Time-Domain (FDTD) method where the perturbative electron oscillations are simply approximated using constitutive parameters. The FDTD approach with constitutive parameters cannot be used when the collective oscillations become non-perturbative.

Our work \cite{PlasmonicsIEEE2021, PlasmonicsSPIE2021, NanofocusingSPIE2022} has adopted the kinetic approach along with Particle-In-Cell (PIC) computational modeling of collective oscillations of the free electrons Fermi gas to account for the nonlinear characteristics of strongly electrostatic plasmons. It is noted that PIC methodology already utilizes the FDTD solver for calculation of electromagnetic fields but utilizes charge and current densities based upon particle tracking. This approach utilizes the collisionless nature of relativistic oscillations of the Fermi gas and does not assume any constitutive parameters as part of the initial conditions. However, initialization and self-consistent evolution of the electron density implicitly accounts for most of the constitutive parameters. Moreover, as relativistic oscillations of the free electron Fermi gas have been experimentally observed to go beyond the Ohm’s law, evolution of these oscillations still remains unaccounted for by collisionless modeling approach. Specifically, conductivity, which is a critical constitutive parameter based upon electron-ion collisions, is not properly understood in the relativistic plasmonic regime. Therefore, heating processes resulting from relativistic plasmonic modes are not fully incorporated in the kinetic modeling approach.

In consideration of this immense promise of PV/m plasmonics using nanomaterials, we call for the support of the modeling community to engage with the our effort and help understand several new challenges that are not part of existing modeling tools.

Below we outline a few of the key challenges identified through our ongoing efforts and bring out the unique requirements our modeling effort:
\begin{itemize}[itemsep=-1mm]
\item Modeling the effect of relativistic energy gain of free electron Fermi gas in the the ionic lattice potential.
\item Understanding the effects of the energy density of plasmonic modes on the energy-band structure.
\item Accounting for self-fields within the ultra-dense electron beams where particle-to-particle interaction may play a relevant role.
\item Incorporating effects from atomistic modeling within the kinetic approach.
\item Devising an approach to handle collisions to determine the longer term effects of electrostatic plasmonic fields and account effects related to conductivity.
\end{itemize}

\subsection{Materials modeling for accelerator design~\label{sec:material}}
\secauthors{Danny Perez}
\recommendation{\RecMaterials}

The quest for ever-higher acceleration gradients leads to stringent demands on the ability of materials to maintain high performance in the harsh environments typical of high-gradient accelerator cavities. Of special interest is the understanding of the origin and properties of RF breakdown precursors, with the goal of eventually using this understanding to develop  breakdown-mitigation strategies. These simulations are typically conducted at multiple scales, ranging from first-principle quantum calculations using methods such as density functional theory (DFT). Such calculations allow one to accurately estimate elastic, thermodynamic, and electronic properties of different alloys. However, the steep scaling of these methods (cubic in the number of electrons) strongly limits possible simulation sizes and times. In order to obtain nanostructural information, such as defect nucleation,  propagation, and reactions, classical molecular dynamics (MD) simulations are instead employed. These simulations are typically extremely scalable ($O(N_{\mathrm{atoms}}$ or $O(N_{\mathrm{atoms}} \log N_{\mathrm{atoms}}$ ). A key challenge with moving from DFT to MD is that electrons are typically integrated out, which implies that electronic physics (e.g., the coupling of electrons with the electric and magnetic fields in the cavity) has to be re-introduced "by hand", which often requires {\em ad hoc} assumptions and is typically extremely time-consuming. While MD simulations are orders of magnitude cheaper than their DFT counterparts, simulations times ($<\mu$s) and sizes ($\sim 10^9$ atoms) remain extremely limited compared to engineering scales. This requires the introduction of yet larger-scale models that operate at the continuum level, including for example crystal plasticity or continuum heat and mass transport codes. While these continuum codes are often extremely scalable, accurately parameterizing the models with materials-specific physics is also extremely tedious and requires a significant level of understanding of the relevant factors, which is often not unambiguously available. We note that the systematic and automated parameterization of continuum models from lower-scale simulations is often not possible, except for ``simple'' properties such as elastic constants. 

The Materials Genome Initiative has pioneered the use of large numbers of quantum simulations (typically at the DFT level) to explore the space of possible material in search of solutions with optimized performance characteristics. These approaches have proved very successful at exploring the space of functional materials (such as materials for batteries), and extensions to structural materials are also possible when suitable figures of merit (i.e., computable from computationally-affordable DFT  calculations) can be defined. First applications to accelerator materials have recently been carried out for dilute binary Cu alloys, which identified new candidates for breakdown-tolerant materials, as well as rationalized the observed performance of certain alloys like Cu/Ag. Extending this exploration to more complex multi-component materials will prevent exhaustive enumeration and is instead likely to require more sophisticated machine-learning-guided active-learning approaches. Similarly, including
complex microstructural effects in the figure of merit (e.g., to also take into account the microstructure of the materials) is likely to require a combination of state-of-the-art computational methods with data-driven ML algorithms.

Further, the development of robust and automated workflows, such as those described in section~\ref{subsec:workflows} ,for the estimation and upscaling of materials properties to inform higher-scale models from lower-scale simulations would be extremely valuable, as it would enable the development of predictive materials models in a practical amount of human effort, compensating with a more aggressive use of high-performance computing resources.

\subsection{Structured plasmas}
\secauthors{Nathan Cook}

\recommendation{\RecPlasmas}

This section summarizes the modeling needs for structured plasmas. A more detailed description is given in [3], Section 2.4.

Structured plasmas, systems for which plasma density, temperature, composition, and ionization state are tailored to enhance an interaction between a beam or laser with that plasma system, are increasingly important for future accelerators. Controllable plasma channels are essential to increasing peak energy and quality across all types of PBAs; they permit the guiding of intense lasers over long distances for laser-driven schemes~\cite{Geddes_2004,Leemans_2006, Ibbotson_2010, Leemans_2014}, control of plasma density over meter-scale distances as required for beam-driven schemes~\cite{Blue_2003, Blumenfeld_2007, Litos_2014}, and the realization of plasma columns for novel positron acceleration schemes~\cite{Gessner_2016,Diederichs_2019}.Structured plasmas have also found application as flexible focusing elements. Discharge capillaries can produce orders-of-magnitude larger magnetic field gradients than traditional electromagnets~\cite{vanTilborg_2015}, subsequently enabling the compact staging of plasma accelerators~\cite{Steinke_2016}. Alternative approaches include passive plasma lenses, consisting of a narrow plasma jet outflow generated by laser pre-ionization~\cite{Lehe_2014,Doss_2019}, which can provide comparable focusing at high densities. Discharge plasmas have been employed as tunable dechirpers to remove correlated energy spreads from GeV-scale electron beams~\cite{DArcy_2019}. Finally, a class of proposed non-destructive diagnostics with high spatiotemporal resolution will rely on controllable plasma densities in concert with electron and laser beams~\cite{Scherkl_2019}.

Although specific needs vary with device type, scale, and application, structured plasmas impose unique modeling requirements as compared with simulations of other accelerator systems. Capillary discharge dynamics require the characterization of a discharge current and corresponding plasma transport properties, including electrical resistivity and thermal conductivity. Magnetohydrodynamic (MHD) codes are well suited to capturing the basic physics of these systems, while maintaining larger timesteps and reduced resolution requirements from kinetic approaches. Significant progress has been made in demonstrating their agreement with 1D analytical models and experimental results for waveguides and active plasma lenses~\cite{Bagdasarov_2017, Pieronek_2021}. The coupling of such MHD codes with laser envelope models~\cite{Benedetti17b} have contributed to record-breaking LPA acceleration of electrons to 8 GeV in 20 cm capillary \cite{Bagdasarov_2021, Gonsalves_2019}. Similarly, active plasma lens studies have benefited from the application of MHD to reproduce species-dependent nonlinearities in current flow and magnetic field~\cite{vanTilborg_2017,Lindstrom_2018,Cook_2020}. MHD codes have also been used to support a new class if structured plasmas predicted on the use of an extended laser focus, known as hydrodynamic optical-field-ionized channels (OFI or HOFI)~\cite{Shalloo_2018,Miao_2020};  Initial investigations into these systems have been supported by 1D hydrodynamic simulations~\cite{Shalloo_2018,Picksley_2020}. Despite these advances, MHD simulations face challenges in accurately capturing dynamics at early timescales, where low ionization rates and large temperature gradients impose constraints on fundamental assumptions of local thermal equilibrium (LTE), and may introduce numerical aberrations resulting from poor convergence of equation of state methods and inaccurate transport models~\cite{Diaw_2022}. Overcoming these limitations can require prohibitive increases in simulation runtime. More work is needed to improve the speed, reliability, and accessibility of MHD codes for modeling these systems.

Pre-ionized plasma sources, such as those used for PWFA stages and passive lenses, present additional modeling challenges. Neutral gas dynamics necessitate hydrodynamic simulation on $\mu$s-timescales to generate proper initial conditions, while the ionization laser requires propagation of an intense electromagnetic fields on fs-timescales. The resulting plasma does not constitute a local thermal equilibrium, therefore the LTE dynamics implemented by most MHD codes is insufficient to capture the ionization and heating dynamics. Moreover, vacuum-plasma interfaces are of interest for matching incident drive beams~\cite{Xu_2016}, and multi-species plasmas have been employed for high-brightness injection scheme~\cite{Deng_2019}. In these cases, kinetic codes, for example PIC techniques, may be coupled with hydrodynamic codes to obtain reasonable approximations, or alternatively, first principles models may be employed~\cite{Xi_2013,Lehe_2014,Manahan_2019,Doss_2019,Gessner_2020}. Similar challenges are faced in understanding the longterm evolution of these systems following laser or beam interactions. More work is needed to improve the interfacing between fluid and kinetic modeling tools to address the multi-physics nature of these systems across disparate spatiotemporal regimes.


\subsection{Superconducting magnets}
\secauthors{Lucas Brouwer and Christian Messe}

\recommendation{\RecMagnets}

Advanced modeling tools are currently utilized across the full range of US Magnet Development Program (US-MDP) research activities~\cite{USMDP}, enabling the design of improved conductors, magnets, and diagnostics. A diverse and challenging set of new modeling tools are required to continue this effort and ultimately improve design time, cost, and performance of future superconducting accelerator magnets.
The new developments include: (1) simulation of conductor and cable, (2) advanced modeling of interfaces and other potential sources of training in stress-managed designs, (3) modeling of LTS/HTS hybrid magnets, and (4) radiation environment thermal effects on magnets~\cite{LOI_magnets}. 

The material laws for superconducting materials are highly non-linear, which implies that a lot of iterations must be made per simulated timestep to achieve a numerically sufficient convergence. To reduce the high computational cost of these material laws cause, a novel finite element formulation called \textbf{h-\textphi} is being discussed within the HTS modeling community \cite{Dular.2019,Dular.2021}. By using the scalar magnetic potential \textbf{\straightphi} rather than the vector potential \textbf{a} that is traditionally used to model the magnetic field in air and vacuum domains. Since the computational effort is roughly proportional to the number of unknowns to the power of two, this means that reducing the magnetic potential form a vector to a scalar can reduce the computational effort up to a factor of nine.

One challenge in sufficiently implementing this novel formulation into a finite element code lies in the hybrid use of node and edge-based elements, the way boundary conditions are imposed and how efficient implementations must be designed. Another challenge lies in efficiently solving the system matrices which are non-symmetric, non-positive-definite and ill-conditioned.

Regarding the  \textbf{h-\straightphi}  formulation, some of the most pressuring questions discussed in the HTS modeling community are:

\begin{itemize}[itemsep=-1mm]
    \item imposing current boundary conditions in a user-friendly fashion\cite{Arsenault.2020y9g, Arsenault.2021}
    \item discretization of tape structures (thin shells) in 2D and 3D space \cite{Alves.2021}
    \item quench prediction (maxwell-thermal coupling)
    \item delamination (maxwell-thermal-mechanical coupling)
\end{itemize}

Most of the published formulations have been implemented on top of commercial toolboxes such as COMSOL and are not yet available to the general public. Moreover, it is in the very nature of closed source codes to limit the access the developer has to the data structure and the knowledge of underlying algorithms.  The desire of having full control over the data structure motivated us to develop a custom codebase that is tailored to the needs of HTS modeling. Having defined the project goals below, we decided to develop a new finite element framework from scratch to achieve them:

\begin{itemize}[itemsep=-1mm]
\item Support \textbf{h-a} and \textbf{h-\straightphi} formulations for 2D and 3D, as well as thin shells, both with first and higher order elements.
\item Support multiphysics, specifically thermal and mechanical coupling, as well as current sharing.
\item Have a text-based user interface that is tailored to the needs of HTS magnet and cable modeling.
\item Use popular open-source data formats, such as HDF5, GMSH, and Exodus II (ParaView)
\item Link against modern sparse linear algebra solvers such as  STRUMPACK, PETSc, PARDISO and MUMPS.
\item Run in parallel using the MPI standard.
\item Be readable, extendable and maintainable.
\end{itemize}

\section{To the next frontier: ultraprecise, ultrafast virtual twins of particle accelerators}
\secauthors{Jean-Luc Vay, Axel Huebl, R\'{e}mi Lehe, Chengkun Huang, Nikita Kuklev, Ji Qiang, David Sagan}
\recommendation{\RecNextFrontier}

The development of future particle accelerators requires predictive modeling tools that will range from 
(i) very detailed, full physics and dimensionality, first-principle kinetic simulation tools that are needed for detailed runs for physics studies (typically based on Monte-Carlo and Particle-In-Cell methods)~\cite{LOI_eva},
to 
(ii) ultrafast simulation tools  that are needed for ensemble runs 
for design studies (using a combination of, e.g., reduced physics, 
low dimensionality, low resolution, artificial intelligence 
(AI)/machine learning (ML) surrogates~\cite{LOI_ML, LOI_ML2}).

\subsection{Interdisciplinary simulations}
\recommendation{\RecInterdisciplinary}

Interdisciplinary simulations are important in a number of areas. In vacuo particle tracking coupled with particle/matter interactions is an example of a growing need. One application is in simulating the radiation induced by "dark current" electrons in accelerating cavities. This radiation may cause damage to cavities which leads to shortened lifetimes of the devices and a radiation safety hazard for the surrounding environment. Dark current induced problems have been observed at many facilities such as the CEBAF\cite{cebaf-dark}, LCLS-II\cite{lcls-dark}, ANL, etc. Sufficient shielding is required to properly contain the radiation which in turn requires a good understanding and prediction of radiation levels through simulations. Another example is the modeling of positron production in a target from the impact of high-energy electron beams accelerated through a linac injector. These simulations require accurate calculations using electromagnetic RF codes for accelerator structures and beam dynamics codes for particle transport in a beamline to characterize the beam profile before it hits the accelerator enclosure or the target. 

Particle/matter simulation codes exist. Examples include Geant4\cite{geant4}, FLUKA\cite{Ferrari:2005zk}, and MARS\cite{Mokhov:2017klc} which have traditionally been used for detector simulation in HEP experiments. However, since these codes and accelerator simulation codes have all been developed without common standards, interfacing them is a laborious task. A seamless simulation requires the proper transfer of field and particle data from accelerator to radiation codes. Communication in a standardized format such as openPMD\cite{openPMD}, which has been adopted in some accelerator codes, would help ensure efficient and error-free field and particle data transfers\cite{LOI_standards}. Another issue with an integrated simulation is in matching of the geometry of the vacuum chamber surface. The surface geometry in accelerator simulations is generally poorly defined if at all. The most comprehensive simulations define the surface using a finite element mesh generated from a CAD model (e.g., both WARP \cite{warp}
and OPAL \cite{adelmann:opal,winklehner:spiral} have some limited capacity for this). In contrast, radiation codes generally employ a faceted representation of the CAD model boundary. A converter for mapping finite element curved surfaces to faceted divisions on an interface boundary is required to accurately determine the location of a particle crossing from one computational domain to another. Much time and effort would be saved if the surface geometry descriptions were standardized so that a single converter module could be used in multiple codes.

Increasingly, accelerator simulation tools are also incorporating more micro-physics models to better describe the complex interplay of the various physics phenomena. One particular example being the emission modeling of a high-brightness electron photocathode gun. A photocathode gun provides a high-brightness electron source for the downstream accelerator beamline where the beam brightness can only be degraded, not improved. Thus, it is essential to understand the cathode emission characteristics and the method to control the beam quality in the gun environment through validated simulations. While Monte Carlo photo-electron emission simulations have been widely employed in studying photocathode performance for dedicated experiments, its potential in integrated simulations has only been explored recently~\cite{LOI_physics_based_injector_modeling}. For such purposes, a tight integration of the micro-physics models into existing gun simulations can be achieved via the best practices and standardization as discussed below.

\subsection{End-to-end Virtual Accelerators (EVA)}
\recommendation{\RecEVA}

The realization of software that are capable of end-to-end virtual accelerator (EVA) modeling 
has been identified as a Grand Challenge of Accelerator and Beam Physics \cite{LOI_ABPRoadmap}.

Thorough modeling of a full particle accelerator in individual parts
(injector, magnets, beam dynamics, etc.) is already an important aspect 
of particle accelerator development; one, without which no project 
nowadays can proceed to the building stage. For example, start-to-end simulations have been used in x-ray light source accelerator designs~\cite{Qiangprab2014,Qiangnapac2016,Qiangnim2022}. However, simulating the 
system in many small units poses two significant challenges: 
(i) The various codes used for the individual parts are often from different
areas, lack the state-of-the-art high-fidelity models for extremely bright and intense beams, written in different languages (C/C++, Fortran, Python, etc.),
and use different standards for data  I/O (particles, fields, etc.), 
slowing down the design and simulation process; 
(ii) Without multiphysics couplings, subtle, but important effects 
(collective effects, halo, coherent synchrotron radiation, etc.) might not
be considered in the design or without sufficient accuracy, ultimately leading to issues during 
commissioning, the need to modify the machine (shimming, 
additional steering, shielding, etc.), and longer downtimes due to added
radiation. Furthermore, opportunities for more efficient working points that can be achieved only from system optimization of the entire accelerator may
be missed entirely.

Thus, there is a clear need for full physics 6-D computer simulations 
of the entire accelerator system that incorporate all components (including both conventional and AAC sections), all pertinent physical 
effects, and that execute fast and reliably. Such EVAs need to be validated with experiments at both the component and system levels. Furthermore, these EVAs
should be able to leverage modern computing infrastructure like HPC clusters and GPU computing, and fully integrate AI/ML tools to maximize efficiency 
for practical applications. 
The development of such tools requires continuous advances in fundamental beam theory and applied mathematics, improvements in mathematical formulations and algorithms, and their optimized implementation on the latest computer architectures. 

\subsection{Virtual twins of particle accelerators}
\recommendation{\RecVirtualTwins}

Ultimately, the availability of multiphysics, end-to-end simulation capabilities will lead to the realization of virtual twins of particle accelerators. These will enable the design and optimization of future accelerators at scale on supercomputers, with unprecedented levels of accuracy and speed.
This capability will dramatically increase the breadth of parameter space that can be explored, thereby enabling the design of particle accelerators that have not been possible before. Ultrafast and ultraprecise tunability will let users simulate as needed: from maximum precision for accurate and realistic accelerator design to maximum speed for online particle accelerator tuning. Modern ML optimization frameworks can be used to take advantage of this tunability via multi-fidelity algorithms, allowing for faster scalable design refinement. Automated uncertainty quantification can be used to provide hints as to where the computational budget should be applied for best overall simulation quality.

\section{Cutting-edge and emerging computing opportunities}
\recommendation{\RecCuttingEdge}

Just as it is essential to continually invest in fundamental studies of accelerator technologies and beam science, it is also essential to do the same for the software and algorithms that enable such studies using computers. This includes the continuation of the study and development of novel algorithms and optimization strategies, as well as leveraging AI/ML and exploring algorithm that will take advantage of upcoming quantum computers. This also includes the development of modeling tools to study and design quantum computers based on storage rings.

\subsection{Advanced algorithms} 
\secauthors{Jean-Luc Vay, Ji Qiang}
\recommendation{\RecAlgorithms}

Computational beam and accelerator physics has spurred the development of many algorithmic advances \cite{Vay1996,Vay1998a,QiangPRE2000,Vayjcp01,Qiangcpc2001,QiangPRAB2002,Vayjcp02,Vay2002a,QiangCPC2004,Vaycpc04,Vaypop04,Qiangcpc2006,Vayprl07,decyk:07,Vaypop2008,Vaypac09,Cohennim2009,Vayjcp2011,Vaypop2011,VayPOPL2011,VayCSD12,GodfreyJCP2013,VayJCP2013,GodfreyJCP2014,GodfreyJCP2014_2,GodfreyIEEE2014,GodfreyCPC2015,LehePRE2016,KirchenPoP2016,VincentiCPC2016,Qiangcpc2016,Blaclard2017,JalasPoP2017,QiangPRAB2017,Qiangnim2017,WarpXEAAC2017,WarpXAAC2018,VincentiCPC2018,QiangPRAB2018,Kallala2019,Shapoval2019,QiangPRAB2019,KirchenPRE2020,li21a,ShapovalPRE2021,zoniARXIV2021_2,LeheARXIV2021_pml,decyk:04,Martinscpc10,decyk:11,XuCPC2013,decyk:14,YuCPC2015,YuCPC2015-Circ,yu:16,FeiCPC2017,li:20a,li:20b,xu:20,vranic:16} that have boosted the computational capabilities tremendously, in some instances by orders of magnitude at a time. Yet, the modeling of particles accelerators remains very demanding, and even extremely challenging in the case of design of future plasma-based or structure-based particle wakefield HEP colliders. It is thus essential to pursue the research on algorithms, from refining the understanding of the properties and bottlenecks of existing algorithms to the elaboration of novel algorithms that exhibit better properties, remove the bottlenecks, and improve the speed and accuracy of accelerator modeling. The discovery of new and better algorithms for beam and accelerator modeling is a fundamental topic that is poised to provide big boosts to the other fundamental topics that are the discovery of new and better accelerator technologies and concepts, and the discoveries of new laws of natures with colliders that they enable.

This also includes the adoption of algorithms that have proven to be very effective into other fields, such as adaptive mesh refinement~\cite{Vaypop2008}, which enables to zoom in on regions that need higher resolution (e.g., sharp edge of high-intensity beam) while zooming out to simulate larger regions (e.g., beam with pipe to include image charges and halo effects). This is discussed in more detail in \cite{ICFAVay2021}.

\subsection{Artificial intelligence, machine learning, and differentiable simulations}
\secauthors{Auralee Edelen, Remi Lehe, Nikita Kuklev, Alexander Scheinker, Ryan Roussel}

\recommendation{\RecAIML}

Machine learning (ML) and artificial intelligence (AI) have revolutionized many computational tasks in recent years, from protein folding prediction to fusion energy control. Speeding up computationally-intensive accelerator simulations and aiding optimization of particle accelerators are two key areas where ML \& AI have also been making substantial contributions. Many ML/AI tools are now technically mature, production-ready, and can be used as part of standard computational toolkits for particle accelerators. Notable examples so far include fast-to-execute accelerator models that can be used online as ``digital twins'', as well as in design and online optimization using Bayesian optimization and other methods, whereby a model that is learned on-the-fly is used to guide sampling of the parameter space. These methods, and tools supporting their use, are beginning to be integrated into dedicated simulation workflows and into control systems for use online that share common standards for interoperability. For example, the open source package LUME \cite{lume} has been used to deploy online ML and simulation models for a variety of systems, and xopt \cite{xopt} has been used to generate datasets (both offline and online) for training ML models. 

So far, most studies in AI/ML for accelerators have focused on  proof-of-principle tests of new methods under narrow operating conditions, rather than integration into dedicated use online where they would be tested under a wide variety of conditions and face additional challenges in robustness. Looking forward, there is great potential to now expand upon the fundamental  methods algorithmically, combine the strengths of different methods in novel ways (for example,  combining neural network models or differentiable simulations with uncertainty quantification and Bayesian optimization), and address the challenges of dedicated deployment that often do not arise in R\&D testing of each new algorithm (e.g. robust uncertainty quantification and adaptation to account for changing conditions over time or exploration of new regions of parameter space, extension to very high dimensional systems with hundreds of freely-tunable accelerator settings, etc.). We highlight some current and future directions below.

\paragraph{Fast-executing, adaptive accelerator models} Physics-based accelerator models can very accurately predict expected beam phase space evolution. These models can be calibrated offline by hand or online based on feedback from diagnostics, such as using energy spread spectra to tune a model to predict the beam's longitudinal phase space at FACET \cite{scheinker2015adaptive}. The major limitation of physics-based simulation models is that they are computationally expensive and therefore cannot give real-time (e.g. at rates of 120 Hz to MHz) predictions in the control room. For physics-based simulations that include the major nonlinear collective beam effects, prediction latencies can be on the order of tens-of-minutes to hours depending on the simulation fidelity. In contrast, trained ML models do have the potential to provide real-time predictions, in some cases providing orders of magnitude improvements to execution speed \cite{edelen2020machine}. Many studies in ML-based modeling of accelerator systems have relied on supervised learning on bulk inputs/outputs and fairly standard neural network architectures, such as multi-layer perceptrons \cite{edelen2016neural,emma2018machine,EdelenNeurIPS2017,EdelenNeurIPS2019,edelen2020machine} and convolutional neural network (CNN) based encoder-decoder architectures \cite{edelen2016first,Gupta_2021,zhu2021high,scheinker2021adaptiveML}, as well as Gaussian process models \cite{duris2020bayesian,ogren2021surrogate}. Common inputs and outputs include accelerator settings such as magnet currents, cavity gradients, and initial conditions such as laser parameters; common outputs include scalar beam parameters such as the emittance and bunch length, and 2D projections of the beam phase space. When trained on large simulation datasets of input-output variables, these models have been demonstrated to have as much as a million times speedup in execution speed with reasonable prediction accuracy \cite{edelen2020machine}. Training on measured data (either from scratch or by adapting a model that is pre-trained on simulations) also enables differences between the as-built machine and the simulation to be represented.

Looking forward, there is a need to further develop methods for making ML-based accelerator models reliable under shifting input-output data distributions. For example, this can include both time-varying changes in inputs from an operational accelerator and deliberate changes in operating conditions or simulation inputs that go beyond the statistical distribution of the training data. Methods to flag when model performance is degraded, as well as methods to adapt models to new conditions are needed. Having accurate uncertainty estimates in addition to model predictions is critical for applications in accelerators, especially in cases where decisions about control actions, analysis, or design are made based on model predictions.  Uncertainty quantification methods, such as the use of Bayesian neural networks\cite{mishra2021uncertainty} and ensembling \cite{convery2021uncertainty}, have been used in accelerators to highlight cases where model uncertainty is high. However, additional work is needed to ensure robustness of uncertainty estimates under changing conditions.  In general, further work incorporating state-of-the-art methods from continual online learning \cite{nagabandi2018deep} and transfer learning \cite{pan2009survey} could be used to maintain model accuracy over time and aid re-use of models in different settings.  This could also help reduce reliance on large training data sets when transferring models between specific accelerators (for example, between injectors with similar designs). Operation in tandem with local feedback algorithms can also help retain model accuracy over time as conditions change. Sample-efficiency is a key consideration for accelerator problems in which the data collection is very time-consuming or detailed beam measurements would interrupt operations. By incorporating physics constraints within ML frameworks, such as developing beam dynamics models based on Taylor maps, it is possible to greatly reduce the amount of data needed for re-training models \cite{Ivanov2020}, as discussed in more detail below. Further work is also needed in developing advanced methods for cases where online re-training is not feasible or practical. Some avenues toward addressing this challenge without re-training include the use of adaptive feedback / local optimization to make the overall ML system more robust to unknown and time-varying system changes and inputs. Some work towards the challenge of ML without re-training is utilizing adaptive feedback within the ML framework by adjusting ML model inputs or internal parameters with gradient-free adaptive feedback algorithms \cite{scheinker2021adaptive,scheinker2021adaptiveML}. In the context of accelerator tuning, it can also be accomplished by combining ML system models for control with local adaptive feedback or optimization algorithms \cite{scheinker2018demonstration}.

\paragraph{Physics-informed and physics-guided ML modeling} Recently there has been interest in incorporating physics information more directly into ML models. This is expected to improve sample efficiency and accurate generalization to unseen regions of parameter space. Such ``physics-informed'' modeling  spans a wide variety of techniques that use physics information more or less directly (see \cite{karniadakis2021physics} for a review). 
In the most broad sense of ``physics informed'', training on very broad, high-fidelity simulation data sets introduces observational biases that allowing ML models to learn functions, vector fields and operators that reflect the physical structure of the data \cite{karniadakis2021physics}. Numerous examples exist in the literature where training on broad accelerator simulation data sets enables enough of the underlying physics to be represented in a sufficiently generalizable form that the learned models can reliably interpolate to unseen configurations or provide qualitative estimates of beam behavior even outside of the training distribution \cite{edelen2020machine,EdelenNeurIPS2019}, for both scalar predictions and beam phase space projections. A recent example highlights that this can also be extended to unobserved  components of the beam parameter space \cite{scheinker2021adaptiveML}. In that work, a convolutional neural network (CNN)-based encoder-decoder trained and then utilized in an unsupervised adaptive approach to predict the 15 unique 2D projections of a beam's 6D phase space for an unknown and time-varying input beam distribution. The approach in  \cite{scheinker2021adaptiveML} is to adaptively tune the low-dimensional latent space of the encoder-decoder to match a measured 2D (longitudinal phase space) projection of an unknown 6D phase space distribution and in doing so accurately predict the other unmeasured 2D projections of the 6D distribution. 

More direct methods of making ``physics-informed'' models include adding physics-based constraints into the learning process \cite{raissi2019physics,karniadakis2021physics}, or structuring the learning algorithm in a way that mathematically resembles the physics problem (e.g., replacing one component of a Hamiltonian update with a neural network \cite{Sanchez-Gonzalez:2019gis}). In accelerator physics this avenue has been less explored but holds significant potential. In one accelerator example, the Hessian from a physics-based or learned model is used to inject expected correlations into a Gaussian process kernel \cite{hanuka2021physics}, which results in more sample-efficient learning and optimization. Another accelerator example uses a differentiable formulation of beam dynamics in a ring, implemented in the tensorflow machine learning library, to enable learning of free parameters using traditional ML workflows \cite{Ivanov2020}. For more efficient predictions, Galerkin-based or symplectic reduced order modeling (ROM) may present new opportunities for nonlinear beam dynamics study beyond the theoretical formulation.


\paragraph{Differentiable simulators} While there is a long history of using automatic differentiation in accelerator physics for higher-order calculations in rings \cite{Berz:1988aj}, accelerator physics simulation codes do not readily support arbitrary computation of gradients. This necessitates the use of numerical estimation of gradients when using gradient-based optimization methods. Having simulation codes that inherently support automatic differentiation would enable gradient-based optimization to be more readily used with accelerator physics simulations. Tying together simulations with gradient-based learning algorithms also has significant promise; for example, this could aid gradient-based learning of free parameters, as was done for accelerators with a differentiable lie map formulation in \cite{Ivanov2020} and for a differential hysteresis model in \cite{roussel2022differentiable}. This is similar to examples using adaptive feedback that approximates gradient descent to optimize free parameters in physics models \cite{scheinker2015adaptive}. Gradient-based learning of external ML algorithms (such as learning a control policy in reinforcement learning) can also be aided by a differentiable forward model, as was done for example in \cite{EdelenNeurIPS2017} but with a differentiable learned ML model rather than a differentiable simulator. Tying together standard simulation tools with ML components (e.g., using a learned model for one machine section or one computation component) would benefit substantially from having end-to-end differentiability. Expanding this concept to detailed differentiable simulations of non-linear collective effects (e.g., with particle-in-cell codes) could enable new capabilities in optimizing particle accelerator simulations and using them in conjunction with AI/ML. One example of integrating optimization with a differentiable physics model is given in \cite{roussel2022differentiable}, where an analytic model of hysteresis is made differentiable and used in conjunction with measured data to efficiently learn an accurate model of a magnet response that includes hysteresis; this is then used for hysteresis-aware Bayesian optimization that can more precisely optimize the resultant beam.

ML/AI frameworks have highly-optimized implementations of automatic-differentiation and other low level computational tools for running efficiently on CPU/GPU. Major libraries (PyTorch, TensorFlow, JAX, Julia) provide low level hooks for these routines. Some frameworks can be used to provide gradients for existing codes written in C or Fortran. This presents two promising avenues toward making differentiable physics simulators: (1) re-writing or translating existing low-level accelerator physics codes into formats compatible with automatic differentiation and (2) wrapping sections of existing code with external software to provide derivatives. This also opens a path toward using efficient AI/ML learning algorithms and frameworks directly with physics simulations.


\paragraph{Model-based optimization} The development and use of advanced algorithms for accelerator design and optimization aims to efficiently solve challenging optimization problems in high dimensional parametric spaces. The so called ``curse of dimensionality", where optimization difficulty scales exponentially with the number of free parameters, limits the number of optimization problems that can be done, especially when running high-fidelity accelerator simulations is computationally expensive.
To make a problem solvable with a limited computational budget, either assumptions are made to simplify the simulation itself or the free parameter space must be heavily restricted.
However, if novel techniques that substantially improve the efficiency of simulated optimization can be implemented, previously unsolvable optimization problems in accelerator physics can be realistically tackled.

Traditional design optimization in accelerators has relied on computationally intensive algorithms such as genetic algorithms (e.g. NSGA-II \cite{deb2002afast}) and particle swarm optimization, 
which require many simulation samples to converge and do not leverage any learned representations of the system. ML-based optimization methods that attempt to learn a model on-the-fly during optimization or use an existing model can improve sample efficiency and convergence speed by enabling more judicious choices of input variables to examine.

In cases where gradient information is not available, known as ``Black Box" problems, model based optimization (such as Bayesian optimization) techniques can be used to maximize optimization efficiency. These algorithms have already been shown to perform well in certain experimental \cite{duris2020bayesian,hanuka2021physics} and simulated \cite{roussel2021multiobj,roussel2021turnkey} contexts. However, there remains substantial room for further innovation and improvement in sample efficiency, scaling, and tailoring to specific applications. For example, multi-fidelity Bayesian optimization \cite{Wu2019Practical}, where evaluations of the optimization objective at a lower fidelity are used as a proxy for high fidelity simulations, is naturally well-suited to simulated optimization of accelerator based problems where multiple fidelities are straightforward to interpret. These algorithms could be used to help balance computational resources and required fidelity automatically during search refinement.
Furthermore, if models used for these algorithms also provide an associated confidence metric in their predictions, they can be used in information-based optimization to either speed up optimization in high dimensional parameter spaces based on model confidence regions \cite{eriksson2019scalable} or improve mutual information gain during nested optimization procedures \cite{neiswanger2021bayesian}. 

Model based algorithms also enable the use of transfer learning based-techniques, where previously created models of similar systems are used to inform algorithms applied to novel optimization problems and aid sample-efficiency. For example, using information about correlated input parameters from a physics simulation to inform the internal model in Bayesian optimization of an operating accelerator \cite{hanuka2021physics,duris2020bayesian} can also be considered a form of kernel-based transfer learning. Physics-informed models or differentiable simulators also have great potential for use in sample-efficient, model-based optimization across different systems. They can be used in conjunction with gradient-based optimization to learn system model parameters on-the-fly with limited data and exploit this information in control. For example, in \cite{roussel2022differentiable}, the parameters of a general differentiable hysteresis model for magnets  is learned from beam measurements and is used in conjunction with Bayesian optimization for hysteresis-aware fine-tuning.


\subsection{Quantum computing}
\secauthors{He Zhang, Ji Qiang}

\recommendation{\RecQIS}

The start-to-end simulation of an accelerator using real beams with billions or more particles remains challenging and expensive even with state-of-the-art exascale machines. The development of quantum computers, which can potentially provide an exponential improvement in efficiency for some classes of simulations, may bring new opportunities to enhance the particle accelerator community’s simulation abilities.

The technique for building quantum computer has entered the Noisy Intermediate Scale Quantum (NISQ) era \cite{preskill2018quantum}. Current state-of-the-art quantum computers have above 50 qubits and quantum supremacy has been demonstrated on some specific problems \cite{arute2019quantum, zhong2020quantum}. Quantum computing is currently available to the public through cloud services provided by some commercial companies such as IBM \cite{IBMQC}, D-Wave \cite{DWaveLeap2}, Amazon \cite{AmazonQSulution} and Microsoft \cite{AzureQ}. Meanwhile, studies on quantum algorithms have also been booming in the past few years. One hot topic is quantum machine learning. Machine learning can be used to optimize the performance of accelerators in design, commissioning and operation. Some time-consuming simulations can also be replaced by well-trained learning systems. Quantum computing can speed up many learning algorithms, including SVM \cite{rebentrost2014quantum,chatterjee2016generalized}, Bayesian networks \cite{low2014quantum}, convolutional neural networks \cite{cong2019quantum}, Bayesian deep learning \cite{zhao2019bayesian}, reinforcement learning \cite{dunjko2016quantum,dunjko2017advances}, etc. 

Another topic that is relevant for accelerator modeling is solving linear systems \cite{harrow2009quantum,clader2013preconditioned,childs2017quantum,bravo2019variational,lee2019hybrid} as well as ordinary differential equations (ODEs) and partial differential equations (PDEs) \cite{leyton2008quantum,berry2014high,arrazola2019quantum,childs2020high,xin2020quantum}. Solving Poisson’s equation and the Vlasov equations — which are often used in the simulation of collective effects such as space charge and beam-beam interactions — with quantum computing is being explored \cite{cao2013quantum,wang2019quantum, engel2019quantum}. From the perspective of programming, quantum computer simulators, as code developing-, debugging-, and testing platforms, are available for almost all mainstream programming languages, e.g. C/C++, Python, Java, Matlab, etc. There also exist some languages specifically designed for quantum computing \cite{MicrosoftQDev}. All these provide the community with the fundamental blocks to build simulation tools for beam and accelerator physics.

To make a problem suitable for quantum simulations, it has to be described by unitary operators so that the system, which may be classical, is mathematically equivalent to a quantum system. Additionally, in quantum computing, all variables are stored in quantum states and reading out the results accurately can be time intensive, which should be carefully managed. Many problems in accelerator modeling and beam dynamic simulations meet the above criteria. The motion of a charged particle, even under collective effects, can often be described as Hamiltonians, the evolution of which can be carried out by unitary operators. Recently, a quantum Schrodinger approach was used to solve the Poisson-Vlasov equations to simulate space-charge effects in high intensity beams~\cite{Qiang2022}. Although particles are used to model beams, in many cases we care more about the macroscopic property of the beam, e.g., emittance, momentum spread, bunch size, luminosity, etc., rather than the individual particle and reading out the results of all the particles can be avoided. 

However, most previous work on quantum algorithms focused on the realization of an algorithm with quantum circuits, but applying the algorithm to solve a practical scientific problem is seldom discussed \cite{dodin2020applications}. Clearly there is a gap between the development of the algorithms and their implementation. To implement the quantum computing technique in accelerator modeling, we need to keep abreast with the latest developments and get involved in algorithm design and program development. We expect, at least in the near future, a quantum computer will not replace a classical computer but will probably work together with one. It is thus probably better for now to focus on how to make new quantum simulation tools that collaborate with the existing pool of accelerator modeling programs. Ideally a protocol will be invented through which the quantum packages could be called by the classical programs. Also needed are innovative analyzing tools to process the simulation results before the reading-out. To achieve these tasks, the accelerator physics problems that will benefit from quantum computing need to be identified and for each of them the proper mathematical model will need to be established, which may be different from the conventional classical one. For this, contributions from experts in both accelerator physics and quantum computing are required.

\subsection{Storage ring quantum computers}
\secauthors{Sandra Biedron, Jean-Luc Vay}

\recommendation{\RecSRQC}

In addition to benefit from quantum computing in the future, storage rings can be used as quantum computing devices themselves \cite{BrownPRAB2020}, with the advantages, over Paul~\cite{PaulRMP1990,JamesAPB1998} or Penning traps~\cite{BrownRMP1986}, of longer coherence time, the ability to store a significantly higher number of ions and the possibility of storing multiple ion chains~\cite{BrownPRAB2020}. 
A storage ring quantum computer (SRQC) could potentially store thousands of ions to form an Ion Coulomb Crystal in a closed trajectory.
For SRQC, we anticipate unique challenges associated with the unprecedented large number of ions in the ICC. For example, on timing the laser pulses to efficiently cool the ICC beam to its equilibrium energy configuration, from which ion manipulations can be initiated.
The solutions to the equilibrium positions of the ICC can be solved numerically for an arbitrary N, but they quickly become lengthy and impractical as N becomes large, from hours to days. A solution that employs AI/ML techniques was described recently that brings the prediction to a few seconds~\cite{HuangIEEE2022}.

SRQC presents a particularly interesting case, where the design of the storage ring can benefit from high-performance computing and AI/ML, to produce quantum computing devices that will open unprecedented capabilities in computing with application to many fields, including beam and accelerator physics and designs themselves.

\section{Computational needs}
\recommendation{\RecComputationalNeeds}

\subsection{Hardware: CPU/GPU time, memory, archive}
\secauthors{Jean-Luc Vay, Axel Huebl}
\recommendation{\RecHardware}





Many activities in accelerator modeling research and design both push the envelope in High-Performance Computing (HPC) as well as High-Throughput Computing (HTC).
Typical HPC problems are spearheaded by plasma-based accelerator modeling, where simulations are limited by the available computational resources, whether in spatial grid resolutions, number of macroparticles, number of time steps, even if using the most powerful supercomputers available at the time of writing~\cite{VayPoP2021}.
Emerging techniques like machine-guided optimization of ensemble studies and AI/ML workflows are often HTC workflows, where many parallel runs need to be coordinated.
Likewise, parallel data analysis tasks are often interactive HTC workflows that furthermore can require interactivity and elasticity of the allocated compute resources (e.g., Jupyter Notebooks).
Over time as computational resources available at leadership-scale computing facilities and individual institutions grow, some prior HPC workflows become HTC workflows as exploration phases evolve into design phases.

Some accelerator software have been or are being ported to GPUs and are being ready to run efficiently on upcoming Exascale supercomputers~\cite{MyersPC2021,Synergia}.
It is expected that alongside GPUs, additional specialized compute hardware such as Field Programmable Gate Arrays (FPGA), reduced instruction set (RISC) multicore-CPUs, tensor-processing units (TPUs) and specialized neural engines will be part of near-term computing architectures.

While the mentioned improvements on GPUs will increase the fidelity and accuracy of the simulations, the bar of fidelity and accuracy to aim at is being raised at the same time. 
The needs in computational hardware are expected to increase in the foreseeable future, whether to simulate up to every real particle, including in the halo, of beams that can be composite of many bunches, to account with high fidelity of collective effects in complex geometries or to simulate beams for very long times in conventional, plasma-based or structure-based wakefield colliders.
While there is an increasing amount of efforts aimed at speeding up simulations with surrogate models, the training of these models will rely on simulations, in addition to experimental data.
Producing the large volume of data from simulations will require tremendous needs in computational power, memory and archiving capabilities.

\subsection{Software performance, portability and scalability}
\secauthors{Axel Huebl, David Sagan}
\recommendation{\RecSoftware}

\addtocontents{toc}{\protect\setcounter{tocdepth}{1}} 
\subsubsection{Performance}
\addtocontents{toc}{\protect\setcounter{tocdepth}{2}} 
Performance increase for modeling software is generally desirable, even outside of large-scale HPC runs, and improves overall scientific productivity when studying particle accelerators at all scales.
For instance, most optimization and machine-learning workflows are GPU-accelerated as well to increase computational throughput for studies, while ultraprecise numerical schemes can often use GPU-accelerated linear algebra and FFT routines; thus GPU acceleration also benefits small model sizes.

Developing software for multiple GPU-platforms in parallel to existing CPU architectures is an undertaking that requires the redesign of many existing algorithms and code-bases.
Programming models and languages evolve rapidly and the need to port algorithms to significantly more fine-grained parallelism is a good incentive to rewrite and adopt modern software practices and popular programming languages.


\addtocontents{toc}{\protect\setcounter{tocdepth}{1}} 
\subsubsection{Portability}
\addtocontents{toc}{\protect\setcounter{tocdepth}{2}} 
Portability of implementations is important as many scientists work with three major platforms (Linux, macOS, Windows), although the former two (Unix-like) variants are certainly dominating with HPC developers and machines.
An additional challenge arises with the diversity in GPU and other compute hardware, which ideally should be addressed with \textit{performance-portable} software design.
In such a design, an algorithm is ideally implemented only in a single, abstract form and then specialized/optimized for the specific hardware needs at compilation-time.

Performance portability relies on using and contributing to standard, industry-supported programming languages to allow such a single-source approach.
Performance portability layers (e.g., Kokkos~\cite{Kokkos}, Alpaka~\cite{Alpaka}, Raja~\cite{Raja}) are currently implemented predominantly as modern C++ libraries. The Julia language natively supports coroutines, multi-threading, distributed and GPU computing.  Other languages lacking significantly in release time, compiler variety and robustness.
It is to be expected that performance portability architecture will evolve into standardized programming language components, such as ISO C++, in coming years.
HEP modeling needs can provide important input to their standardization by openly sharing critical community use cases and codes and interacting directly with computer scientists for development.

Common numerical and data management libraries (e.g., AMReX~\cite{AMReX}, CoPA~\cite{Copa}) build on such performance portability layers and only as the next steps, domain-specific libraries and applications are implemented to benefit from all aforementioned developments.

\addtocontents{toc}{\protect\setcounter{tocdepth}{1}} 
\subsubsection{Parallel scalability}
\addtocontents{toc}{\protect\setcounter{tocdepth}{2}} 
Parallel scalability on leadership-scale machines continues to be only achievable with additional parallelization over multiple coupled compute nodes (each powered with CPUs or GPUs).
Methods that (a) split the computational domain into local sub-problems, and (b) ideally overlap communications (usually of data in guard cell regions) between collaborating nodes with computation within each node, are essential for such capability simulation runs.
Furthermore, at large-scale, load balancing of particle-based methods as well as refinement of fidelity needs (e.g., adaptive mesh-refinement) of the simulation domain become important to make optimal use of available resources.


Looking at the rapidly evolving HPC landscape, many themes in advancing accelerator and beam modeling software are common, e.g., Poisson solvers, robust particle pushers, need to model hybrid particle accelerators combining conventional and advanced accelerator elements, etc.
It is thus evident that collaborative efforts between conventional and advanced accelerator development can benefit the community, relying for instance on similar or compatible low-level parallelization layers, but also coordination when implementing and sharing domain-specific numerics can ensure that scalable solution can be readily extended for research needs.

\subsection{Scalable I/O and in-situ analysis}
\secauthors{Axel Huebl, Jean-Luc Vay}
\recommendation{\RecIO}

While modern programming models, software design and advanced algorithms ensure that simulations can execute efficiently, one further challenge emerges to scientific productivity on HPC hardware:
When comparing the computational peak performance evolution to the parallel system storage bandwidth of HPC machines as in Table \ref{table:olcf}, the gap between potential simulation fidelity that can be computed and data that can be stored for analysis widens with every new machine generation.

\begin{table}[!ht]
 \centering
 \begin{tabular}{||c c c c||} 
 \hline
 System Specs & Titan & Summit & Frontier \\ [0.5ex] 
 \hline\hline
 Peak Performance & 27 PFLOP & 200 PFLOP & $>$1.5 EFLOP \\ 
 \hline
 Storage Bandwidth & 1 TB/s & 2.5 TB/s & 2-4x Summit \\
 \hline
 Ratio & 27:1 & 80:1 & 150-300:1 \\
 \hline
 \end{tabular}
 \caption{Evolution of leadership-scale machines at the Oak Ridge Leadership Computing Facility with respect to peak performance and storage bandwidth, according to \href{https://web.archive.org/web/20210715000401/https://www.olcf.ornl.gov/frontier/}{https://www.olcf.ornl.gov/frontier/}.}
 \label{table:olcf}
\end{table}

Following this trend, modeling applications that run at a system's capability and follow the traditional workflow of simulation + post-processing will spend more of their time in large-scale data input and output, which eventually can dominate the time-to-solution.
Driven by this trend, several steps can be taken by simulation developers.

As a first step, adopting scalable, parallel I/O libraries with adequate data layouts \cite{Wan2021} can extent the applicability to truly make optimal use of parallel filesystems \cite{Huebl2017}.
Implementing these libraries in both data producing as well as data consuming codes opens another possibility to standardize on and share the community solutions, as demonstrated in the openPMD project \cite{openPMD}.

As a second step and long-term solution, data analysis and processing need to be transitioned to in situ-processing, running online with the modeling simulation and avoiding traditional, file-based long-term storage for raw data.
This approach essentially moves the design of data analysis to the setup and design phase of a simulation that, similar to an experimental setup, needs to plan the locations, acceptance and dynamic ranges of ``virtual diagnostics'' before even starting a high-fidelity modeling run.
Virtual diagnostics can reduce data sizes and thus data throughput needs by orders of magnitudes, e.g., by calculating beam momenta, phase space histograms, spectra and even 3D videos as a simulation runs instead of storing realistically trillions of particles in regular intervals to disk for later analysis.
While in-situ data processing and analysis are not new, and have even been mainstream in some accelerator codes or frameworks (e.g., Warp~\cite{VayCSD12}, PIConGPU~\cite{PIConGPU2013}), the discrepancy between peak performance and storage bandwidth reported in Table \ref{table:olcf} triggers the development of community libraries that enable a more systematic description of new diagnostics in an in-situ approach, ensuring high efficiency and scalability.

A challenge arises in the rapid setup of in situ processing pipelines, since typical approaches of implementing a virtual detector in the simulation code itself can be more complicated compared to scripted (serial) analysis workflows on files.
Innovative software design can potentially alleviate this challenge, with approaches such as streaming data with close-to-file-like analysis syntax \cite{Poeschel2021} or embedding of interpreters into running simulations to execute scripts on in-memory data.
Continued research in this area and close collaboration with computer science and engineering teams will be essential to ensure that flexible, science-case specific analysis can be designed by accelerator and beam physicists.

\section{Sustainability, reliability, user support, training}
\secauthors{Axel Huebl, R\'{e}mi Lehe, David Sagan, Christopher Mayes, Jean-Luc  Vay, Kiersten Ruisard, Steven Lund}

\recommendation{\RecSustainability}

An important aspect of the development of scientific modeling tools is to ensure that the corresponding software is robust, easy to use and can be extended for new science cases. This is made more difficult by the fact that these tools are oftentimes constantly evolving, with new capabilities being continuously added to meet ever changing computational needs. Here we mention a few important practices \cite{LOI_BestPractices} that facilitate this process.

Accelerator simulation is a large, complex topic, and much time and effort has been spent in developing simulation software. Nevertheless, in the field of accelerator physics, simulation code development has often been a haphazard affair. Due to developers retiring or moving on to other projects, numerous simulation programs have been completely abandoned or are seldom used.

As simulation programs become more complex due to the ever-increasing demands placed upon machine performance, the situation will become worse if not addressed. Such demands include the accelerator and beam physics Grand Challenges that have been identified recently by the community~\cite{LOI_ABPRoadmap,LOI_eva}
\begin{itemize}[itemsep=-1mm]
\item  Increasing beam intensities by orders of magnitude.
\item  Increasing the beam phase-space density by orders of
magnitude, towards the quantum degeneracy limit.
\item Complete and highly accurate start-to-end “virtual particle accelerators” simulations.
\item Fast and accurate multi-objective optimization methods to speed up the design process.
\end{itemize}

Accelerator and beam modeling software development should allow for extensive testing of new functionality while preserving demonstrated capabilities on previously validated scenarios\cite{LOI_BestPractices}. Performance and interoperability must be constantly improved to increase understanding (multi-physics problems) and optimization (machine learning). One must also ensure a transfer of knowledge over generations of scientists in form of formal education and easy accessibility to the tools.

Addressing these Grand Challenges will require a community effort to coordinate and modernize the current set of modeling tools, with capabilities that extend far beyond what the current software can do, including interdisciplinary simulations and advanced models for virtual prototyping of complete accelerator systems.

\subsection{Code robustness, validation \& verification, benchmarking, reproducibility}

\recommendation{\RecRobustness}

\begin{figure}[htb]
    \centering
    \includegraphics[width=5in]{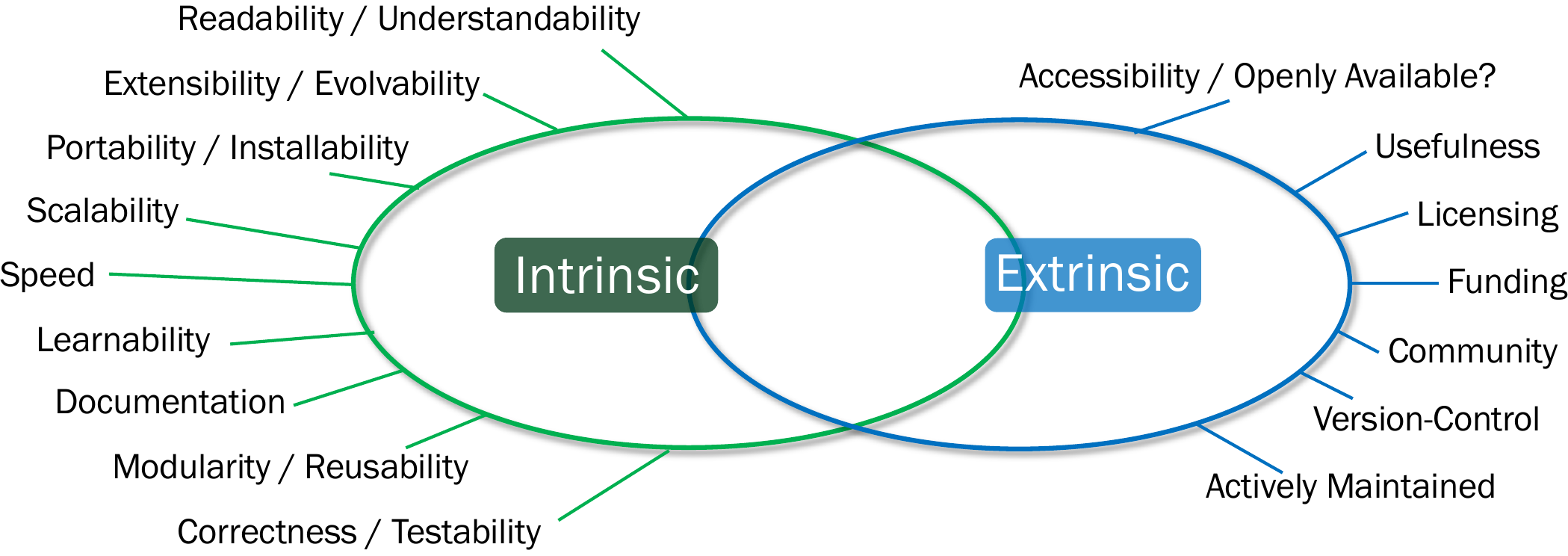}
    \caption{There are a number of aspects that make software sustainable. Broadly, they can be grouped into the ``intrinsic'' characteristics of the software itself and the ``extrinsic'' environment in which the software is developed and used.}
    \label{f:ss}
\end{figure}

There are several aspects that must be addressed to enable the development of the quality software that will be needed for the machines of tomorrow. One facet can be put under the rubric of ``software sustainability'' which can be defined as~\cite{katz}:
\begin{quotation}
``the capacity of the software to endure. In other words, sustainability means that the software will continue to be available in the future, on new platforms, meeting new needs.”
\end{quotation}

There are many aspects to software sustainability, as illustrated in Fig~\ref{f:ss}. Broadly, these aspects can be grouped into the ``intrinsic'' characteristics of the software itself and the ``extrinsic'' environment in which the software is developed and used. Software sustainability has been studied academically and there is even a Software Sustainability Institute~\cite{ssi}, which promotes ``the advancement of software in research by cultivating better, more sustainable, research software to enable world-class research''.

As mentioned above, many software packages developed for simulating accelerators,
while showing excellent results for their specific application and their era,
are not ``sustainable'' in the long run. However, software sustainability is extremely important given the limited resources 
that the accelerator community has for code development in conjunction with the even more limited resources 
for maintaining codes. To meet future needs, it is imperative that there is a community wide effort to promote sustainable practices.

Because modeling tools are being continuously improved and extended by developers, there is always a risk that a given change to the source code may introduce bugs and produce erroneous results in some specific cases. It is therefore paramount that the developers track the changes to the source code (with a version control tool such as \texttt{git}), and that they simultaneously maintain a comprehensive suite of tests that the code should always successfully pass. In the case of scientific modeling codes, these tests can consist of a number of physical setups for which there are well-known theoretical predictions. It can thus be checked that the results of the code conform to these predictions. In order for these tests to be most effective, they need to be run \emph{automatically} whenever the source code is modified -- a process known as \emph{continuous integration}.

In addition to validating against theoretical predictions, it is also critical to benchmark against experimental observations. 
The relevance of code predictions to accelerator performance and output may be limited by a number of factors, such as: 
\begin{itemize}[itemsep=-1mm]
    \item results from experiments that are not fully characterized (e.g., magnet misalignments)
    \item initial beam distribution is unknown or incompletely known (e.g., beam phase-space projections from experiments may not allow full reconstruction of beam 6-D phase-space accurately)
    \item diagnostics or simulations with insufficient range or resolution to fully characterize relevant parameters (e.g., phase space diagnostics that are sensitive to halo / beam loss levels~\cite{LOI_halo_loss})
    \item missing or incompletely described physics in simulations
\end{itemize}
Proper benchmarking of codes against theory and experiments is time-consuming and requires dedicated resources. In general, there are few resources dedicated to code benchmarking at experimental facilities and in simulation teams. As the complexity of codes and the computers that they run on increase, it is essential to ensure enough resources for proper benchmarkings, including dedicated beam time and test stands for well-characterized experimental results~\cite{LOI_halo_loss}.

In addition, the development of large-scale scientific tools nowadays involve a whole community of developers and users rather than a single, isolated developer. Therefore, the robustness and quality of the software is generally greatly improved by the adoption of collaborative development platforms. These online platforms allow the developers to easily and efficiently review each other's changes to the source code, before these changes are incorporated in the mainline version of the code. They also provide a central venue for users to raise issues and questions, and generally communicate their needs with the developers.

Finally, another aspect of a code's robustness is its interface with the user. As the code evolves, the interface with the user often needs to be modified so as to enable functionalities or generalize existing ones. It is not uncommon that these changes break some of the users' scientific workflows. Thus, it is important for the developers to regularly publish releases of the source code, where these changes are clearly documented. The different releases of a given software can be made citable and available through an online archival service, so that users can roll back to a previous release if needed and scrutinize changes in a reproducible manner.
\subsection{Usability, user support and maintenance}

\recommendation{\RecUsability}

\addtocontents{toc}{\protect\setcounter{tocdepth}{1}} 
\subsubsection{Usability}
\addtocontents{toc}{\protect\setcounter{tocdepth}{2}} 
In order to maximize the impact of a given software tool on its scientific community, it is also key that this tool be user-friendly. For example, a common obstacle to the adoption of some scientific tools is their complex installation procedure, especially if many dependencies are involved. 
The developers can drastically simplify the installation procedure of a given software tool by making it available through modern package managers \cite{spack, conda, pip}. These package managers can automatically download and compile the source code as well as its set of dependencies. They can also ensure that compatible versions of the dependencies are being used. Importantly, some of these package managers can allow users to have the exact same installation workflow on large-scale HPC resources as on their desktop workstation.

Another key aspect of the user experience is the code documentation. For scientific modeling tools, the documentation usually includes sections on how to install and use the code, as well as a description of algorithm being used along with relevant references. Since, again, scientific tools are often continuously evolving, it can be beneficial that the documentation be generated automatically, whenever the source code is modified or whenever a new release is published.
Just as the modeling tools themselves, documentation should be easily accessible for the whole community and easy to improve from developers and users alike.

\addtocontents{toc}{\protect\setcounter{tocdepth}{1}} 
\subsubsection{User support}
\addtocontents{toc}{\protect\setcounter{tocdepth}{2}} 
In order for scientific codes to continuously improve and meet the 
needs of the community, it is important to have a communication channel for users to provide feedback~\cite{LOI_BestPractices}. This feedback may include potential bug
reports, questions on installation issues, and inquiries on typical usage.
Again, many tools exist, including issue trackers (e.g., GitHub issues \cite{GithubIssues}), chat rooms (e.g., Gitter \cite{Gitter} or Slack \cite{Slack}), mailing list, forums, stack-overflow, etc.
These different tools have different purposes and target different segments of the user pool. The aim is thus certainly not for a developer to embrace all of them simultaneously, but rather to select one or two channels that are best suited for their particular code. These communication channels should also be clearly identified in the documentation.

It can also be helpful for these communications channels to be openly accessible and easily searchable, so as to avoid that multiple users report the same issues.
In this regard, some of the conversations in chat rooms may sometimes have to be transferred to an issue tracker, or summarized in the documentation.

Open user support channels establish significant synergies not only across related accelerator modeling projects, but with the computational physics community as a whole.
For instance, documented computational and numerical issues, testing approaches, chosen conventions and implemented automation frameworks add significantly to the effectivity of the community, in terms of development speed, issue triage, installation, solution/work-around propagation, communication with high-performance computing centers and adoption of best-practices, to name a few.

\addtocontents{toc}{\protect\setcounter{tocdepth}{1}} 
\subsubsection{Maintenance}
\addtocontents{toc}{\protect\setcounter{tocdepth}{2}} 

\paragraph{Version control and software releases -}

It is now widely accepted, in the scientific community, that any significant software development project should be managed with a version control system (e.g., \texttt{git} \cite{git}). However, a less common practice is that of regularly publishing new releases of the code.

Code releases are important because scientific codes are constantly evolving and improving, and, as part of this process, the code interface with the user may change, bugs may be fixed, and new bugs may occasionally be introduced. It is therefore not uncommon for a new version of the code to suddenly break a workflow that a user previously relied on. Well-documented releases (e.g., with a \texttt{CHANGELOG} document) help the user understand how the code has evolved with each release, and how to update their workflow to adapt to these changes. Archiving each release (e.g., with tools such as Zenodo \cite{zenodo}) may also allow the user to roll back to a previous version of the code, if a new version is altogether incompatible with a prior workflow.
Also here, readily available services can speed up the generation of \texttt{CHANGELOG} documents and publication with package managers through so-called automated deployments.

\paragraph{Access and licensing -} over the last decade, it emerged that scrutinizable implementations and low-barrier contributions to accelerator modeling software are highly desirable for scientific productivity.
Consequently, a significant part of the community considers open source development \cite{LOI_open,LOI_reproducibility,Katz_SWSustainability2020} and permissive open source licensing \cite{OpenSource,FreeSoftware} the preferred way for software access, contribution, teaching, collaboration and reuse \cite{LOI_open,LOI_BestPractices,LOI_barriers,Chen2018}.
For a few modules where such a model is not possible, e.g., due to export control constrains, other community sharing methods such as limited access through a memorandum of understanding (MoU) exist.

\paragraph{Automated testing -}

In order to minimize the risk of new bugs, it is important to verify that the code still works as expected, whenever the source code is modified. 
In the case of scientific codes, this can be done for instance by comparing the results of the code against known solutions, for a number of analytically-tractable problems. This process has long been done by hand, but experience shows that it is time-consuming and may in practice be often omitted by the developers.

Instead, these tests can be done automatically and systematically, whenever the source code is modified. Automated tests usually also results in faster code development, since the different contributors to the code can be rapidly assured that their changes do not break previous functionalities. A number of tools that are free to open source software allow to easily setup those automated tests, including GitHub Actions \cite{githubactions}, Azure Pipelines \cite{azure} and Travis-CI \cite{travisCI}. Automated test on exotic hardware or HPC platforms is intricate, but addressed for instance by the E4S \cite{e4s} effort within the Exascale Computing Project.
\subsection{Training and education}

\recommendation{\RecTraining}

Training and education are key to the long term viability of positions put forth in this white paper. University classes cover the core of essential programming skills, numerical algorithms, AI/ML, and some software development and maintenance tools. Traditionally workers in the field rely heavily on online documentation and training materials as well as help from colleagues. Much benefit is derived from the larger plasma and EM modeling communities. Accelerator science and engineering also relies on specialized training provided by the US Particle Accelerator School (USPAS). The USPAS convenes two, intensive-format sessions per year to provide graduate-level training in topics that are not practical to teach regularly at universities. Recommendations for enhancing the USPAS are covered in a separate Education, Outreach, and Diversity white paper~\cite{Snowmass_WP_EOD}. Specific to accelerator modeling, USPAS offerings have been primarily centered on code demonstrations and applications in various accelerator dynamics, design, and applications classes. Recent classes have covered optimization and ML, python/matlab applications, and levels of self-consistent modeling in beams and plasmas. It would benefit the community if more offerings were regularly available in the USPAS. The USPAS classes are also an excellent opportunity to foster community cooperation via team teaching efforts and to help ingrain productive collaborative work approaches in the younger generation. Focused topical collaborations centered on classes needed could also be used to generate needed tutorial materials as well as disseminate and archive useful code tools while demonstrating the advantages of effective use in classes. Tutorials and archives of materials from the courses would improve modeling efforts and encourage increased productivity via sharing.

\section{Toward community ecosystems \& data repositories}
\secauthors{Jean-Luc Vay, Axel Huebl, R\'{e}mi Lehe, Warren~B.~Mori, Frank Tsung, David Bruhwiler, Ji Qiang, David Sagan}

\recommendation{\RecEcosystem}

\subsection{Loose integration: Integrated workflows}\label{subsec:workflows}

\recommendation{\RecWorkflows}

Integrated workflows are often needed to answer complex contemporary research questions, such as the start-to-end description of a hybrid accelerator with conventional and plasma elements.
Furthermore, all optimization, AI/ML training as well as exploratory studies benefit from establishing and maintenance of workflows.
These workflows might (a) link several codes together to solve a larger problem or solve a problem with more physical processes, (b) benchmark and validate the various codes against each other and known solutions, and (c) establish reproducibility for new scientific results.

\subsubsection{I/O standardization}
At the moment, many simulation codes are developed independently with little coordination between various groups.
In order to establish workflows that span multiple applications, a typical approach so far is to rely on file-based data exchange and code-specific high-level wrapping of input options.
Along these lines, the Consortium for Advanced Modeling of Particle Accelerators (CAMPA~\cite{CAMPA}) supports (among other activities) the development of standardized input and output formats through the openPMD~\cite{openPMD} and PICMI~\cite{PICMI} projects.

The Open Standard for Particle-Mesh Data (openPMD) is a meta-data standard for research data, adding scientific self-description on top of modern, portable, scalable file-formats such as ADIOS~\cite{ADIOS2} and HDF5~\cite{HDF5}.
openPMD achieves this by collaboratively defining an open standard document that evolves in versions, similar to software.
Community members then implement openPMD data handling in their simulation codes and analysis types either directly against supported file formats or use an open source library layer, openPMD-api~\cite{openPMDapi}.

The standard input format for Particle-In-Cell codes (PICMI) strives to unify the usage of accelerator modeling codes by standardization on an API for simulation design.
Conceptually, PICMI defines a common, explicit PIC modeling interface (API) for a wide range of numerical and initialization options.
A PICMI script can be run with multiple codes, given that both implement the requested features, and simplifies comparability of physics cases tremendously.
Code ``backend'' choices can be picked based on performance or numerical needs, without breaking context for the user and relearning the simulation design for each specific application.

\begin{figure}[tb]
    \centering
    \includegraphics[width=6in]{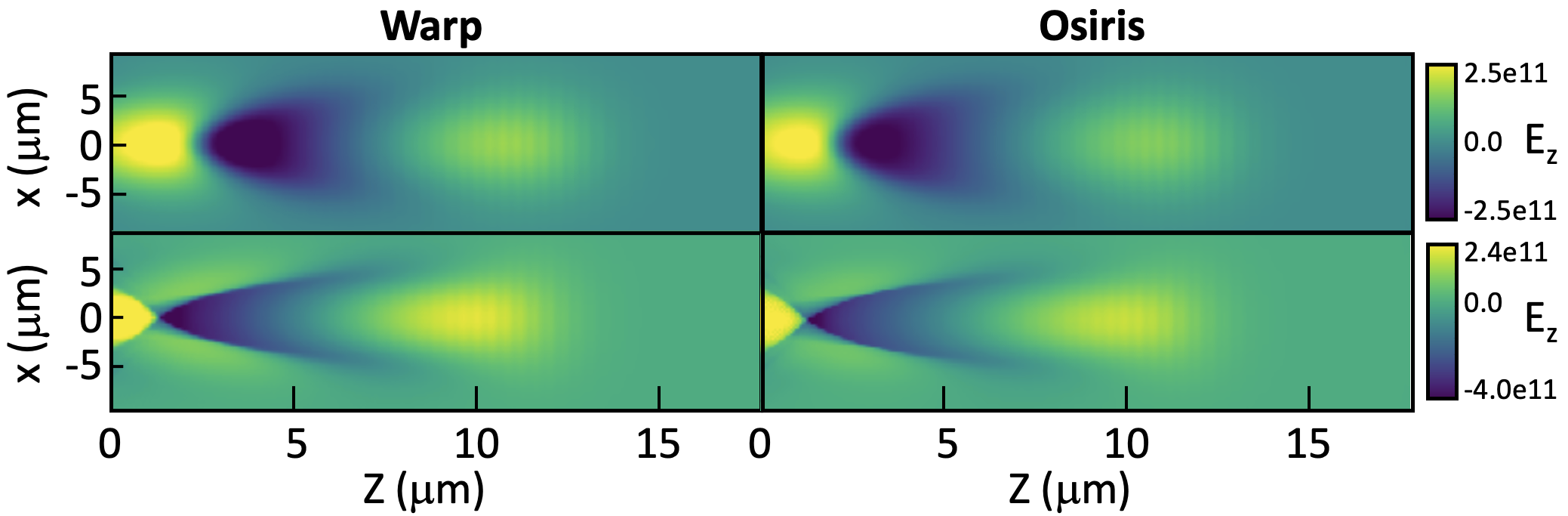}
    \caption{Longitudinal electric field (in V/m) in a laser-driven plasma acceleration stage at two times (top: $t\approx 300$fs, bottom: $t\approx 600$fs) along the laser propagation
from 2-D PIC simulations with: (left) Warp; (right) Osiris. Plots are based on rendering from the openPMD-viewer.}
    \label{f:warp_osiris_openpmd}
\end{figure}

As part of the CAMPA efforts to foster compatibility and coordination between particle accelerator modeling applications, a code-validation workflow between the Warp~\cite{VayCSD12} and Osiris~\cite{Osiris} code has been carried out.
In the study on the two codes in Figure~\ref{f:warp_osiris_openpmd}, both independent code bases modeled the same physical setup of an LWFA (with a laser driving a wake in a plasma column) and used standardized openPMD output and the same, community-developed post-processing tool (openPMD-viewer) to analyze the data.
A one-to-one comparison to machine precision between the two codes was beyond the scope of the study because: (a) while using the same physical parameters to initialize the simulations with each code, some secondary parameters were left unspecified such as, e.g., the exact phase of the laser oscillations, (b) some numerical aspects are different such as, e.g., the method of initialization of the laser in the boosted frame.
Nonetheless, it is rewarding to observe on Figure~\ref{f:warp_osiris_openpmd} that the two codes predict very similar wake evolutions and dephasing of the laser as it propagates through the plasma column.

Another series of comparisons is underway that also incorporates a common Python input script using PICMI, and including more codes, in the workflow, to validate the workflow concept for further adoption and development.

Another example is the integrated electromagnetics and beam dynamics simulation on high performance computers using the ACE3P and the IMPACT codes~\cite{bizzozero2021}. In this case, the design of the RF cavity using the ACE3P and the beam dynamics simulation using the IMPACT were integrated into an optimization workflow for final beam quality optimization.

\subsection{Tighter integration: Integrated frameworks}

\recommendation{\RecFrameworks}

Integrated developments of frameworks could implement more tightly coupled workflows on a library level instead of relying on high-level application coupling.
As presented at the beginning of the section in the concept of an modular software ecosystem in~\cite{ICFAVay2021}, modular, cross-cutting software designs can provide reusability, efficient development investments, and compatibility.
Built-in features of lower-level software such as GPU-support, multi-node parallelism, translation to AI/ML frameworks and multi-physics support can, if evolved together, be inherited by a wide community.
In particular, common or compatible in-memory data structures have not been attempted in accelerator and beam modeling so far, but would provide significant performance and software maintenance benefits for tightly-coupled, co-developed physics modules for integrated modeling frameworks that address grand challenges.

Several community examples exist to demonstrate modular development.
For instance, the codes WarpX~\cite{VayPoP2021}, HiPACE++~\cite{Diederichs2022} and PIConGPU~\cite{PIConGPU2013} as well as several analysis frameworks share the development of the I/O library openPMD-api~\cite{openPMDapi}, which provides access to high-performant, portable low-level I/O libraries and a standardized metadata schema \cite{openPMD}.
WarpX, FBPIC implement PICMI assisted by a shared, central interface library, with implementation underway in other codes (e.g., OSIRIS).
WarpX, HiPACE++~\cite{HipacePP} and related codes share the data structures from AMReX~\cite{AMReX} and parts of their code bases.
\subsubsection{Community development}
The adoption of a more coordinated and collaborative approach is also driven by the need to transition a large body of software from CPUs to GPUs, which is more disruptive than most past transitions, including the one from serial to parallel codes. Furthermore, this latest transition is anticipated to be one of many transitions to come to adapt to the rapidly evolving computer hardware.

More realisms, better numerics and advanced algorithms are continuously being added to many accelerator codes, and code developers have developed best practices for adding new modules while maintaining the performance needed to run on state-of-the-art HPC centers around the world. 
Thus, instead of porting solely to GPUs, accelerator modeling teams already teamed up with computer scientists and embrace libraries that abstract computer hardware specific details for performance and sustainability.
Establishing more community standards, funding reliable software dependencies and reusable physics modules are the logical steps to speed up development of code bases further.

An efficient mean for guiding community development is the collaborative setting of and adherence to common policies.
Rather than starting from scratch, one can adopt an existing libraries and codes and build a beam and accelerator modeling ecosystem on it.
As a blueprint, one could build on community development concepts of the Interoperable Design of Extreme-scale Application Software (IDEAS) project \cite{IDEAS} and its Extreme-scale Scientific Software Development Kit (xSDK) \cite{xSDK}.

The main goal of the IDEAS project is to ``help move scientific software development toward an approach of building new applications as a composition of reusable, robust, and scalable software components and libraries, using the best available practices and tools.''\cite{IDEAS}, while the 
xSDK project is being developed to ``provide a coordinated infrastructure for independent mathematical libraries to support the productive and efficient development of high-quality applications'' \cite{xSDK}:

\begin{quote}
\small
Rapid, efficient production of high-quality, sustainable extreme-scale scientific applications is best accomplished using a rich ecosystem of state-of-the art reusable libraries, tools, lightweight frameworks, and defined software methodologies, developed by a community of scientists who are striving to identify, adapt, and adopt best practices in software engineering. The vision of the xSDK is to provide infrastructure for and interoperability of a collection of related and complementary software elements—developed by diverse, independent teams throughout the high-performance computing (HPC) community—that provide the building blocks, tools, models, processes, and related artifacts for rapid and efficient development of high-quality applications.
\end{quote}

Although xSDK targets ``extreme-scale scientific applications'' of relevance to exascale supercomputing, its derived community policies apply well to all scales of computing, from laptops to clusters or Cloud computing.  Hence, the paradigm and tools are readily applicable to the full set of modeling needs of the particle accelerator community.

Individual packages in such an ecosystem are composable libraries, frameworks, and domain components developed by individual groups in the community.
Each package publishes technical design documents, documentation (reference, tutorials, how-to guides), API definitions, and a concrete implementation \cite{LOI_BestPractices}.  In order to solve a specific physics case, a typical application uses functionalities from several packages, often in an innovative way.  

There are many advantages to adopting policies and tools such as the IDEAS/xSDK:
\begin{itemize}[itemsep=-1mm]
   \item Community policies have been established over years by teams of specialists in scientific software development and computing, and they include best practices in software development.
    \item For reusable components, the policies require the use of {\it permissive open source licenses} (``Non-critical optional dependencies can use any OSI-approved license.'') that allow reuse of source and binary code, including for commercial and proprietary applications, fostering collaborations across laboratories, academia and industrial partners.
   \item A wide-ranging set of open source, interoperable tools from a variety of backgrounds (including numerical solvers, e.g., Hypre, multi-parameter optimizer, e.g., LibEnsemble, and mesh-refinement frameworks, e.g., AMReX) can be combined and used as foundation of accelerator and beam physics toolkits and codes.
    \item A wide community of developers that can help improve software capability and sustainability across projects.
    \item Partial overlap in functionality is not problematic and in fact improves the diversity of the whole.
    \item Time of development and maintenance of domain-specific software is greatly reduced, as the domain scientists can concentrate on the domain-specific functionalities while lower-level, cross-cutting, numerical packages are maintained by dedicated specialists.
    \item Portability across platforms (CPUs, GPUs, etc.) of low-level libraries ensures portability of the software that is built upon them.  Building new tools on these low-level libraries greatly reduces the overall burden on the community for porting codes to new platforms.
    \item The domain-specific packages that are built, following established policies are portable across platforms and interoperable, and can  be further combined to form larger toolkits and codes.
    \item Since packages are reused across software and duplication is minimized, bugs and inefficiencies are spotted earlier.
    \item Many codes, libraries and packages exist in the community that can be reused as building blocks, and progressively modernized and blended in a module that is portable and runs efficiently on modern CPUs and GPUs. Hence, not all functionalities have to be rewritten from scratch at once, offering a progressive path from the current status to an ecosystem of accelerator modeling tools. 

\end{itemize}

It is to be emphasized that the goal is not to propose that only one code be developed to study, e.g., RF accelerators or plasma-based accelerators. 
On the contrary,
the goal is to provide a coordinated infrastructure that enables inclusive and collaborative development of modeling tools for the community, and for these tools, to follow modern practices for scientific software: be portable, efficient, robust and leading to consistently accurate and reproducible results.

\subsection{Data repositories}

\recommendation{\RecRepositories}

In current community practice, the reporting and archival of many modeling results are often limited to prose and figures in papers.
As accelerator systems and support for experimental campaigns become more complex, it is desirable to establish reuse of modeling results, especially if simulations themselves made use of significant HPC resources.

Similar to needs in HEP detector data \cite{SM_whitepaper_HEP_DATA}, it is desirable to systematically categorize and share accelerator data, such as beam distributions, source information, field maps, among others.
The benefits of such archives are manifold: sharing data enables preservation, recasting, reinterpretation \cite{SM_whitepaper_HEP_DATA} and meta-analysis of published results; archives establish training sets for AI/ML workflows, calibration and benchmark cases for new models and theories, beam transport from one model (and code) into another, design of hybrid particle accelerators combining conventional and plasma elements/sources, accurate input for modeling at interaction points, etc.
The standardization of data and meta-data (see previous section) for machine-readability, findability, accessibility and reuse would maximize the productivity when sharing data over such repositories.

Following Open Science practices \cite{OpenScienceUNSECO,OpenScienceEU}, such data archives and repositories should be established in an open access manner \cite{OpenAccessUNSECO}, similar to existing data repositories in particle physics experiments \cite{CERNdata,FAIRdata}.
Lastly, open science compatible data policies need to be advanced on the national, laboratory and university level.
For instance, as of today data sharing in the U.S. Department of Energy laboratories is bound to first assert (often individually) copyright for all but raw data, which can produce significant workflow overheads for scientists and entry burdens to contribute to community data repositories.
The policies for universities vary with individual states.
Clear advise also needs to be established for corner-cases such as publishing interactive notebooks and code snippets, which traditionally fall under software copyright and publishing policies, with data sets in a single and time-effective authorization and license.


\subsection{Centers \& consortia, collaborations with industry}

\recommendation{\RecCenters}

\addtocontents{toc}{\protect\setcounter{tocdepth}{1}} 
\subsubsection{Centers \& consortia}
\addtocontents{toc}{\protect\setcounter{tocdepth}{2}} 

It quickly becomes clear that, in order to achieve what is described and 
proposed in the earlier sections, a coherent and consolidated
effort is needed. This is best achieved in the form of dedicated Centers 
for Accelerator and Beam Physics Modeling \cite{LOI_center}.
Other areas of computer science have already
embraced this. New colleges for computing are established at universities
to consolidate the dispersed computing efforts of the various departments 
(e.g., MIT's Schwarzman College of Computing \cite{schwarzman}), 
and new centers for Quantum Computing \cite{quics, ibm:quantum} 
have been built. Exascale computing has been embraced through the Exascale 
Computing Project \cite{exascale}. 
The US Department of Energy (DOE) has founded SciDAC \cite{scidac} to accelerate progress in
scientific computing across the different programs supported by DOE: Advanced Scientific Computing Research, Basic Energy Sciences, Biological and Environmental Research, Fusion Energy Sciences, High-Energy Physics, and Nuclear Physics.

The respective communities have benefited strongly from these new centers and the partnerships across disciplines.

Accelerator and Beam Physics Modeling would no doubt benefit similarly. The 
centers can be at a given location or distributed geographically and among 
institutions across laboratories, academia and industrial partners. 
They would bring together domain scientists (computational accelerator and 
beam physicists), applied mathematicians, computer scientists and software engineers 
with collaborations across the full landscape of accelerator modeling. 
In addition, some of the computer science centers mentioned above are already 
supporting accelerator modeling efforts on which the Centers for Accelerator and Beam Physics Modeling could build.

Depending on the overall size, the centers could enable part or all of the following:
\begin{itemize}[itemsep=-1mm]
    \item Community development and maintenance of codes using industry-standard quality processes by dedicated, specialized teams \cite{LOI_BestPractices}.
    \item Collect libraries for field solvers, particle trackers, and other modules.
    \item Provide a modular community ecosystem for multiphysics particle accelerator 
          modeling and design \cite{LOI_ecosystem}.
    \item Standardize input scripts, output data, lattice description and 
          start-to-end workflows \cite{LOI_standards}.
    \item Provide compatibility layers to use the same libraries and modules in a 
          number of programming languages.
    \item Development and maintenance of End-to-end Virtual Accelerators (EVA) \cite{LOI_eva}. 
    \item User support, high-quality and detailed documentations, 
          online tutorials, and training.
    \item Easy-to-use, standardized, user interfaces for preparation and analysis of simulations.
    \item Automated tools for ensemble simulations for optimization with builtin AI/ML support. 
    \item Suite of test problems with well-characterized solutions for benchmarking, quality assurance and regression testing.
    \item Development, analysis and efficient implementation of novel 
          algorithms and numerical methods (e.g., high-order solvers, symplectic multiparticle tracking, Fast Multipole Methods,
          adaptive mesh refinement).
    \item Providing a space to meet (physically or virtually) for the integration 
          of developments from contributors into larger codes, such as PhD projects 
          from external groups, organizing development hackathons, 
          knowledge-transfer, and onboarding.
    \item Developing and organizing workshops for developers and users of codes alike.
          Inviting national and international speakers/developers (travel/hosting funds).
    \item Interacting with existing schools, by developing and maintaining 
          state-of-the-art educational resources (e.g., tutorials, lectures) on codes.
    \item Exploration of novel use of machine learning for accelerator modeling, 
          and, further in the future, of quantum computing \cite{LOI_ML}.
\end{itemize}

Multiple Centers can be organized through a Consortium (e.g., CAMPA \cite{CAMPA}). Except for special restrictions such as export control, it would be desirable for the software developed by the Center to be open source, enabling crosschecking, testing and contribution by the community at large, beyond the participants to the Center(s) \cite{LOI_open}.

\addtocontents{toc}{\protect\setcounter{tocdepth}{1}} 
\subsubsection{Collaborations with industry}
\addtocontents{toc}{\protect\setcounter{tocdepth}{2}} 

Effective long-term collaboration between national laboratories, academia and industry will lead to important benefits for the entire HEP community. Labs and universities will have access to better software with lower lifecycle costs~\cite{LOI_industry}. Companies will be strengthened by knowledge transfer from labs and universities. Computational scientists will be better able to concentrate on core competencies, without spending time on user interface design, ease of use, cloud computing, etc. Society will reap the benefits of better science, more innovation, and stronger businesses. State-of-the-art simulation codes will become readily available to students. Training time and associated costs will be reduced, as new team members will become productive more quickly. This will contribute to equity, diversity and inclusion (EDI), as barriers to entry are removed for scientists in developing countries and for those at US institutions with less federal funding and no direct access to code developers.

In order to facilitate the necessary collaborations and to provide the entire community with confidence that the software will be widely available and adequately supported over decades, an open source license is required for the industry software and can be very helpful for the entire software ecosystem~\cite{LOI_open}. This imposes an open source business model on the corresponding businesses, at least with regard to this specific activity. The software design objectives must include seamless integration with legacy codes, low barrier to entry for new users, easily moving between GUI and command-line modes, cataloging of provenance to aid reproducibility, and simplified collaboration through multi-modal sharing.

\section{Outlook}
Computer simulations will continue to be essential to particle accelerator research, design and operation. Its relative importance is even expected to grow, thanks to improvements in algorithms, computer hardware, and new opportunities in machine learning and quantum computing. These will enable accelerator modeling capabilities that include more physics that is integrated self-consistently to model accelerators with ever increasing fidelity and accuracy, toward the ultimate realization of the grand challenge of virtual twins of particle accelerators. A more collaborative and coordinated approach that enables the development of community ecosystems, adopting best practices in software developments and maintenance, is needed to meet the challenge in a realistic budget envelope and timeframe.

\printbibliography

@article{VayJCP2013,
    title = {{A domain decomposition method for pseudo-spectral electromagnetic simulations of plasmas}},
    year = {2013},
    journal = {Journal of Computational Physics},
    author = {Vay, Jean Luc and Haber, Irving and Godfrey, Brendan B.},
    pages = {260--268},
    volume = {243}
}

@article{Kallala2019,
    title = {{A generalized massively parallel ultra-high order FFT-based Maxwell solver}},
    year = {2019},
    journal = {Computer Physics Communications},
    author = {Kallala, Haithem and Vay, Jean Luc and Vincenti, Henri},
    month = {11},
    pages = {25--34},
    volume = {244},
    publisher = {Elsevier B.V.},
    doi = {10.1016/j.cpc.2019.07.009},
    issn = {00104655},
    arxivId = {1812.07357}
}

@article{Vay1996,
    title = {{A three-dimensional electromagnetic particle-in-cell code to simulate heavy ion beam propagation in the reaction chamber}},
    year = {1996},
    journal = {Fusion Engineering and Design},
    author = {Vay, J.-L. and Deutsch, C.},
    number = {1-4},
    volume = {32-33},
    doi = {10.1016/S0920-3796(96)00502-9},
    issn = {09203796}
}

@article{NagaitsevARXIV_ABP2021,
    title = {{Accelerator and Beam Physics Research Goals and Opportunities}},
    year = {2021},
    author = {Nagaitsev, S. and Huang, Z. and Power, J. and Vay, J. -L. and Piot, P. and Spentzouris, L. and Rosenzweig, J. and Cai, Y. and Cousineau, S. and Conde, M. and Hogan, M. and Valishev, A. and Minty, M. and Zolkin, T. and Huang, X. and Shiltsev, V. and Seeman, J. and Byrd, J. and Hao, Y. and Dunham, B. and Carlsten, B. and Seryi, A. and Patterson, R.},
    month = {1},
    url = {https://arxiv.org/abs/2101.04107},
    arxivId = {2101.04107}
}

@article{JalasPoP2017,
    title = {{Accurate modeling of plasma acceleration with arbitrary order pseudo-spectral particle-in-cell methods}},
    year = {2017},
    journal = {Physics of Plasmas},
    author = {Jalas, S. and Dornmair, I. and Lehe, R. and Vincenti, H. and Vay, J.-L. and Kirchen, M. and Maier, A. R.},
    number = {3},
    month = {3},
    pages = {033115},
    volume = {24},
    publisher = {AIP Publishing LLC},
    url = {http://aip.scitation.org/doi/10.1063/1.4978569},
    doi = {10.1063/1.4978569},
    issn = {1070-664X}
}

@misc{ACE3P,
    title = {{ACE3P}},
    url = {https://portal.slac.stanford.edu/sites/ard_public/acd/Pages/Default.aspx}
}

@misc{AAC_Roadmap_2016,
    title = {{Advanced Accelerator Concepts Research Roadmap Workshop Report}},
    year = {2016},
    url = {https://science.energy.gov/~/media/hep/pdf/accelerator-rd-stewardship/Advanced_Accelerator_Development_Strategy_Report.pdf}
}

@article{Alpaka,
    title = {{Alpaka - An Abstraction Library for Parallel Kernel Acceleration}},
    year = {2016},
    journal = {Proceedings - 2016 IEEE 30th International Parallel and Distributed Processing Symposium, IPDPS 2016},
    author = {Zenker, Erik and Worpitz, Benjamin and Widera, René and Huebl, Axel and Juckeland, Guido and Kn{\"{u}}pfer, Andreas and Nagel, Wolfgang E. and Bussmann, Michael},
    month = {2},
    pages = {631--640},
    publisher = {Institute of Electrical and Electronics Engineers Inc.},
    url = {https://arxiv.org/abs/1602.08477v1},
    doi = {10.1109/ipdpsw.2016.50},
    arxivId = {1602.08477}
}

@misc{AMReX,
    title = {{AMReX}},
    url = {https://amrex-codes.github.io/}
}

@article{Vayjcp01,
    title = {{An Extended Fdtd Scheme For The Wave Equation: Application To Multiscale Electromagnetic Simulation}},
    year = {2001},
    journal = {Journal of Computational Physics},
    author = {Vay, J.-L.},
    number = {1},
    month = {2},
    pages = {72--98},
    volume = {167},
    issn = {0021-9991}
}

@article{Cohennim2009,
    title = {{An Implicit ``Drift-Lorentz{\{}''{\}} Mover For Plasma And Beam Simulations}},
    year = {2009},
    journal = {Nuclear Instruments {\&} Methods In Physics Research Section A-Accelerators Spectrometers Detectors And Associated Equipment},
    author = {Cohen, R H and Friedman, A and Grote, D P and Vay, J -L.},
    number = {1-2},
    month = {7},
    pages = {53--55},
    volume = {606},
    institution = {Tokyo Inst Technol, Res Lab Nucl Reactors; Japan Soc Plasma Sci {\&} Nucl Fus Res; Particle Accelerator Soc Japan},
    doi = {10.1016/J.Nima.2009.03.083},
    issn = {0168-9002}
}

@article{Vaypop04,
    title = {{Application Of Adaptive Mesh Refinement To Particle-In-Cell Simulations Of Plasmas And Beams}},
    year = {2004},
    journal = {Physics Of Plasmas},
    author = {Vay, J.-L. and Colella, P and Kwan, J W and Mccorquodale, P and Serafini, D B and Friedman, A and Grote, D P and Westenskow, G and Adam, J.-C. and Heron, A and Haber, I},
    number = {5},
    month = {5},
    pages = {2928--2934},
    volume = {11},
    doi = {10.1063/1.1689669},
    issn = {1070-664X}
}

@inproceedings{Vaypac09,
    title = {{Application Of The Reduction Of Scale Range In A Lorentz Boosted Frame To The Numerical Simulation Of Particle Acceleration Devices}},
    year = {2009},
    booktitle = {Proc. Particle Accelerator Conference},
    author = {Vay, J.-L. and Fawley, W. M. and Geddes, C. G. R. and Cormier-Michel, E. and Grote, D. P.},
    url = {https://accelconf.web.cern.ch/PAC2009/papers/tu1pbi04.pdf},
    address = {Vancouver, Canada}
}

@techreport{JamesAPB1998,
    title = {{Applied Physics B Lasers and Optics Quantum dynamics of cold trapped ions with application to quantum computation}},
    year = {1998},
    booktitle = {Appl. Phys. B},
    author = {James, D F V},
    pages = {181--190},
    volume = {66}
}

@article{HuangIEEE2022,
    title = {{Artificial Intelligence-Assisted Design and Virtual Diagnostic for the Initial Condition of a Storage-Ring-Based Quantum Information System}},
    year = {2022},
    journal = {IEEE Access},
    author = {Huang, Bohong and Gonzalez-Zacarias, Clio and Guitron, Salvador Sosa and Aslam, Aasma and Biedron, Sandra G. and Brown, Kevin and Bolin, Trudy},
    pages = {14350--14358},
    volume = {10},
    publisher = {Institute of Electrical and Electronics Engineers Inc.},
    doi = {10.1109/ACCESS.2022.3147727},
    issn = {21693536}
}

@article{Vayjcp02,
    title = {{Asymmetric Perfectly Matched Layer For The Absorption Of Waves}},
    year = {2002},
    journal = {Journal of Computational Physics},
    author = {Vay, J.-L.},
    number = {2},
    month = {12},
    pages = {367--399},
    volume = {183},
    doi = {10.1006/Jcph.2002.7175},
    issn = {0021-9991}
}

@article{Vaycpc04,
    title = {{Asymmetric Pml For The Absorption Of Waves. Application To Mesh Refinement In Electromagnetic Particle-In-Cell Plasma Simulations}},
    year = {2004},
    journal = {Computer Physics Communications},
    author = {Vay, J.-L. and Adam, J.-C. and Heron, A},
    number = {1-3},
    month = {12},
    pages = {171--177},
    volume = {164},
    doi = {10.1016/J.Cpc.2004.06.026},
    issn = {0010-4655}
}

@article{Vay1998a,
    title = {{Charge compensated ion beam propagation in a reactor sized chamber}},
    year = {1998},
    journal = {Physics of Plasmas},
    author = {Vay, J.L. and Deutsch, C.},
    number = {4},
    volume = {5},
    doi = {10.1063/1.872648},
    issn = {1070664X}
}

@article{FeiCPC2017,
    title = {{Controlling the numerical Cerenkov instability in PIC simulations using a customized finite difference Maxwell solver and a local FFT based current correction}},
    year = {2017},
    journal = {Computer Physics Communications},
    author = {Li, Fei and Yu, Peicheng and Xu, Xinlu and Fiuza, Frederico and Decyk, Viktor K. and Dalichaouch, Thamine and Davidson, Asher and Tableman, Adam and An, Weiming and Tsung, Frank S. and Fonseca, Ricardo A. and Lu, Wei and Mori, Warren B.},
    month = {5},
    pages = {6--17},
    volume = {214},
    publisher = {North-Holland},
    doi = {10.1016/J.CPC.2017.01.001},
    issn = {0010-4655},
    arxivId = {1605.01496}
}

@article{VincentiCPC2016,
    title = {{Detailed analysis of the effects of stencil spatial variations with arbitrary high-order finite-difference Maxwell solver}},
    year = {2016},
    journal = {Computer Physics Communications},
    author = {Vincenti, H. and Vay, J.-L.},
    month = {3},
    pages = {147--167},
    volume = {200},
    publisher = {ELSEVIER SCIENCE BV, PO BOX 211, 1000 AE AMSTERDAM, NETHERLANDS},
    url = {https://apps.webofknowledge.com/full_record.do?product=UA&search_mode=GeneralSearch&qid=1&SID=1CanLFIHrQ5v8O7cxqV&page=1&doc=2},
    doi = {10.1016/j.cpc.2015.11.009},
    issn = {00104655}
}

@article{VayPOPL2011,
    title = {{Effects Of Hyperbolic Rotation In Minkowski Space On The Modeling Of Plasma Accelerators In A Lorentz Boosted Frame}},
    year = {2011},
    journal = {Physics Of Plasmas},
    author = {Vay, Jl and Geddes, C G R and Cormier-Michel, E and Grote, D P},
    number = {3},
    month = {3},
    pages = {30701},
    volume = {18},
    doi = {10.1063/1.3559483}
}

@article{PaulRMP1990,
    title = {{Electromagnetic traps for charged and neutral particles}},
    year = {1990},
    journal = {Reviews of Modern Physics},
    author = {Paul, Wolfgang},
    number = {3},
    month = {7},
    pages = {531},
    volume = {62},
    publisher = {American Physical Society},
    url = {https://journals.aps.org/rmp/abstract/10.1103/RevModPhys.62.531},
    doi = {10.1103/RevModPhys.62.531},
    issn = {00346861}
}

@article{LehePRE2016,
    title = {{Elimination of numerical Cherenkov instability in flowing-plasma particle-in-cell simulations by using Galilean coordinates}},
    year = {2016},
    journal = {Physical Review E},
    author = {Lehe, Remi and Kirchen, Manuel and Godfrey, Brendan B. and Maier, Andreas R. and Vay, Jean-Luc},
    number = {5},
    month = {11},
    pages = {053305},
    volume = {94},
    publisher = {American Physical Society},
    url = {https://link.aps.org/doi/10.1103/PhysRevE.94.053305},
    doi = {10.1103/PhysRevE.94.053305},
    issn = {2470-0045}
}

@article{YuCPC2015,
    title = {{Elimination of the numerical Cerenkov instability for spectral EM-PIC codes}},
    year = {2015},
    journal = {Computer Physics Communications},
    author = {Yu, Peicheng and Xu, Xinlu and Decyk, Viktor K. and Fiuza, Frederico and Vieira, Jorge and Tsung, Frank S. and Fonseca, Ricardo A. and Lu, Wei and Silva, Luis O. and Mori, Warren B.},
    month = {7},
    pages = {32--47},
    volume = {192},
    publisher = {ELSEVIER SCIENCE BV, PO BOX 211, 1000 AE AMSTERDAM, NETHERLANDS},
    url = {https://apps.webofknowledge.com/full_record.do?product=UA&search_mode=GeneralSearch&qid=2&SID=1CanLFIHrQ5v8O7cxqV&page=1&doc=3},
    doi = {10.1016/j.cpc.2015.02.018},
    issn = {00104655}
}

@article{BrownRMP1986,
    title = {{Geonium theory: Physics of a single electron or ion in a Penning trap}},
    year = {1986},
    journal = {Reviews of Modern Physics},
    author = {Brown, Lowell S. and Gabrielse, Gerald},
    number = {1},
    month = {1},
    pages = {233},
    volume = {58},
    publisher = {American Physical Society},
    url = {https://journals.aps.org/rmp/abstract/10.1103/RevModPhys.58.233},
    doi = {10.1103/RevModPhys.58.233},
    issn = {00346861}
}

@misc{Raja,
    title = {{GitHub - LLNL/RAJA: RAJA Performance Portability Layer (C++)}},
    url = {https://github.com/LLNL/RAJA}
}

@article{GodfreyCPC2015,
    title = {{Improved numerical Cherenkov instability suppression in the generalized PSTD PIC algorithm}},
    year = {2015},
    journal = {Computer Physics Communications},
    author = {Godfrey, Brendan B. and Vay, Jean Luc},
    pages = {221--225},
    volume = {196},
    publisher = {Elsevier}
}

@misc{Kokkos,
    title = {{Kokkos {\textperiodcentered} GitHub}},
    url = {https://github.com/kokkos}
}

@article{Vay2002a,
    title = {{Mesh refinement for particle-in-cell plasma simulations: Applications to and benefits for heavy ion fusion}},
    year = {2002},
    journal = {Laser and Particle Beams},
    author = {Vay, J.-L. and Colella, P. and McCorquodale, P. and Van Straalen, B. and Friedman, A. and Grote, D.P.},
    number = {4},
    volume = {20},
    doi = {10.1017/S0263034602204139},
    issn = {02630346}
}

@article{YuCPC2015-Circ,
    title = {{Mitigation of numerical Cerenkov radiation and instability using a hybrid finite difference-FFT Maxwell solver and a local charge conserving current deposit}},
    year = {2015},
    journal = {Computer Physics Communications},
    author = {Yu, Peicheng and Xu, Xinlu and Tableman, Adam and Decyk, Viktor K. and Tsung, Frank S. and Fiuza, Frederico and Davidson, Asher and Vieira, Jorge and Fonseca, Ricardo A. and Lu, Wei and Silva, Luis O. and Mori, Warren B.},
    month = {12},
    pages = {144--152},
    volume = {197},
    publisher = {ELSEVIER SCIENCE BV, PO BOX 211, 1000 AE AMSTERDAM, NETHERLANDS},
    url = {https://apps.webofknowledge.com/full_record.do?product=UA&search_mode=GeneralSearch&qid=2&SID=1CanLFIHrQ5v8O7cxqV&page=1&doc=2},
    doi = {10.1016/j.cpc.2015.08.026},
    issn = {00104655}
}

@article{Vaypop2011,
    title = {{Modeling Of 10 Gev-1 Tev Laser-Plasma Accelerators Using Lorentz Boosted Simulations}},
    year = {2011},
    journal = {Physics Of Plasmas},
    author = {Vay, J -L. and Geddes, C G R and Esarey, E and Schroeder, C B and Leemans, W P and Cormier-Michel, E and Grote, D P},
    number = {12},
    month = {12},
    volume = {18},
    doi = {10.1063/1.3663841},
    issn = {1070-664X}
}

@article{VayPoP2021,
    title = {{Modeling of a chain of three plasma accelerator stages with the WarpX electromagnetic PIC code on GPUs}},
    year = {2021},
    journal = {Physics of Plasmas},
    author = {Vay, J. L. and Huebl, A. and Almgren, A. and Amorim, L. D. and Bell, J. and Fedeli, L. and Ge, L. and Gott, K. and Grote, D. P. and Hogan, M. and Jambunathan, R. and Lehe, R. and Myers, A. and Ng, C. and Rowan, M. and Shapoval, O. and Th{\'{e}}venet, M. and Vincenti, H. and Yang, E. and Za{\"{i}}m, N. and Zhang, W. and Zhao, Y. and Zoni, E.},
    number = {2},
    month = {2},
    pages = {23105},
    volume = {28},
    publisher = {American Institute of Physics Inc.},
    url = {https://doi.org/10.1063/5.0028512},
    doi = {10.1063/5.0028512},
    issn = {10897674}
}

@article{Osiris,
    title = {{No Title}},
    year = {2002},
    journal = {Lec. Notes In Comp. Sci.},
    author = {Fonseca, R. A.},
    pages = {342},
    volume = {2329}
}

@article{Vayprl07,
    title = {{Noninvariance Of Space- And Time-Scale Ranges Under A Lorentz Transformation And The Implications For The Study Of Relativistic Interactions}},
    year = {2007},
    journal = {Physical Review Letters},
    author = {Vay, J.-L.},
    number = {13},
    pages = {1--4},
    volume = {98},
    issn = {0031-9007}
}

@article{Sprangleprl90,
    title = {{Nonlinear-Theory Of Intense Laser-Plasma Interactions}},
    year = {1990},
    journal = {Physical Review Letters},
    author = {Sprangle, P and Esarey, E and Ting, A},
    number = {17},
    month = {4},
    pages = {2011--2014},
    volume = {64},
    issn = {0031-9007}
}

@article{VayCSD12,
    title = {{Novel methods in the particle-in-cell accelerator code-framework warp}},
    year = {2012},
    journal = {Computational Science and Discovery},
    author = {Vay, J.-L. and Grote, D P and Cohen, R H and Friedman, A},
    number = {1},
    pages = {014019 (20 pp.)},
    volume = {5},
    issn = {1749-4680}
}

@article{XuCPC2013,
    title = {{Numerical instability due to relativistic plasma drift in EM-PIC simulations}},
    year = {2013},
    journal = {Computer Physics Communications},
    author = {Xu, Xinlu and Yu, Peicheng and Martins, Samual F and Tsung, Frank S and Decyk, Viktor K and Vieira, Jorge and Fonseca, Ricardo A and Lu, Wei and Silva, Luis O and Mori, Warren B},
    number = {11},
    pages = {2503--2514},
    volume = {184},
    url = {http://www.sciencedirect.com/science/article/pii/S0010465513002312},
    doi = {http://dx.doi.org/10.1016/j.cpc.2013.07.003},
    issn = {0010-4655}
}

@article{Vayjcp2011,
    title = {{Numerical Methods For Instability Mitigation In The Modeling Of Laser Wakefield Accelerators In A Lorentz-Boosted Frame}},
    year = {2011},
    journal = {Journal of Computational Physics},
    author = {Vay, J L and Geddes, C G R and Cormier-Michel, E and Grote, D P},
    number = {15},
    month = {7},
    pages = {5908--5929},
    volume = {230},
    doi = {10.1016/J.Jcp.2011.04.003}
}

@article{Martinscpc10,
    title = {{Numerical Simulations Of Laser Wakefield Accelerators In Optimal Lorentz Frames}},
    year = {2010},
    journal = {Computer Physics Communications},
    author = {Martins, Samuel F and Fonseca, Ricardo A and Silva, Luis O and Lu, Wei and Mori, Warren B},
    number = {5},
    month = {5},
    pages = {869--875},
    volume = {181},
    doi = {10.1016/J.Cpc.2009.12.023},
    issn = {0010-4655}
}

@article{GodfreyJCP2014,
    title = {{Numerical stability analysis of the pseudo-spectral analytical time-domain {\{}PIC{\}} algorithm}},
    year = {2014},
    journal = {Journal of Computational Physics},
    author = {Godfrey, Brendan B and Vay, Jean-Luc and Haber, Irving},
    number = {0},
    pages = {689--704},
    volume = {258},
    url = {http://www.sciencedirect.com/science/article/pii/S0021999113007298},
    doi = {http://dx.doi.org/10.1016/j.jcp.2013.10.053},
    issn = {0021-9991}
}

@article{GodfreyIEEE2014,
    title = {{Numerical stability improvements for the pseudospectral EM PIC algorithm}},
    year = {2014},
    journal = {IEEE Transactions on Plasma Science},
    author = {Godfrey, Brendan B. and Vay, Jean Luc and Haber, Irving},
    number = {5},
    pages = {1339--1344},
    volume = {42},
    publisher = {Institute of Electrical and Electronics Engineers Inc.}
}

@article{GodfreyJCP2013,
    title = {{Numerical stability of relativistic beam multidimensional {\{}PIC{\}} simulations employing the Esirkepov algorithm}},
    year = {2013},
    journal = {Journal of Computational Physics},
    author = {Godfrey, Brendan B and Vay, Jean-Luc},
    number = {0},
    pages = {33--46},
    volume = {248},
    url = {http://www.sciencedirect.com/science/article/pii/S0021999113002556},
    doi = {http://dx.doi.org/10.1016/j.jcp.2013.04.006},
    issn = {0021-9991}
}

@article{ShapovalPRE2021,
    title = {{Overcoming timestep limitations in boosted-frame particle-in-cell simulations of plasma-based acceleration}},
    year = {2021},
    journal = {Physical Review E},
    author = {Shapoval, Olga and Lehe, Remi and Th{\'{e}}venet, Maxence and Zoni, Edoardo and Zhao, Yinjian and Vay, Jean-Luc},
    number = {5},
    month = {11},
    pages = {055311},
    volume = {104},
    publisher = {American Physical Society (APS)},
    url = {https://journals.aps.org/pre/abstract/10.1103/PhysRevE.104.055311},
    doi = {10.1103/PHYSREVE.104.055311/FIGURES/6/MEDIUM},
    issn = {2470-0045},
    arxivId = {2104.13995}
}

@article{LifschitzJCP2009,
    title = {{Particle-in-Cell modelling of laser plasma interaction using Fourier decomposition}},
    year = {2009},
    journal = {Journal of Computational Physics},
    author = {Lifschitz, A F and Davoine, X and Lefebvre, E and Faure, J and Rechatin, C and Malka, V},
    number = {5},
    pages = {1803--1814},
    volume = {228},
    url = {http://www.sciencedirect.com/science/article/pii/S0021999108005950},
    doi = {http://dx.doi.org/10.1016/j.jcp.2008.11.017},
    issn = {0021-9991}
}

@misc{PICMI,
    title = {{PICMI}},
    url = {https://github.com/picmi-standard}
}

@book{Birdsalllangdon,
    title = {{Plasma Physics Via Computer Simulation}},
    year = {1991},
    author = {Birdsall, C K and Langdon, A B},
    pages = {Xxvi+479 Pp.},
    publisher = {Adam-Hilger},
    isbn = {0 07 005371 5}
}

@article{MyersPC2021,
    title = {{Porting WarpX to GPU-accelerated platforms}},
    year = {2021},
    journal = {Parallel Computing},
    author = {Myers, A. and Almgren, A. and Amorim, L. D. and Bell, J. and Fedeli, L. and Ge, L. and Gott, K. and Grote, D. P. and Hogan, M. and Huebl, A. and Jambunathan, R. and Lehe, R. and Ng, C. and Rowan, M. and Shapoval, O. and Th{\'{e}}venet, M. and Vay, J. L. and Vincenti, H. and Yang, E. and Za{\"{i}}m, N. and Zhang, W. and Zhao, Y. and Zoni, E.},
    month = {12},
    pages = {102833},
    volume = {108},
    publisher = {North-Holland},
    doi = {10.1016/J.PARCO.2021.102833},
    issn = {0167-8191},
    arxivId = {2101.12149}
}

@article{Blaclard2017,
    title = {{Pseudospectral Maxwell solvers for an accurate modeling of Doppler harmonic generation on plasma mirrors with particle-in-cell codes}},
    year = {2017},
    journal = {Physical Review E},
    author = {Blaclard, G. and Vincenti, H. and Lehe, R. and Vay, J. L.},
    number = {3},
    month = {9},
    pages = {033305},
    volume = {96},
    publisher = {American Physical Society},
    url = {https://link.aps.org/doi/10.1103/PhysRevE.96.033305},
    doi = {10.1103/PhysRevE.96.033305},
    issn = {2470-0045}
}

@article{KirchenPRE2020,
    title = {{Scalable spectral solver in Galilean coordinates for eliminating the numerical Cherenkov instability in particle-in-cell simulations of streaming plasmas}},
    year = {2020},
    journal = {Physical Review E},
    author = {Kirchen, Manuel and Lehe, Remi and Jalas, Soeren and Shapoval, Olga and Vay, Jean Luc and Maier, Andreas R.},
    number = {1},
    month = {7},
    pages = {13202},
    volume = {102},
    publisher = {American Physical Society},
    url = {https://journals.aps.org/pre/abstract/10.1103/PhysRevE.102.013202},
    doi = {10.1103/PhysRevE.102.013202},
    issn = {24700053},
    pmid = {32794957}
}

@article{QiangPRE2000,
    title = {{Second-order stochastic leapfrog algorithm for multiplicative noise Brownian motion}},
    year = {2000},
    journal = {Physical Review E},
    author = {Qiang, Ji and Habib, Salman},
    number = {5},
    month = {11},
    pages = {7430},
    volume = {62},
    publisher = {American Physical Society},
    url = {https://journals.aps.org/pre/abstract/10.1103/PhysRevE.62.7430},
    doi = {10.1103/PhysRevE.62.7430},
    issn = {1063651X},
    arxivId = {physics/9912055}
}

@inproceedings{Furman2007,
    title = {{Self-consistent 3D modeling of electron cloud dynamics and beam response}},
    year = {2007},
    booktitle = {Proceedings of the IEEE Particle Accelerator Conference},
    author = {Furman, M.A. and Celata, C.M. and Kireeff-Covo, M. and Sonnad, K.G. and Vay, J.-L. and Venturini, M. and Cohen, R. and Friedman, A. and Grote, D. and Molvik, A. and Stoltz, P.},
    isbn = {1424409179},
    doi = {10.1109/PAC.2007.4441093}
}

@article{Vaypop2008,
    title = {{Simulation Of Beams Or Plasmas Crossing At Relativistic Velocity}},
    year = {2008},
    journal = {Physics Of Plasmas},
    author = {Vay, J L},
    number = {5},
    month = {5},
    pages = {56701},
    volume = {15},
    doi = {10.1063/1.2837054}
}

@article{KirchenPoP2016,
    title = {{Stable discrete representation of relativistically drifting plasmas}},
    year = {2016},
    journal = {Physics of Plasmas},
    author = {Kirchen, M. and Lehe, R. and Godfrey, B. B. and Dornmair, I. and Jalas, S. and Peters, K. and Vay, J.-L. and Maier, A. R.},
    number = {10},
    month = {10},
    pages = {100704},
    volume = {23},
    publisher = {AIP Publishing LLC},
    url = {http://aip.scitation.org/doi/10.1063/1.4964770},
    doi = {10.1063/1.4964770},
    issn = {1070-664X}
}

@article{QiangPRAB2002,
    title = {{Strong-strong beam-beam simulation using a Green function approach}},
    year = {2002},
    journal = {Physical Review Special Topics - Accelerators and Beams},
    author = {Qiang, Ji and Furman, Miguel A. and Ryne, Robert D.},
    number = {10},
    month = {10},
    pages = {60--66},
    volume = {5},
    publisher = {American Physical Society},
    url = {https://journals.aps.org/prab/abstract/10.1103/PhysRevSTAB.5.104402},
    doi = {10.1103/PHYSREVSTAB.5.104402/FIGURES/9/MEDIUM},
    issn = {10984402}
}

@article{GodfreyJCP2014_2,
    title = {{Suppressing the numerical Cherenkov instability in {\{}FDTD{\}} {\{}PIC{\}} codes}},
    year = {2014},
    journal = {Journal of Computational Physics},
    author = {Godfrey, Brendan B and Vay, Jean-Luc},
    number = {0},
    pages = {1--6},
    volume = {267},
    url = {http://www.sciencedirect.com/science/article/pii/S0021999114001429},
    doi = {http://dx.doi.org/10.1016/j.jcp.2014.02.022},
    issn = {0021-9991}
}

@article{QiangPRAB2017,
    title = {{Symplectic multiparticle tracking model for self-consistent space-charge simulation}},
    year = {2017},
    journal = {Physical Review Accelerators and Beams},
    author = {Qiang, Ji},
    number = {1},
    month = {1},
    pages = {014203},
    volume = {20},
    publisher = {American Physical Society},
    doi = {10.1103/PhysRevAccelBeams.20.014203},
    issn = {24699888}
}

@misc{Synergia,
    title = {{Synergia}},
    url = {http://web.fnal.gov/sites/synergia}
}

@book{godfrey1985iprop,
    title = {{The IPROP Three-Dimensional Beam Propagation Code}},
    year = {1985},
    author = {Godfrey, B B and NM., MISSION RESEARCH CORP ALBUQUERQUE},
    publisher = {Defense Technical Information Center},
    url = {https://books.google.com/books?id=hos_OAAACAAJ}
}

@inproceedings{Warp,
    title = {{The Warp Code: Modeling High Intensity Ion Beams}},
    year = {2005},
    booktitle = {Aip Conference Proceedings},
    author = {Grote, D P and Friedman, A and Vay, J.-L. and Haber, I},
    number = {749},
    pages = {55--58},
    issn = {0094-243X}
}

@article{QiangCPC2004,
    title = {{Three-dimensional Poisson solver for a charged beam with large aspect ratio in a conducting pipe}},
    year = {2004},
    journal = {Computer Physics Communications},
    author = {Qiang, Ji and Gluckstern, Robert L.},
    number = {2},
    month = {7},
    pages = {120--128},
    volume = {160},
    publisher = {North-Holland},
    doi = {10.1016/J.CPC.2004.03.002},
    issn = {0010-4655}
}

@article{Qiang2006,
    title = {{Three-dimensional quasistatic model for high brightness beam dynamics simulation}},
    year = {2006},
    journal = {Physical Review Special Topics - Accelerators and Beams},
    author = {Qiang, Ji and Lidia, Steve and Ryne, Robert D. and Limborg-Deprey, Cecile},
    number = {4},
    month = {4},
    pages = {044204},
    volume = {9},
    publisher = {American Physical Society},
    url = {https://link.aps.org/doi/10.1103/PhysRevSTAB.9.044204},
    doi = {10.1103/PhysRevSTAB.9.044204},
    issn = {1098-4402}
}

@inproceedings{WarpXAAC2018,
    title = {{Toward plasma wakefield simulations at exascale}},
    year = {2019},
    booktitle = {2018 IEEE Advanced Accelerator Concepts Workshop, ACC 2018 - Proceedings},
    author = {Vay, J. L. and Almgren, A. and Bell, J. and Lehe, R. and Myers, A. and Park, J. and Shapoval, O. and Thevenet, M. and Zhang, W. and Grote, D. P. and Hogan, M. and Ge, L. and Ng, C.},
    month = {3},
    publisher = {Institute of Electrical and Electronics Engineers Inc.},
    isbn = {9781538677216},
    doi = {10.1109/AAC.2018.8659392}
}

@article{ALEGRO,
    title = {{Towards an Advanced Linear International Collider}},
    year = {2019},
    author = {{ALEGRO}},
    month = {1},
    url = {http://arxiv.org/abs/1901.10370},
    arxivId = {1901.10370}
}

@article{BrownPRAB2020,
    title = {{Towards storage rings as quantum computers}},
    year = {2020},
    journal = {Physical Review Accelerators and Beams},
    author = {Brown, K. A. and Roser, T.},
    number = {5},
    month = {5},
    pages = {054701},
    volume = {23},
    publisher = {American Physical Society},
    url = {https://journals.aps.org/prab/abstract/10.1103/PhysRevAccelBeams.23.054701},
    doi = {10.1103/PHYSREVACCELBEAMS.23.054701/FIGURES/4/MEDIUM},
    issn = {24699888}
}

@article{Shapoval2019,
    title = {{Two-step perfectly matched layer for arbitrary-order pseudo-spectral analytical time-domain methods}},
    year = {2019},
    journal = {Computer Physics Communications},
    author = {Shapoval, Olga and Vay, Jean-Luc and Vincenti, Henri},
    month = {2},
    pages = {102--110},
    volume = {235},
    publisher = {North-Holland},
    url = {https://www.sciencedirect.com/science/article/pii/S0010465518303291?via%3Dihub},
    doi = {10.1016/J.CPC.2018.09.015},
    issn = {0010-4655}
}

@article{VincentiCPC2018,
    title = {{Ultrahigh-order Maxwell solver with extreme scalability for electromagnetic PIC simulations of plasmas}},
    year = {2018},
    journal = {Computer Physics Communications},
    author = {Vincenti, H. and Vay, J.-L.},
    doi = {10.1016/j.cpc.2018.03.018},
    issn = {00104655}
}

@article{WarpXEAAC2017,
    title = {{Warp-X: A new exascale computing platform for beam–plasma simulations}},
    year = {2018},
    journal = {Nuclear Instruments and Methods in Physics Research Section A: Accelerators, Spectrometers, Detectors and Associated Equipment},
    author = {Vay, J.-L. and Almgren, A. and Bell, J. and Ge, L. and Grote, D.P. and Hogan, M. and Kononenko, O. and Lehe, R. and Myers, A. and Ng, C. and Park, J. and Ryne, R. and Shapoval, O. and Th{\'{e}}venet, M. and Zhang, W.},
    month = {1},
    publisher = {North-Holland},
    url = {https://www.sciencedirect.com/science/article/pii/S0168900218300524},
    doi = {10.1016/J.NIMA.2018.01.035},
    issn = {0168-9002}
}

@misc{WarpX,
    title = {{WarpX}},
    url = {https://github.com/ECP-WarpX/WarpX}
}

@article{Pieronek_2021,
author = {Pieronek,C. V.  and Gonsalves,A. J.  and Benedetti,C.  and Bulanov,S. S.  and van Tilborg,J.  and Bin,J. H.  and Swanson,K. K.  and Daniels,J.  and Bagdasarov,G. A.  and Bobrova,N. A.  and Gasilov,V. A.  and Korn,G.  and Sasorov,P. V.  and Geddes,C. G. R.  and Schroeder,C. B.  and Leemans,W. P.  and Esarey,E. },
title = {Laser-heated capillary discharge waveguides as tunable structures for laser-plasma acceleration},
journal = {Physics of Plasmas},
volume = {27},
number = {9},
pages = {093101},
year = {2020},
doi = {10.1063/5.0014961},

URL = { 
        https://doi.org/10.1063/5.0014961
    
},
eprint = { 
        https://doi.org/10.1063/5.0014961
    
}

}

@article{Leemans_2014,
  title = {Multi-GeV Electron Beams from Capillary-Discharge-Guided Subpetawatt Laser Pulses in the Self-Trapping Regime},
  author = {Leemans, W. P. and Gonsalves, A. J. and Mao, H.-S. and Nakamura, K. and Benedetti, C. and Schroeder, C. B. and T\'oth, Cs. and Daniels, J. and Mittelberger, D. E. and Bulanov, S. S. and Vay, J.-L. and Geddes, C. G. R. and Esarey, E.},
  journal = {Phys. Rev. Lett.},
  volume = {113},
  issue = {24},
  pages = {245002},
  numpages = {5},
  year = {2014},
  month = {Dec},
  publisher = {American Physical Society},
  doi = {10.1103/PhysRevLett.113.245002},
  url = {https://link.aps.org/doi/10.1103/PhysRevLett.113.245002}
}

@article{Benedetti_2010,
	author = {C. Benedetti and C. B. Schroeder and E. Esarey and C. G. R. Geddes and W. P. Leemans},
	journal = {AIP conference proceedings},
	number = {1},
	pages = {250--255},
	title = {Efficient modeling of laser‐plasma accelerators with {INF\&RNO}},
	volume = {1299},
	year = {2010}}

@article{Mehrling_2014,
	doi = {10.1088/0741-3335/56/8/084012},
	url = {https://doi.org/10.1088/0741-3335/56/8/084012},
	year = 2014,
	month = {jul},
	publisher = {{IOP} Publishing},
	volume = {56},
	number = {8},
	pages = {084012},
	author = {T Mehrling and C Benedetti and C B Schroeder and J Osterhoff},
	title = {{HiPACE}: a quasi-static particle-in-cell code},
	journal = {Plasma Physics and Controlled Fusion}
}

@article{Benedetti17b,
	doi = {10.1088/1361-6587/aa8977},
	url = {https://doi.org/10.1088/1361-6587/aa8977},
	year = 2017,
	month = {oct},
	publisher = {{IOP} Publishing},
	volume = {60},
	number = {1},
	pages = {014002},
	author = {C Benedetti and C B Schroeder and C G R Geddes and E Esarey and W P Leemans},
	title = {An accurate and efficient laser-envelope solver for the modeling of laser-plasma accelerators},
	journal = {Plasma Physics and Controlled Fusion}
}

@article{LOI_ML,
title = {{Machine learning and surrogates models for simulation-based optimization of accelerator design}},
author = {R. Lehe and others},
journal = {Snowmass21 LOI},
year = 2020,
url = {https://www.snowmass21.org/docs/files/summaries/CompF/SNOWMASS21-CompF2_CompF3-AF1_AF6_Lehe-075.pdf}
}

@article{LOI_ML2,
title = {{Application of Machine Learning to Particle Accelerator Simulations}},
author = {Winklehner, D. and Adelmann, A.},
journal = {Snowmass21 LOI},
year = 2020,
url = {https://www.snowmass21.org/docs/files/summaries/CompF/SNOWMASS21-CompF3_CompF0-AF1_AF0_Winklehner-108.pdf}
}

@article{LOI_center,
title = {{Center(s) for Accelerator and Beam Physics Modeling}},
author = {Jean-Luc Vay and David Bruhwiler and David Sagan and Axel Huebl and Remi Lehe and Cho-Kuen Ng and Ji Qiang and Robert Ryne and Maxence Thevenet and Henri Vincenti and Alex Thomas},
journal = {Snowmass21 LOI},
url = {https://www.snowmass21.org/docs/files/summaries/CompF/SNOWMASS21-CompF2_CompF0-AF1_AF0_Vay-069.pdf},
year={2020}
}

@article{LOI_eva,
title = {{End-to-End Virtual Accelerators (EVA)}},
author = {J.-L. Vay and D. Sagan and A. Huebl and M. Th\'evenet and R. Lehe and C.-K. Ng and H. Vincenti and M. Bussmann and A. Debus and R. Pausch and J. Qiang},
journal = {Snowmass21 LOI},
year={2020},
url = {https://www.snowmass21.org/docs/files/summaries/CompF/SNOWMASS21-CompF2_CompF0-AF1_AF0_Vay-067.pdf }
}

@article{LOI_standards,
title = {Develop/integrate data standards and start-to-end workflows},
author = {A. Huebl and J.-L. Vay and Lehe, R. and Th\'evenet, M. and Mayes, C. and Sagan, D. and Y.-D. Tsai and  J. C. E and  F.~Tsung and  H. Vincenti and  A. Ferran Pousa and  N. M. Cook and  S. J. Gessner and  F. Poeschel and  M.~Bussmann and  D.~P. Grote and  N. A. Murphy and  R. Schmitz and  C. H. Yoon and  D.~L. Bruhwiler and  K.~Cranmer and  S.~R. Yoffe and  B. Cros and  A. L. Edelen and  G. Stark},
journal = {Snowmass21 LOI},
url = {https://www.snowmass21.org/docs/files/summaries/CompF/SNOWMASS21-CompF2_CompF7-AF1_AF0_Huebl-079.pdf},
year = {2020}
}

@article{LOI_open,
title = {{Aspiration for Open Science in Accelerator \& Beam Physics Modeling }},
author = {A. Huebl and J.-L. Vay and R. Lehe and C. Mayes and Y.-D. Tsai and A. Friedman and M. Th´evenet and H. Vincenti and D. L. Bruhwiler and A. Sauers and N. M. Cook and S. J. Gessner and M. Bussmann and D. P. Grote and R. Schmitz and B. Cros and D. Sagan and S. R. Yoffe and A. L. Edelen and A. Ferran Pousa and G. Stark},
journal = {Snowmass21 LOI},
year = 2020,
url = {https://www.snowmass21.org/docs/files/summaries/CompF/SNOWMASS21-CompF2_CompF7-AF1_AF0_Huebl-081.pdf}
}

@article{Chen2018,
  doi = {10.1038/s41567-018-0342-2},
  % url = {https://doi.org/10.1038/s41567-018-0342-2},
  year = {2018},
  month = nov,
  publisher = {Springer Science and Business Media {LLC}},
  volume = {15},
  number = {2},
  pages = {113--119},
  author = {Xiaoli Chen and S\"{u}nje Dallmeier-Tiessen and Robin Dasler and Sebastian Feger and Pamfilos Fokianos and Jose Benito Gonzalez and Harri Hirvonsalo and Dinos Kousidis and Artemis Lavasa and Salvatore Mele and Diego Rodriguez Rodriguez and Tibor {\v{S}}imko and Tim Smith and Ana Trisovic and Anna Trzcinska and Ioannis Tsanaktsidis and Markus Zimmermann and Kyle Cranmer and Lukas Heinrich and Gordon Watts and Michael Hildreth and Lara Lloret Iglesias and Kati Lassila-Perini and Sebastian Neubert},
  title = {Open is not enough},
  journal = {Nature Physics}
}

@misc{OpenSource,
  title = {{The Open Source Definition}},
  author = {{Open Source Initiative}},
  url = {https://opensource.org/docs/osd},
  year = {2004},
}

@misc{FreeSoftware,
  title = {{The Free Software Definition}},
  author = {{Free Software Foundation}},
  url = {https://www.gnu.org/philosophy/free-sw.en.html},
  year = {1990}
}

@misc{OpenScienceUNSECO,
  title = {{Open Science Movement}},
  author = {{United Nations Educational, Scientific and Cultural Organization (UNESCO)}},
  url = {http://www.unesco.org/new/en/communication-and-information/portals-and-platforms/goap/open-science-movement/}
}

@misc{OpenAccessUNSECO,
  title = {{What is Open Access?}},
  author = {{United Nations Educational, Scientific and Cultural Organization (UNESCO)}},
  url = {https://en.unesco.org/open-access/what-open-access}
}

@misc{OpenScienceEU,
  title = {{Open Science Definition}},
  author = {{FOSTER Plus (European Union funded open project in Horizon 2020 and beyond)}},
  url = {https://www.fosteropenscience.eu/foster-taxonomy/open-science-definition}
}

@article{LOI_halo_loss,
title = {{Loss prediction through modeling of high dynamic range beam distributions}},
author = {Kiersten Ruisard and Alexander Aleksandrov and Sarah Cousineau},
journal = {Snowmass21 LOI},
year = 2020,
url = {https://www.snowmass21.org/docs/files/summaries/AF/SNOWMASS21-AF1_AF2_Kiersten_Ruisard-145.pdf}
}

@article{LOI_reproducibility,
title = {{Long Term Reproducibility and Sustainability of Scientific Software}},
author = {Matthew Feickert and Giordon Stark and Steven Gardiner and Yu-Dai Tsai},
journal = {Snowmass21 LOI},
year = 2020,
url = {https://www.snowmass21.org/docs/files/summaries/CompF/SNOWMASS21-CompF7_CompF0_Matthew_Feickert-107.pdf}
}

@article{LOI_barriers,
title = {{Barriers to Entry in Physics Computing}},
author = {R. Schmitz and T. Eichlersmith and A. Huebl},
journal = {Snowmass21 LOI},
year = 2020,
url = {https://www.snowmass21.org/docs/files/summaries/CompF/SNOWMASS21-CompF0_CompF0_Ryan_Schmitz-104.pdf}
}

@article{Dular.2021,
	author = {Dular, Julien and Harutyunyan, Mane and Bortot, Lorenzo and Schops, Sebastian and Vanderheyden, Benoit and Geuzaine, Christophe},
	doi = {10.1109/tasc.2021.3098724},
	eprint = {2106.00313},
	issn = {1051-8223},
	journal = {IEEE Transactions on Applied Superconductivity},
	number = {6},
	pages = {1--12},
	title = {On the stability of mixed finite-element formulations for high-temperature superconductors},
	volume = {31},
	year = {2021},
	url = {https://doi.org/10.1109/tasc.2021.3098724}}

@article{Dular.2019,
	author = {Dular, Julien and Geuzaine, Christophe and Vanderheyden, Benot},
	doi = {10.1109/tasc.2019.2935429},
	issn = {1051-8223},
	journal = {IEEE Transactions on Applied Superconductivity},
	number = {3},
	pages = {1--13},
	title = {{Finite-Element formulations for systems with high-temperature superconductors}},
	volume = {30},
	year = {2019},
	url = {https://doi.org/10.1109/tasc.2019.2935429}}

@article{Arsenault.2020y9g,
	author = {Arsenault, Alexandre and Sirois, Frdric and Grilli, Francesco},
	doi = {10.1109/tasc.2020.3033998},
	eprint = {2006.13784},
	issn = {1051-8223},
	journal = {IEEE Transactions on Applied Superconductivity},
	number = {2},
	pages = {1--11},
	title = {{Implementation of the h-$\phi$ formulation in COMSOL Multiphysics for simulating the magnetization of bulk superconductors and comparison with the h-formulation}},
	volume = {31},
	year = {2021},
	url = {https://doi.org/10.1109/tasc.2020.3033998}}

@article{Arsenault.2021,
	author = {Arsenault, Alexandre and Alves, Bruno de Sousa and Sirois, Frdric},
	doi = {10.1109/tasc.2021.3097245},
	issn = {1051-8223},
	journal = {IEEE Transactions on Applied Superconductivity},
	number = {6},
	pages = {1--9},
	title = {{COMSOL Implementation of the h-$\phi$-formulation with thin cuts for modeling superconductors with transport currents}},
	volume = {31},
	year = {2021},
	url = {https://doi.org/10.1109/tasc.2021.3097245}}

@article{Alves.2021,
	author = {Alves, Bruno de Sousa and Lahtinen, Valtteri and Laforest, Marc and Sirois, Fr\`ed\`eric},
	eprint = {2108.08828},
	journal = {arXiv},
	note = {h-phi thin},
	title = {{Thin-shell approach for modeling superconducting tapes in the $h$-$\phi$ finite-element formulation}},
	year = {2021}}

@misc{Katz_SWSustainability2020,
  author       = {Katz, Daniel S. and
                  Malik, Sudhir and
                  Neubauer, Mark S. and
                  Stewart, Graeme A. and
                  Assamagan, K\'{e}t\'{e}vi A. and
                  Becker, Erin A. and
                  Chue Hong, Neil P. and
                  Cosden, Ian A. and
                  Meehan, Samuel and
                  Moyse, Edward J. W. and
                  Price-Whelan, Adrian M. and
                  Sexton-Kennedy, Elizabeth and
                  Evans, Meirin Oan and
                  Feickert, Matthew and
                  Lange, Clemens and
                  Lieret, Kilian and
                  Quick, Rob and
                  S\'{a}nchez Pineda, Arturo and
                  Tunnell, Christopher},
  title        = {Software Sustainability \& High Energy Physics},
  month        = oct,
  year         = 2020,
  publisher    = {Zenodo},
  doi          = {10.5281/zenodo.4095837},
  url          = {https://arxiv.org/abs/2010.05102}
}

@article{LOI_industry,
title = {{Collaboration between industry and the HEP community}},
author = {David Bruhwiler and Robert Nagler and Paul Moeller and Robert O’Bara and Dan Abell and Stanislav Baturin and Evan Carlin and Stephen Coleman and Nathan Cook and Jonathan Edelen and Callie Federer and Ahmad Habib and Christopher Hall and Thomas Heinemann and Bernhard Hidding and Axel Huebl and Michael Keilman and Remi Lehe and Evan Carlin and Paula Messamer and Boaz Nash and Cho-Kuen Ng and Chong-Shik Park and Philippe Piot and Ilya Pogorelov and Emily Poore and Aaron Sauers and Paul Scherkl and John Tourtellott and Jean-Luc Vay and Stephen Webb},
journal = {Snowmass21 LOI},
year = 2020,
url = {https://www.snowmass21.org/docs/files/summaries/CommF/SNOWMASS21-CommF1_CommF0-AF0_AF1_Bruhwiler-066.pdf}
}

@article{LOI_ecosystem,
title = {A modular community ecosystem for multiphysics particle accelerator modeling and design},
author ={J.-L. Vay and  A. Huebl and D. Sagan and D. Bruhwiler and R. Lehe and C.-K. Ng and A. Liu and Ji Qiang and R. Ryne and E. Stern and A. Friedman and M. Th\'evenet and H. Vincenti and S. Gessner and B. Cowan and A. Adelmann and M. Bussmann and S. Bastrakov},
journal = {Snowmass21 LOI},
year={2020},
url={https://www.snowmass21.org/docs/files/summaries/CompF/SNOWMASS21-CompF2_CompF0-AF1_AF0_Vay-070.pdf}
}

@article{SM_whitepaper_HEP_DATA,
title = {{Data and Analysis Preservation, Recasting, and Reinterpretation}},
author = {{Bierlich, Christian and Buckley, Andy Cranmer, Kyle and others}},
journal = {Snowmass21 Whitepaper to TF07 and CompF7},
year = 2022
}

@article{davidson:15,
author = "A. Davidson and A. Tableman and W. An and F. S. Tsung and W. Lu annd J. Vieira and R. A. Fonnseca, annd L. O. Silva and W. B. Mori",
    title = {{Implementation of a hybrid particle code with a PIC description in r–z and a gridless description in {$\phi$} into OSIRIS}},
journal = {Journal of Computational Physics},
volume=281,
year=2015,
pages = 1063
}

@article{mora:97,
author = {P. Mora and T. M. Antonsen},
title={Kinetic modeling of intense, short laser pulses propagating in tenuous plasmas},
journal = {Physics of Plasmas},
volumes = 4,
pages = 217,
year = 1997

}

@article{LOI_magnets,
title = {{Numerical Modeling for Superconducting Accelerator Magnets}},
author = {D. Arbelaez and E. Barzi and L. Brouwer and B. Cowan and D. Davis and R. Gupta and V. Kashikhin and V.
Marinozzi and T. Shen and M. Sumption and R. Teyber and G. Vallone and X. Wang},
journal = {Snowmass21 LOI},
year = 2020,
url = {https://www.snowmass21.org/docs/files/summaries/CompF/SNOWMASS21-CompF2_CompF0-AF7_AF0-027.pdf}
}

@article{USMDP,
title = {{The US Magnet Development Program Plan}},
author = {S. Gourlay and S. Prestemon and A. Zlobin and L. Cooley and D. Larbalestier},
year = 2016,
url = {https://osti.gov/scitech/servlets/purl/1306334}
}

@article{ICFASagan2021,
	doi = {10.1088/1748-0221/16/10/t10002},
	url = {https://doi.org/10.1088/1748-0221/16/10/t10002},
	year = 2021,
	month = {10},
	publisher = {{IOP} Publishing},
	volume = {16},
	number = {10},
	pages = {T10002},
	author = {D. Sagan and M. Berz and N.M. Cook and Y. Hao and G. Hoffstaetter and A. Huebl and C.-K. Huang and M.H. Langston and C.E. Mayes and C.E. Mitchell and C.-K. Ng and J. Qiang and R.D. Ryne and A. Scheinker and E. Stern and J.-L. Vay and D. Winklehner and H. Zhang},
	title = {Simulations of future particle accelerators: issues and mitigations},
	journal = {Journal of Instrumentation}
}

@article{ICFAVay2021,
	doi = {10.1088/1748-0221/16/10/t10003},
	url = {https://doi.org/10.1088/1748-0221/16/10/t10003},
	year = 2021,
	month = {10},
	publisher = {{IOP} Publishing},
	volume = {16},
	number = {10},
	pages = {T10003},
	author = {J.-L. Vay and A. Huebl and R. Lehe and N.M. Cook and R.J. England and U. Niedermayer and P. Piot and F. Tsung and D. Winklehner},
	title = {Modeling of advanced accelerator concepts},
	journal = {Journal of Instrumentation}
}

@article{huang06,
author = "C. K. Huang and V. K. Decyk and C. Ren and M. Zhou and W. Lu and W. B. Mori and J. H. Cooley and T. M. Antonsen and T. Katsouleas",
title = {QUICKPIC: A highly efficient particle-in-cell code for modeling wakefield acceleration in plasmas},
journal = "Journal of Computational Physics",
volume = 217,
pages = 658,
year = 2006
}

@article{decyk:07,
author = "V. K. Decyk",
title = "UPIC: A framework for massively parallel particle-in-cell codes",
journal = {Computer Physics Communications},
volume = 177,
pages = 95,
year = 2007
}

@article{li21a,
author = {F. Li and W. An and V. K. Decyk and X. Xu and M. J. Hogan and W. B. Mori}, 
title = {A quasi-static particle-in-cell algorithm based on an azimuthal Fourier decomposition for highly efficient simulations of plasma-based 
acceleration: QPAD},
journal = {Computer Physics Communications},
volume = 261,
pages = 107784,
year = 2021

}

@article{decyk:04,
author ="V. K. Decyk annd C. D. Norton",
title = "UCLA Parallel PIC Framework",
journal = {Computer Physics Communications},
volume = 164,
pages = 80,
year = 2004
}

@article{decyk:11,
author = "V. K. Decyk and T. V. Singh",
title = "Adaptable Particle-in-Cell algorithms for graphical processing units", 
journal = {Computer Physics Communications},
volume = 182,
pages = 641,
year = 2011

}

@article{decyk:14,
author = "V. K. Decyk and T. V. Singh",
title = "Particle-in-Cell algorithms for emerging computer architectures",
journal = {Computer Physics Communications},
volume = 185,
pages = 708,
year = 2014

}

@article{yu:16,
author = {P. Yu and X. Xu and A. Davidson and A. Tableman and T. Dalichaouch and F. Li and M. D. Meyers and W. An and F. S. Tsung and V. K. Decyk and F. Fiuza and J. Vieira and R. A. Fonseca and W. Lu and
L. O. Silva and W. B. Mori},
title = "Enabling Lorentz boosted frame particle-in-cell simulations of laser wakefield acceleration in quasi-3D geometry",
journal = {Journal of Computational Physics},
volume = 316,
pages = 747,
year = 2016
}

@article{an:13,
author ={W. An ad V. K. Decyk and W. B. Mori and T. M. Atonsen},
title={An improved iteration loop for thre three dimensionnal quasi-static particle-in-cell algorithm: QuickPIC},
journal = {Jour. Comp. Phys.},
volume = 250,
pages = {165-177},
year = 2013,
}

@article{li:20a, 
author = {F. Li and K. G. Miller and X. L. Xu and F. S. Tsung and V. K. Decyk and W. An and R. A. Fonseca and W. B. Mori},
title={A new field solver for modeling of relativistic 
particle-laser interactions using the particle-in-cell algorithm\url{https://arxiv.org/abs/2004.03754}},
year = 2020
}

@article{li:20b,
author ={ F. Li and V. K. Decyk and K. G. Miller and A. 
Tableman and F. S. Tsung and M. Vranic and R. A. Fonseca and W. B. Mori},
title={Accurately simulating nine-dimensional phase space 
of relativistic particles in strong fields},
journal = {Jour. Comp. Phys.},
volume = 438,
pages=110367,
year= 2021,
}

@article{xu:20,
author = {X. L. Xu and F. S. Tsung, and T. Dalichaouch and W. An and H. Wen and V. K. Decyk and R. A. Fonnseca ad
M. J. Hogan and W. B. Mori},
title ={On nnumerical errors to the fields surrounding a relativistically moving particle in PIC codes},
journal = {Jour. Comp. Phys.},
volume = 413,
pages = 109451,
year = 2020,
}

@article{vranic:16,
author = {M. Vranic and J. L. Martin and R. A> Fonseca and L. O. Silva},
title= {Classical radiation reaction in particle-in-cell simulations},
journal = {Comp. Phys. Comm.},
volume = 204,
pages= {141-151},
year=2016,
}

@misc{zoniARXIV2021_2,
  author = {Edoardo Zoni and Remi Lehe and Olga Shapoval and Daniel Belkin and Neil Zaim and Luca Fedeli and Henri Vincenti and Jean-Luc Vay},
  title = {A Hybrid Nodal-Staggered Pseudo-Spectral Electromagnetic
Particle-In-Cell Method with Finite-Order Centering \url{https://arxiv.org/abs/2106.12919v1}},
  year = {2021}
}

@misc{LeheARXIV2021_pml,
  author = {Remi Lehe and Aurore Blelly and Lorenzo Giacomel and Revathi Jambunathan and Jean-Luc Vay},
  title = {Absorption of charged particles in Perfectly-Matched-Layers by optimal damping of the deposited current \url{https://arxiv.org/abs/2201.09084}},
  year = {2022}
}

@article{goble2014,
	title        = {Better Software, Better Research},
	author       = {Goble, Carole},
	year         = 2014,
	journal      = {IEEE Internet Computing},
	volume       = 18,
	number       = 5,
	pages        = {4--8},
	doi          = {10.1109/MIC.2014.88}
}

@misc{ssi,
    author = {Software Sustainability Institute },
    address = {The University of Edinburgh},
    url = {https://www.software.ac.uk/}
}

@misc{katz,
    title = {Scientific Software Challenges and Community Responses},
    author = {Daniel S. Katz},
    url = {https://www.slideshare.net/danielskatz/scientific-software-challenges-and-community-responses},
    note = {Talk given at RTI International on 7 December 2015, discussing 12 scientific software challenges and how the scientific software community is responding to them},
    year = {2015}
}

@misc{quics,
	title = {{QuICS}},
	url = {https://quics.umd.edu/},
	urldate = {2021-06-29},
}

@misc{ibm:quantum,
	title = {{IBM} {\textbar} Quantum Computing},
	rights = {Copyright {IBM} Corp. 2020},
	url = {https://www.ibm.com/quantum-computing//},
	author = {Fisher, Chris},
	urldate = {2021-06-29},
	langid = {english}
}

@article{geant4,
title = "Recent developments in Geant4",
journal = "Nuclear Instruments and Methods in Physics Research Section A: Accelerators, Spectrometers, Detectors and Associated Equipment",
volume = "835",
pages = "186 - 225",
year = "2016",
author = "J. Allison and K. Amako and J. Apostolakis and P. Arce and M. Asai and T. Aso and E. Bagli and A. Bagulya and S. Banerjee and G. Barrand and B.R. Beck and A.G. Bogdanov and D. Brandt and J.M.C. Brown and H. Burkhardt and Ph. Canal and D. Cano-Ott and S. Chauvie and K. Cho and G.A.P. Cirrone and G. Cooperman and M.A. Cortés-Giraldo and G. Cosmo and G. Cuttone and G. Depaola and L. Desorgher and X. Dong and A. Dotti and V.D. Elvira and G. Folger and Z. Francis and A. Galoyan and L. Garnier and M. Gayer and K.L. Genser and V.M. Grichine and S. Guatelli and P. Guèye and P. Gumplinger and A.S. Howard and I. Hřivnáčová and S. Hwang and S. Incerti and A. Ivanchenko and V.N. Ivanchenko and F.W. Jones and S.Y. Jun and P. Kaitaniemi and N. Karakatsanis and M. Karamitros and M. Kelsey and A. Kimura and T. Koi and H. Kurashige and A. Lechner and S.B. Lee and F. Longo and M. Maire and D. Mancusi and A. Mantero and E. Mendoza and B. Morgan and K. Murakami and T. Nikitina and L. Pandola and P. Paprocki and J. Perl and I. Petrović and M.G. Pia and W. Pokorski and J.M. Quesada and M. Raine and M.A. Reis and A. Ribon and A. {Ristić Fira} and F. Romano and G. Russo and G. Santin and T. Sasaki and D. Sawkey and J.I. Shin and I.I. Strakovsky and A. Taborda and S. Tanaka and B. Tomé and T. Toshito and H.N. Tran and P.R. Truscott and L. Urban and V. Uzhinsky and J.M. Verbeke and M. Verderi and B.L. Wendt and H. Wenzel and D.H. Wright and D.M. Wright and T. Yamashita and J. Yarba and H. Yoshida",
}

@misc{CAMPA,
  title = {{CAMPA: Consortium for Advanced Modeling of Particle Accelerators}},
  url = {http://campa.lbl.gov}
}

@misc{schwarzman,
	title = {{MIT} Schwarzman College of Computing},
	url = {https://computing.mit.edu/},
	titleaddon = {{MIT} Schwarzman College of Computing},
	urldate = {2021-06-23},
	langid = {american}
}

@misc{scidac,
	title = {Scientific Discovery through Advanced Computing},
	url = {https://www.scidac.gov/},
	urldate = {2021-06-23},
}

@misc{exascale,
	title = {Exascale Computing Project},
	url = {https://www.exascaleproject.org/},
	titleaddon = {Exascale Computing Project},
	urldate = {2021-06-23},
	langid = {american},
}

@article{arute2019quantum,
  title={Quantum supremacy using a programmable superconducting processor},
  author={Arute, Frank and Arya, Kunal and Babbush, Ryan and Bacon, Dave and Bardin, Joseph C and Barends, Rami and Biswas, Rupak and Boixo, Sergio and Brandao, Fernando GSL and Buell, David A and others},
  journal={Nature},
  volume={574},
  number={7779},
  pages={505--510},
  year={2019},
  publisher={Nature Publishing Group}
}

@misc{DWaveLeap2,
title = {Leap$^2$},
year = {2020}, 
howpublished = "\url{https://www.dwavesys.com/}"
}

@misc{AmazonQSulution,
title = {Amazon Quantum Solutions Lab},
year = {2020}, 
howpublished = "\url{https://aws.amazon.com/quantum-solutions-lab/}"
}

@misc{AzureQ,
title = {Azure Quantum},
year = {2020}, 
howpublished = "\url{https://azure.microsoft.com/en-us/services/quantum/}"
}

@article{harrow2009quantum,
  title={Quantum algorithm for linear systems of equations},
  author={Harrow, Aram W and Hassidim, Avinatan and Lloyd, Seth},
  journal={Physical review letters},
  volume={103},
  number={15},
  pages={150502},
  year={2009},
  publisher={APS}
}

@article{clader2013preconditioned,
  title={Preconditioned quantum linear system algorithm},
  author={Clader, B David and Jacobs, Bryan C and Sprouse, Chad R},
  journal={Physical review letters},
  volume={110},
  number={25},
  pages={250504},
  year={2013},
  publisher={APS}
}

@article{childs2017quantum,
  title={Quantum algorithm for systems of linear equations with exponentially improved dependence on precision},
  author={Childs, Andrew M and Kothari, Robin and Somma, Rolando D},
  journal={SIAM Journal on Computing},
  volume={46},
  number={6},
  pages={1920--1950},
  year={2017},
  publisher={SIAM}
}

@article{bravo2019variational,
  title={Variational quantum linear solver: A hybrid algorithm for linear systems},
  author={Bravo-Prieto, Carlos and LaRose, Ryan and Cerezo, Marco and Subasi, Yigit and Cincio, Lukasz and Coles, Patrick J},
  journal={arXiv preprint arXiv:1909.05820},
  year={2019}
}

@article{lee2019hybrid,
  title={Hybrid quantum linear equation algorithm and its experimental test on ibm quantum experience},
  author={Lee, Yonghae and Joo, Jaewoo and Lee, Soojoon},
  journal={Scientific reports},
  volume={9},
  number={1},
  pages={4778},
  year={2019},
  publisher={Nature Publishing Group}
}

@article{leyton2008quantum,
  title={A quantum algorithm to solve nonlinear differential equations},
  author={Leyton, Sarah K and Osborne, Tobias J},
  journal={arXiv preprint arXiv:0812.4423},
  year={2008}
}

@article{berry2014high,
  title={High-order quantum algorithm for solving linear differential equations},
  author={Berry, Dominic W},
  journal={Journal of Physics A: Mathematical and Theoretical},
  volume={47},
  number={10},
  pages={105301},
  year={2014},
  publisher={IOP Publishing}
}

@article{arrazola2019quantum,
  title={Quantum algorithm for nonhomogeneous linear partial differential equations},
  author={Arrazola, Juan Miguel and Kalajdzievski, Timjan and Weedbrook, Christian and Lloyd, Seth},
  journal={Physical Review A},
  volume={100},
  number={3},
  pages={032306},
  year={2019},
  publisher={APS}
}

@article{childs2020high,
  title={High-precision quantum algorithms for partial differential equations},
  author={Childs, Andrew M and Liu, Jin-Peng and Ostrander, Aaron},
  journal={arXiv preprint arXiv:2002.07868},
  year={2020}
}

@article{xin2020quantum,
  title={Quantum algorithm for solving linear differential equations: Theory and experiment},
  author={Xin, Tao and Wei, Shijie and Cui, Jianlian and Xiao, Junxiang and Arrazola, I{\~n}igo and Lamata, Lucas and Kong, Xiangyu and Lu, Dawei and Solano, Enrique and Long, Guilu},
  journal={Physical Review A},
  volume={101},
  number={3},
  pages={032307},
  year={2020},
  publisher={APS}
}

@article{cao2013quantum,
  title={Quantum algorithm and circuit design solving the Poisson equation},
  author={Cao, Yudong and Papageorgiou, Anargyros and Petras, Iasonas and Traub, Joseph and Kais, Sabre},
  journal={New Journal of Physics},
  volume={15},
  number={1},
  pages={013021},
  year={2013},
  publisher={IOP Publishing}
}

@article{dodin2020applications,
  title={On applications of quantum computing to plasma simulations},
  author={Dodin, Ilya Y and Startsev, Edward A},
  journal={arXiv preprint arXiv:2005.14369},
  year={2020}
}

@article{engel2019quantum,
  title={Quantum algorithm for the Vlasov equation},
  author={Engel, Alexander and Smith, Graeme and Parker, Scott E},
  journal={Physical Review A},
  volume={100},
  number={6},
  pages={062315},
  year={2019},
  publisher={APS}
}

@article{zhong2020quantum,
  title={Quantum computational advantage using photons},
  author={Zhong, Han-Sen and Wang, Hui and Deng, Yu-Hao and Chen, Ming-Cheng and Peng, Li-Chao and Luo, Yi-Han and Qin, Jian and Wu, Dian and Ding, Xing and Hu, Yi and others},
  journal={Science},
  volume={370},
  number={6523},
  pages={1460--1463},
  year={2020},
  publisher={American Association for the Advancement of Science}
}

@article{AMUNDSON2006229,
title = "Synergia: An accelerator modeling tool with 3-D space charge",
journal = "Journal of Computational Physics",
volume = "211",
number = "1",
pages = "229 - 248",
year = "2006",
issn = "0021-9991",
doi = "https://doi.org/10.1016/j.jcp.2005.05.024",
author = "J. Amundson and P. Spentzouris and J. Qiang and R. Ryne"
}

@article{QIANG2000434,
title = "An Object-Oriented Parallel Particle-in-Cell Code for Beam Dynamics Simulation in Linear Accelerators",
journal = "Journal of Computational Physics",
volume = "163",
number = "2",
pages = "434 - 451",
year = "2000",
issn = "0021-9991",
doi = "https://doi.org/10.1006/jcph.2000.6570",
author = "Ji Qiang and Robert D. Ryne and Salman Habib and Viktor Decyk"
}

@article{ZHANG2011338,
title = "The fast multipole method in the differential algebra framework",
journal = "Nuclear Instruments and Methods in Physics Research Section A: Accelerators, Spectrometers, Detectors and Associated Equipment",
volume = "645",
number = "1",
pages = "338 - 344",
year = "2011",
note = "The Eighth International Conference on Charged Particle Optics",
issn = "0168-9002",
doi = "https://doi.org/10.1016/j.nima.2011.01.053",
author = "He Zhang and Martin Berz"
}

@misc{CERNdata,
  title = {{CERN Data Portal}},
  url = {http://opendata.cern.ch}
}

@misc{FAIRdata,
  title = {{FAIR Principles}},
  url = {https://www.go-fair.org/fair-principles/}
}

@article{openPMD,
  author       = {Huebl, Axel and
                  Lehe, Rémi and
                  Vay, Jean-Luc and
                  Grote, David P. and
                  Sbalzarini, Ivo and
                  Kuschel, Stephan and
                  Sagan, David and
                  Pérez, Frédéric and
                  Koller, Fabian and
                  Bussmann, Michael},
  title        = {openPMD: A meta data standard for particle and mesh based data},
  year         = {2015},
  publisher    = {Zenodo},
  doi          = {10.5281/zenodo.591699},
  url          = {https://doi.org/10.5281/zenodo.591699}
}

@techreport{libEnsemble,
  author      = {Stephen Hudson and Jeffrey Larson and Stefan M. Wild and
                 David Bindel and John-Luke Navarro},
  title       = {{libEnsemble} Users Manual},
  institution = {Argonne National Laboratory},
  number      = {Revision 0.7.0},
  year        = {2020},
  url         = {https://buildmedia.readthedocs.org/media/pdf/libensemble/latest/libensemble.pdf}
}

@misc{xopt,
     author = {Mayes, Christopher},
     title = {xopt: Simulation optimization, based on DEAP},
     note = {https://github.com/ChristopherMayes/xopt}
}

@misc{IDEAS,
  title = {{IDEAS: Interoperable Design of Extreme-scale Application Software}},
  url = {https://ideas-productivity.org}
}

@misc{xSDK,
  title = {{xSDK: Extreme-scale Scientific Software Development Kit}},
  url = {http://xsdk.info}
}

@inproceedings{lume,
    author         = "C. E. Mayes and P. H. Fuoss and J. R. Garrahan and H. Slepicka and A. Halavanau and J. Krzywinski and A. L. Edelen and F. Ji, W. Lou and N. R. Neveu and A. Huebl and R. Lehe and L. Gupta and C. M. Gulliford and D. C. Sagan and J. C. E and C. Fortmann-Grote",
    title          = "Lightsource unified modeling environment (LUME), a start-to-end simulation ecosystem",
    booktitle      = "Proc. of IPAC",
    year           = "2021",
    pages       = "THPAB217"
}

@article{LOI_ABPRoadmap,
title = {{Accelerator and Beam Physics: Grand Challenges and Research Opportunities}},
author = {Nagaitsev, S. and Huang, Z. and Power J. and Vay, J.-L. and Piot, P. and Spentzouris, L. and Rosenzweig, J. and Cai, Y and Cousineau, S. and Conde, M. and Hogan, M. and Valishev, A. and Minty, M. and Zolkin, T. and Huang, X. and Shiltsev, V. and  Seeman, J. and Byrd, J. and Patterson, J.R.},
journal = {Snowmass21 LOI},
year = {2020},
url = {https://www.snowmass21.org/docs/files/summaries/AF/SNOWMASS21-AF1_AF7_S_Nagaitsev-056.pdf}
}

@article{ADIOS2,
title = {ADIOS 2: The Adaptable Input Output System. A framework for high-performance data management},
journal = {SoftwareX},
volume = {12},
pages = {100561},
year = {2020},
issn = {2352-7110},
doi = {https://doi.org/10.1016/j.softx.2020.100561},
url = {https://www.sciencedirect.com/science/article/pii/S2352711019302560},
author = {William F. Godoy and Norbert Podhorszki and Ruonan Wang and Chuck Atkins and Greg Eisenhauer and Junmin Gu and Philip Davis and Jong Choi and Kai Germaschewski and Kevin Huck and Axel Huebl and Mark Kim and James Kress and Tahsin Kurc and Qing Liu and Jeremy Logan and Kshitij Mehta and George Ostrouchov and Manish Parashar and Franz Poeschel and David Pugmire and Eric Suchyta and Keichi Takahashi and Nick Thompson and Seiji Tsutsumi and Lipeng Wan and Matthew Wolf and Kesheng Wu and Scott Klasky}
}

@misc{HDF5,
    author = {{The HDF Group}},
    title = "{Hierarchical data format version 5}",
    url = {http://www.hdfgroup.org/HDF5}
}

@misc{HYPRE,
    author = {{The HYPRE Team}},
    title = "{HYPRE}",
    url = {https://computing.llnl.gov/projects/hypre-scalable-linear-solvers-multigrid-methods}
}

@InProceedings{Huebl2017,
author="Huebl, Axel
and Widera, Ren{\'e}
and Schmitt, Felix
and Matthes, Alexander
and Podhorszki, Norbert
and Choi, Jong Youl
and Klasky, Scott
and Bussmann, Michael",
editor="Kunkel, Julian M.
and Yokota, Rio
and Taufer, Michela
and Shalf, John",
title="On the Scalability of Data Reduction Techniques in Current and Upcoming HPC Systems from an Application Perspective",
booktitle="High Performance Computing",
year="2017",
publisher="Springer International Publishing",
address="Cham",
pages="15--29",
isbn="978-3-319-67630-2"
}

@article{LOI_physics_based_injector_modeling,
title = {{Physics-based high-fidelity modeling of high brightness beam injectors}},
author = {Huang, C.-K. and Kwan, T.J.T. and Pavlenko, Vitaly and Ng, Cho-Kuen and Wang, Erdong},
journal = {Snowmass21 LOI},
url = {https://www.snowmass21.org/docs/files/summaries/AF/SNOWMASS21-AF7_AF1-CompF2_CompF0_Huang-183.pdf},
year={2020}
}

@inproceedings{cebaf-dark,
title = {Operation of the CEBAF 100 MV Cryomodules},
author = {C. Hovater and T. Allison and G. Biallas and R. Bachimanchi and E. Daly and M. Drury and A. Freyberger and
 R. L. Geng and G. Lahti and R. Legg and C. Mounts and R. Nelson and T. Plawski and T. Powers},
  booktitle    = {Proc. of Linear Accelerator Conference (LINAC'16),
                  East Lansing, MI, USA, 25-30 September 2016},
  pages        = {65--67},
  paper        = {MOOP11},
  language     = {english},
  venue        = {East Lansing, MI, USA},
  series       = {Linear Accelerator Conference},
  number       = {28},
  publisher    = {JACoW},
  address      = {Geneva, Switzerland},
  month        = {5},
  year         = {2017},
  isbn         = {978-3-95450-169-4},
  doi          = {https://doi.org/10.18429/JACoW-LINAC2016-MOOP11},
  url          = {http://jacow.org/linac2016/papers/moop11.pdf}
}

@InProceedings{lcls-dark,
    author = {J. R. Lewandowski and R. C. Field and A. S. Fisher and H.-D. Nuhn and J. J. Welch},
    title = {RF Gun Dark Current Suppression with a Transverse Deflecting Cavity at LCLS},
    booktitle = {Proc. 37th Int. Free Electron Laser Conf. (FEL'15)},
    pages = {583--586},
    paper = {WEP001},
    venue = {Daejeon, Korea},
    publisher = {JACoW Publishing},
    month = {8},
    year = {2015},
    doi = {doi:10.18429/JACoW-FEL2015-WEP001},
    note = {https://doi.org/10.18429/JACoW-FEL2015-WEP001},
    language = {english}
}

@article{scheinker2015adaptive,
  title={Adaptive method for electron bunch profile prediction},
  author={Scheinker, Alexander and Gessner, Spencer},
  journal={Physical Review Special Topics-Accelerators and Beams},
  volume={18},
  number={10},
  pages={102801},
  year={2015},
  doi={https://doi.org/10.1103/PhysRevSTAB.18.102801},
  publisher={APS}
}

@article{scheinker2021adaptive,
  title={An adaptive approach to machine learning for compact particle accelerators},
  author={Scheinker, Alexander and Cropp, Frederick and Paiagua, Sergio and Filippetto, Daniele},
  journal={Scientific reports},
  volume={11},
  number={1},
  pages={1--11},
  year={2021},
  publisher={Nature Publishing Group},
  url={https://doi.org/10.1038/s41598-021-98785-0}
}

@article{Mokhov:2017klc,
    author = "Mokhov, Nikolai V. and James, Catherine C.",
    title = "{The MARS Code System User's Guide Version 15(2016)}",
    reportNumber = "FERMILAB-FN-1058-APC",
    doi = "10.2172/1462233",
    month = "2",
    year = "2017"
}

@article{Ferrari:2005zk,
    author = "Ferrari, Alfredo and Sala, Paola R. and Fasso, Alberto and Ranft, Johannes",
    title = "{FLUKA: A multi-particle transport code (Program version 2005)}",
    reportNumber = "CERN-2005-010, SLAC-R-773, INFN-TC-05-11, CERN-2005-10",
    doi = "10.2172/877507",
    month = "10",
    year = "2005"
}

@ONLINE{Copa,
    author = {{The Copa Team}},
    title = "{Copa}",
    url = {https://github.com/ECP-copa}
}

@article{LOI_BestPractices,
title = {{Embracing modern software tools and user-friendly practices, when distributing scientific codes}},
author = {
Lehe, R. and Huebl, A. and Vay, J.-L. and Friedman, A. and Th\'evenet, M. and Mitchell, C. and Bruhwiler, D. and Grote, D. and Cowan, B. and Vincenti, H. and Hanuka, A. and Cros, B. and Yoffe, S. and Widera, R. and Bussmann, M. and Edelen, A.},
journal = {Snowmass21 LOI},
year = 2020,
url = {https://www.snowmass21.org/docs/files/summaries/CompF/SNOWMASS21-CompF2_CompF0_Lehe-076.pdf}
}

@article{openPMDapi,
  author       = {Huebl, Axel and
                  Poeschel, Franz and
                  Koller, Fabian and
                  Gu, Junmin},
  title        = {{openPMD-api: C++ \& Python API for Scientific I/O with openPMD}},
  year         = 2018,
  publisher    = {Rodare},
  doi          = {10.14278/rodare.27},
  url          = {https://github.com/openPMD/openPMD-api}
}

@misc{HipacePP,
  author       = {Th{\'e}venet, Maxence and
                  Diederichs, Severin and
                  Sinn, Alexander and
                  Lehe, Remi and
                  Huebl, Axel and
                  Myers, Andrew and
                  Zhang, Weiqun and
                  Vay, Jean-Luc},
  title        = {{HiPACE++: Highly efficient Plasma Accelerator Emulation, quasistatic particle-in-cell code}},
  year         = 2021,
  url          = {https://github.com/Hi-PACE/hipace}
}

@inproceedings{PIConGPU2013,
 author = {Bussmann, M. and Burau, H. and Cowan, T. E. and Debus, A. and Huebl, A. and Juckeland, G. and Kluge, T. and Nagel, W. E. and Pausch, R. and Schmitt, F. and Schramm, U. and Schuchart, J. and Widera, R.},
 title = {Radiative Signatures of the Relativistic Kelvin-Helmholtz Instability},
 booktitle = {Proceedings of the International Conference on High Performance Computing, Networking, Storage and Analysis},
 series = {SC '13},
 year = {2013},
 isbn = {978-1-4503-2378-9},
 location = {Denver, Colorado},
 pages = {5:1--5:12},
 articleno = {5},
 numpages = {12},
 url = {http://doi.acm.org/10.1145/2503210.2504564},
 doi = {10.1145/2503210.2504564},
 acmid = {2504564},
 publisher = {ACM},
 address = {New York, NY, USA},
}

@misc{spack,
     title = {Spack},
     note = {\url{https://spack.io/}}
}

@misc{conda,
     title = {Conda},
     note = {\url{https://docs.conda.io/en/latest/}}
}

@misc{pip,
     title = {The Phython package Installer},
     note = {\url{https://pip.pypa.io/en/stable/}}
}

@article{Wan2021,
author = {Wan, Lipeng and Huebl, Axel and Gu, Junmin and
Poeschel, Franz and Gainaru, Ana and Wang, Ruonan and Chen, Jieyang and Liang, Xin and Ganyushin, Dmitry and Munson, Todd and Foster, Ian and Vay, Jean-Luc and Podhorszki, Norbert and Wu, Kesheng and Klasky, Scott},
journal = {accepted in IEEE Transactions on Parallel and Distributed Systems},
title = {{Improving I/O Performance for Exascale Applications through Online Data Layout Reorganization}},
url = {https://arxiv.org/abs/2107.07108},
year = {2021}
}

@article{Poeschel2021,
author = {Poeschel, Franz and E, Juncheng and Godoy, William F. and Podhorszki, Norbert and Klasky, Scott and Eisenhauer, Greg and Davis, Philip E. and Wan, Lipeng and Gainaru, Ana and Gu, Junmin and Koller, Fabian and Widera, Rene and Bussmann, Michael and Huebl, Axel},
journal = {submitted},
title = {{Transitioning from file-based HPC workflows to streaming data pipelines with openPMD and ADIOS2}},
year = {2021}
}

@misc{Slack,
     title = {Slack},
     note = {\url{https://slack.com/}}
}

@misc{GithubIssues,
     title = {GitHub Issues},
     note = {\url{https://guides.github.com/features/issues/}}
}

@misc{Gitter,
     title = {Gitter},
     note = {\url{https://gitter.im/}}
}

@misc{zenodo,
     title = {Zenodo},
     note = {\url{https://zenodo.org/}}
}

@misc{git,
     title = {Git},
     note = {\url{https://git-scm.com/}}
}

@misc{githubactions,
     title = {GitHub actions},
     note = {\url{https://github.com/features/actions}}
}

@misc{azure,
     title = {Azure pipelines},
     note = {\url{https://azure.microsoft.com/en-us/services/devops/pipelines/}}
}

@misc{travisCI,
     title = {{Travis CI}},
     note = {\url{https://travis-ci.org/}}
}

@misc{e4s,
     title = {{E4S}},
     note = {\url{https://e4s-project.github.io/}}
}

@TechReport{usdoe-aac-2016,
   title={Advanced Accelerator Development Strategy Report: {DOE} Advanced Accelerator Concepts Research Roadmap Workshop},
   year={2016},
   address={United States},
   note={Research Org.: USDOE Office of Science, Washington, DC (United States)},
   note={Sponsor Org.: USDOE Office of Science (SC), High Energy Physics (HEP)},
   doi={10.2172/1358081},
   url={https://www.osti.gov/biblio/1358081},
   url={https://doi.org/10.2172/1358081},
   language={English}
}

@article{JING201872,
title = {Electron acceleration through two successive electron beam driven wakefield acceleration stages},
journal = {Nuclear Instruments and Methods in Physics Research Section A: Accelerators, Spectrometers, Detectors and Associated Equipment},
volume = {898},
pages = {72-76},
year = {2018},
issn = {0168-9002},
doi = {https://doi.org/10.1016/j.nima.2018.05.007},
url = {https://www.sciencedirect.com/science/article/pii/S0168900218305928},
author = {C. Jing and S. Antipov and M. Conde and W. Gai and G. Ha and W. Liu and N. Neveu and J.G. Power and J. Qiu and J. Shi and D. Wang and E. Wisniewski}
}

@misc{1GWpower,
   author={Picard, J. and al.},
     title={Generation of 1 GW of 11.7~{GHz} Power using a Metamaterial-based Power Extractor for Structure-based Wakefield Acceleration},
    note={Presented at the APS April Meeting 2021, Volume 66, Number 5},
    url = {https://meetings.aps.org/Meeting/APR21/Session/T08.1},
    year={2021}}

@Article{OShea2016,
   author={O'Shea, B. D. 
   and Andonian, G. 
   and Barber, S. K.
   and Fitzmorris, K. L.
   and Hakimi, S.
   and Harrison, J.
   and Hoang, P. D.
   and Hogan, M. J.
   and Naranjo, B.
   and Williams, O. B.
  and Yakimenko, V.
  and Rosenzweig, J. B.},
  title={Observation of acceleration and deceleration in gigaelectron-volt-per-metre gradient dielectric wakefield accelerators},
   journal={Nature Communications},
   year={2016},
   month={9},
   day={14}, 
   volume={7},
   number={1},
   pages={12763},
  issn={2041-1723},
  doi={10.1038/ncomms12763},
  url={https://doi.org/10.1038/ncomms12763}
}

@article{PhysRevLett.124.044802,
  title = {Single-Shot Characterization of High Transformer Ratio Wakefields in Nonlinear Plasma Acceleration},
  author = {Roussel, R. and Andonian, G. and Lynn, W. and Sanwalka, K. and Robles, R. and Hansel, C. and Deng, A. and Lawler, G. and Rosenzweig, J. B. and Ha, G. and Seok, J. and Power, J. G. and Conde, M. and Wisniewski, E. and Doran, D. S. and Whiteford, C. E.},
  journal = {Phys. Rev. Lett.},
  volume = {124},
  issue = {4},
  pages = {044802},
  numpages = {6},
  year = {2020},
  month = {1},
  publisher = {American Physical Society},
  doi = {10.1103/PhysRevLett.124.044802},
  url = {https://link.aps.org/doi/10.1103/PhysRevLett.124.044802}
}

@article{PhysRevLett.120.114801,
  title = {Observation of High Transformer Ratio of Shaped Bunch Generated by an Emittance-Exchange Beam Line},
  author = {Gao, Q. and Ha, G. and Jing, C. and Antipov, S. P. and Power, J. G. and Conde, M. and Gai, W. and Chen, H. and Shi, J. and Wisniewski, E. E. and Doran, D. S. and Liu, W. and Whiteford, C. E. and Zholents, A. and Piot, P. and Baturin, S. S.},
  journal = {Phys. Rev. Lett.},
  volume = {120},
  issue = {11},
  pages = {114801},
  numpages = {5},
  year = {2018},
  month = {3},
  publisher = {American Physical Society},
  doi = {10.1103/PhysRevLett.120.114801},
  url = {https://link.aps.org/doi/10.1103/PhysRevLett.120.114801}
}

@article{PhysRevAccelBeams.21.031301,
  title = {Stability condition for the drive bunch in a collinear wakefield accelerator},
  author = {Baturin, S. S. and Zholents, A.},
  journal = {Phys. Rev. Accel. Beams},
  volume = {21},
  issue = {3},
  pages = {031301},
  numpages = {11},
  year = {2018},
  month = {3},
  publisher = {American Physical Society},
  doi = {10.1103/PhysRevAccelBeams.21.031301},
  url = {https://link.aps.org/doi/10.1103/PhysRevAccelBeams.21.031301}
}

@misc{abp-roadmap,
      title={Accelerator and Beam Physics Research Goals and Opportunities}, 
      author={S. Nagaitsev and Z. Huang and J. Power and J. -L. Vay and P. Piot and L. Spentzouris and J. Rosenzweig and Y. Cai and S. Cousineau and M. Conde and M. Hogan and A. Valishev and M. Minty and T. Zolkin and X. Huang and V. Shiltsev and J. Seeman and J. Byrd and Y. Hao and B. Dunham and B. Carlsten and A. Seryi and R. Patterson},
      year={2021},
      eprint={2101.04107},
      archivePrefix={arXiv},
      primaryClass={physics.acc-ph}
}

@article{Geddes_2004,
	Author = {Geddes, C. G. R. and Toth, Cs. and van Tilborg, J. and Esarey, E. and Schroeder, C. B. and Bruhwiler, D. and Nieter, C. and Cary, J. and Leemans, W. P.},
	Date = {2004-09-30},
	Day = {30},
	Issn = {0028-0836},
	Journal = {Nature},
	M3 = {10.1038/nature02900},
	Month = {09},
	Number = {7008},
	Pages = {538--541},
	Title = {High-quality electron beams from a laser wakefield accelerator using plasma-channel guiding},
	Ty = {JOUR},
	Url = {http://dx.doi.org/10.1038/nature02900},
	Volume = {431},
	Year = {2004}}

@article{Leemans_2006,
	Author = {Leemans, W. P. and Nagler, B. and Gonsalves, A. J. and Toth, Cs. and Nakamura, K. and Geddes, C. G. R. and Esarey, E. and Schroeder, C. B. and Hooker, S. M.},
	Date = {2006-10},
	Issn = {1745-2473},
	Journal = {Nat Phys},
	M3 = {10.1038/nphys418},
	Month = {10},
	Number = {10},
	Pages = {696--699},
	Title = {GeV electron beams from a centimetre-scale accelerator},
	Ty = {JOUR},
	Url = {http://dx.doi.org/10.1038/nphys418},
	Volume = {2},
	Year = {2006}}

@article{Ibbotson_2010,
	Author = {Ibbotson, T. P. A. and Bourgeois, N. and Rowlands-Rees, T. P. and Caballero, L. S. and Bajlekov, S. I. and Walker, P. A. and Kneip, S. and Mangles, S. P. D. and Nagel, S. R. and Palmer, C. A. J. and Delerue, N. and Doucas, G. and Urner, D. and Chekhlov, O. and Clarke, R. J. and Divall, E. and Ertel, K. and Foster, P. S. and Hawkes, S. J. and Hooker, C. J. and Parry, B. and Rajeev, P. P. and Streeter, M. J. V. and Hooker, S. M.},
	Doi = {10.1103/PhysRevSTAB.13.031301},
	Issue = {3},
	Journal = {Phys. Rev. ST Accel. Beams},
	Month = {3},
	Numpages = {4},
	Pages = {031301},
	Publisher = {American Physical Society},
	Title = {Laser-wakefield acceleration of electron beams in a low density plasma channel},
	Url = {http://link.aps.org/doi/10.1103/PhysRevSTAB.13.031301},
	Volume = {13},
	Year = {2010}}

@article{Blue_2003,
	Author = {Blue, B. E. and Clayton, C. E. and O'Connell, C. L. and Decker, F.-J. and Hogan, M. J. and Huang, C. and Iverson, R. and Joshi, C. and Katsouleas, T. C. and Lu, W. and Marsh, K. A. and Mori, W. B. and Muggli, P. and Siemann, R. and Walz, D.},
	Doi = {10.1103/PhysRevLett.90.214801},
	Issue = {21},
	Journal = {Phys. Rev. Lett.},
	Month = {5},
	Numpages = {4},
	Pages = {214801},
	Publisher = {American Physical Society},
	Title = {Plasma-Wakefield Acceleration of an Intense Positron Beam},
	Url = {https://link.aps.org/doi/10.1103/PhysRevLett.90.214801},
	Volume = {90},
	Year = {2003}}

@article{Blumenfeld_2007,
	Author = {Blumenfeld, Ian and Clayton, Christopher E. and Decker, Franz-Josef and Hogan, Mark J. and Huang, Chengkun and Ischebeck, Rasmus and Iverson, Richard and Joshi, Chandrashekhar and Katsouleas, Thomas and Kirby, Neil and Lu, Wei and Marsh, Kenneth A. and Mori, Warren B. and Muggli, Patric and Oz, Erdem and Siemann, Robert H. and Walz, Dieter and Zhou, Miaomiao},
	Date = {2007-02-15},
	Day = {15},
	Journal = {Nature},
	L3 = {10.1038/nature05538; https://www.nature.com/articles/nature05538#supplementary-information},
	Month = {02},
	Pages = {741 EP -},
	Publisher = {Nature Publishing Group SN -},
	Title = {Energy doubling of 42 GeV electrons in a metre-scale plasma wakefield accelerator},
	Ty = {JOUR},
	Url = {http://dx.doi.org/10.1038/nature05538},
	Volume = {445},
	Year = {2007}}

@article{Litos_2014,
	Author = {Litos, M. and Adli, E. and An, W. and Clarke, C. I. and Clayton, C. E. and Corde, S. and Delahaye, J. P. and England, R. J. and Fisher, A. S. and Frederico, J. and Gessner, S. and Green, S. Z. and Hogan, M. J. and Joshi, C. and Lu, W. and Marsh, K. A. and Mori, W. B. and Muggli, P. and Vafaei-Najafabadi, N. and Walz, D. and White, G. and Wu, Z. and Yakimenko, V. and Yocky, G.},
	Date = {2014-11-05},
	Day = {05},
	Journal = {Nature},
	L3 = {10.1038/nature13882; https://www.nature.com/articles/nature13882#supplementary-information},
	Month = {11},
	Pages = {92 EP -},
	Publisher = {Nature Publishing Group, a division of Macmillan Publishers Limited. All Rights Reserved. SN -},
	Title = {High-efficiency acceleration of an electron beam in a plasma wakefield accelerator},
	Ty = {JOUR},
	Url = {http://dx.doi.org/10.1038/nature13882},
	Volume = {515},
	Year = {2014}}

@article{Gessner_2016,
	Author = {Gessner, Spencer and Adli, Erik and Allen, James M. and An, Weiming and Clarke, Christine I. and Clayton, Chris E. and Corde, Sebastien and Delahaye, J. P. and Frederico, Joel and Green, Selina Z. and Hast, Carsten and Hogan, Mark J. and Joshi, Chan and Lindstr{\o}m, Carl A. and Lipkowitz, Nate and Litos, Michael and Lu, Wei and Marsh, Kenneth A. and Mori, Warren B. and O'Shea, Brendan and Vafaei-Najafabadi, Navid and Walz, Dieter and Yakimenko, Vitaly and Yocky, Gerald},
	Date = {2016-06-02},
	Day = {02},
	Journal = {Nature Communications},
	L3 = {10.1038/ncomms11785; https://www.nature.com/articles/ncomms11785#supplementary-information},
	M3 = {Article},
	Month = {06},
	Pages = {11785 EP -},
	Publisher = {The Author(s) SN -},
	Title = {Demonstration of a positron beam-driven hollow channel plasma wakefield accelerator},
	Ty = {JOUR},
	Url = {http://dx.doi.org/10.1038/ncomms11785},
	Volume = {7},
	Year = {2016}}

@article{Diederichs_2019,
	Author = {Diederichs, S. and Mehrling, T. J. and Benedetti, C. and Schroeder, C. B. and Knetsch, A. and Esarey, E. and Osterhoff, J.},
	Doi = {10.1103/PhysRevAccelBeams.22.081301},
	Issue = {8},
	Journal = {Phys. Rev. Accel. Beams},
	Month = {8},
	Numpages = {6},
	Pages = {081301},
	Publisher = {American Physical Society},
	Title = {Positron transport and acceleration in beam-driven plasma wakefield accelerators using plasma columns},
	Url = {https://link.aps.org/doi/10.1103/PhysRevAccelBeams.22.081301},
	Volume = {22},
	Year = {2019}}

@article{vanTilborg_2015,
	Author = {van Tilborg, J. and Steinke, S. and Geddes, C. G. R. and Matlis, N. H. and Shaw, B. H. and Gonsalves, A. J. and Huijts, J. V. and Nakamura, K. and Daniels, J. and Schroeder, C. B. and Benedetti, C. and Esarey, E. and Bulanov, S. S. and Bobrova, N. A. and Sasorov, P. V. and Leemans, W. P.},
	Doi = {10.1103/PhysRevLett.115.184802},
	Issue = {18},
	Journal = {Phys. Rev. Lett.},
	Month = {10},
	Numpages = {5},
	Pages = {184802},
	Publisher = {American Physical Society},
	Title = {Active Plasma Lensing for Relativistic Laser-Plasma-Accelerated Electron Beams},
	Url = {http://link.aps.org/doi/10.1103/PhysRevLett.115.184802},
	Volume = {115},
	Year = {2015}}

@article{Steinke_2016,
	Author = {Steinke, S. and van Tilborg, J. and Benedetti, C. and Geddes, C. G. R. and Schroeder, C. B. and Daniels, J. and Swanson, K. K. and Gonsalves, A. J. and Nakamura, K. and Matlis, N. H. and Shaw, B. H. and Esarey, E. and Leemans, W. P.},
	Date = {2016-02-11},
	Day = {11},
	Issn = {0028-0836},
	Journal = {Nature},
	L3 = {10.1038/nature16525},
	M3 = {Letter},
	Month = {02},
	Number = {7589},
	Pages = {190--193},
	Publisher = {Nature Publishing Group, a division of Macmillan Publishers Limited. All Rights Reserved.},
	Title = {Multistage coupling of independent laser-plasma accelerators},
	Ty = {JOUR},
	Url = {http://dx.doi.org/10.1038/nature16525},
	Volume = {530},
	Year = {2016}}

@article{Lehe_2014,
	Author = {Lehe, R. and Thaury, C. and Guillaume, E. and Lifschitz, A. and Malka, V.},
	Doi = {10.1103/PhysRevSTAB.17.121301},
	Issue = {12},
	Journal = {Phys. Rev. ST Accel. Beams},
	Month = {12},
	Numpages = {9},
	Pages = {121301},
	Publisher = {American Physical Society},
	Title = {Laser-plasma lens for laser-wakefield accelerators},
	Url = {https://link.aps.org/doi/10.1103/PhysRevSTAB.17.121301},
	Volume = {17},
	Year = {2014}}

@article{DArcy_2019,
	Author = {D'Arcy, R. and Wesch, S. and Aschikhin, A. and Bohlen, S. and Behrens, C. and Garland, M. J. and Goldberg, L. and Gonzalez, P. and Knetsch, A. and Libov, V. and de la Ossa, A. Martinez and Meisel, M. and Mehrling, T. J. and Niknejadi, P. and Poder, K. and R\"ockemann, J.-H. and Schaper, L. and Schmidt, B. and Schr\"oder, S. and Palmer, C. and Schwinkendorf, J.-P. and Sheeran, B. and Streeter, M. J. V. and Tauscher, G. and Wacker, V. and Osterhoff, J.},
	Doi = {10.1103/PhysRevLett.122.034801},
	Issue = {3},
	Journal = {Phys. Rev. Lett.},
	Month = {1},
	Numpages = {6},
	Pages = {034801},
	Publisher = {American Physical Society},
	Title = {Tunable Plasma-Based Energy Dechirper},
	Url = {https://link.aps.org/doi/10.1103/PhysRevLett.122.034801},
	Volume = {122},
	Year = {2019}}

@misc{Scherkl_2019,
	Archiveprefix = {arXiv},
	Author = {Paul Scherkl and Alexander Knetsch and Thomas Heinemann and Andrew Sutherland and Ahmad Fahim Habib and Oliver Karger and Daniel Ullmann and Andrew Beaton and Gavin Kirwan and Grace Manahan and Yunfeng Xi and Aihua Deng and Michael Dennis Litos and Brendan D. OShea and Selina Z. Green and Christine I. Clarke and Gerard Andonian and Ralph Assmann and Dino A. Jaroszynski and David L. Bruhwiler and Jonathan Smith and John R. Cary and Mark J. Hogan and Vitaly Yakimenko and James B. Rosenzweig and Bernhard Hidding},
	Eprint = {1908.09263},
	Primaryclass = {physics.plasm-ph},
	Title = {Plasma-photonic spatiotemporal synchronization of relativistic electron and laser beams},
	Year = {2019}}

@article{Cook_2020,
	doi = {10.1088/1742-6596/1596/1/012063},
	url = {https://doi.org/10.1088/1742-6596/1596/1/012063},
	year = 2020,
	month = {9},
	publisher = {{IOP} Publishing},
	volume = {1596},
	pages = {012063},
	author = {N M Cook and J Carlsson and P Moeller and R Nagler and P Tzeferacos},
	title = {Modeling of capillary discharge plasmas for wakefield acceleration and beam transport},
	journal = {Journal of Physics: Conference Series}
}

@article{Picksley_2020,
	author = {Picksley, A. and Alejo, A. and Cowley, J. and Bourgeois, N. and Corner, L. and Feder, L. and Holloway, J. and Jones, H. and Jonnerby, J. and Milchberg, H. M. and Reid, L. R. and Ross, A. J. and Walczak, R. and Hooker, S. M.},
	doi = {10.1103/PhysRevAccelBeams.23.081303},
	issue = {8},
	journal = {Phys. Rev. Accel. Beams},
	month = {Aug},
	numpages = {8},
	pages = {081303},
	publisher = {American Physical Society},
	title = {Guiding of high-intensity laser pulses in 100-mm-long hydrodynamic optical-field-ionized plasma channels},
	url = {https://link.aps.org/doi/10.1103/PhysRevAccelBeams.23.081303},
	volume = {23},
	year = {2020}}

@article{Shalloo_2018,
	author = {Shalloo, R. J. and Arran, C. and Corner, L. and Holloway, J. and Jonnerby, J. and Walczak, R. and Milchberg, H. M. and Hooker, S. M.},
	doi = {10.1103/PhysRevE.97.053203},
	issue = {5},
	journal = {Phys. Rev. E},
	month = {May},
	numpages = {8},
	pages = {053203},
	publisher = {American Physical Society},
	title = {Hydrodynamic optical-field-ionized plasma channels},
	url = {https://link.aps.org/doi/10.1103/PhysRevE.97.053203},
	volume = {97},
	year = {2018}}

@article{Diaw_2022,
author = {Diaw,A.  and Coleman,S. J.  and Cook,N. M.  and Edelen,J. P.  and Hansen,E. C.  and Tzeferacos,P. },
title = {Impact of electron transport models on capillary discharge plasmas},
journal = {Physics of Plasmas},
volume = {29},
number = {6},
pages = {063101},
year = {2022},
doi = {10.1063/5.0091809},
URL = {https://doi.org/10.1063/5.0091809},
eprint = {https://doi.org/10.1063/5.0091809}}

@article{Miao_2020,
	author = {Miao, B. and Feder, L. and Shrock, J. E. and Goffin, A. and Milchberg, H. M.},
	doi = {10.1103/PhysRevLett.125.074801},
	issue = {7},
	journal = {Phys. Rev. Lett.},
	month = {Aug},
	numpages = {7},
	pages = {074801},
	publisher = {American Physical Society},
	title = {Optical Guiding in Meter-Scale Plasma Waveguides},
	url = {https://link.aps.org/doi/10.1103/PhysRevLett.125.074801},
	volume = {125},
	year = {2020}}

@article{Gonsalves_2019,
	Author = {Gonsalves, A. J. and Nakamura, K. and Daniels, J. and Benedetti, C. and Pieronek, C. and de Raadt, T. C. H. and Steinke, S. and Bin, J. H. and Bulanov, S. S. and van Tilborg, J. and Geddes, C. G. R. and Schroeder, C. B. and T\'oth, Cs. and Esarey, E. and Swanson, K. and Fan-Chiang, L. and Bagdasarov, G. and Bobrova, N. and Gasilov, V. and Korn, G. and Sasorov, P. and Leemans, W. P.},
	Doi = {10.1103/PhysRevLett.122.084801},
	Issue = {8},
	Journal = {Phys. Rev. Lett.},
	Month = {2},
	Numpages = {6},
	Pages = {084801},
	Publisher = {American Physical Society},
	Title = {Petawatt Laser Guiding and Electron Beam Acceleration to 8 GeV in a Laser-Heated Capillary Discharge Waveguide},
	Url = {https://link.aps.org/doi/10.1103/PhysRevLett.122.084801},
	Volume = {122},
	Year = {2019}}

@article{Bagdasarov_2017,
	Author = {Bagdasarov, G. A. and Sasorov, P. V. and Gasilov, V. A. and Boldarev, A. S. and Olkhovskaya, O. G. and Benedetti, C. and Bulanov, S. S. and Gonsalves, A. and Mao, H. -S. and Schroeder, C. B. and van Tilborg, J. and Esarey, E. and Leemans, W. P. and Levato, T. and Margarone, D. and Korn, G.},
	Booktitle = {Physics of Plasmas},
	Date = {2017-08-07},
	Doi = {10.1063/1.4997606},
	Issn = {1070-664X},
	Journal = {Physics of Plasmas},
	M3 = {doi: 10.1063/1.4997606},
	Month = {8},
	Number = {8},
	Pages = {083109},
	Publisher = {American Institute of Physics},
	Title = {Laser beam coupling with capillary discharge plasma for laser wakefield acceleration applications},
	Ty = {JOUR},
	Url = {https://doi.org/10.1063/1.4997606},
	Volume = {24},
	Year = {2017}
}

@article{Bagdasarov_2021,
author = {Bagdasarov,G. A.  and Bobrova,N. A.  and Olkhovskaya,O. G.  and Gasilov,V. A.  and Benedetti,C.  and Bulanov,S. S.  and Gonsalves,A. J.  and Pieronek,C. V.  and van Tilborg,J.  and Geddes,C. G. R.  and Schroeder,C. B.  and Sasorov,P. V.  and Bulanov,S. V.  and Korn,G.  and Esarey,E. },
title = {Creation of an axially uniform plasma channel in a laser-assisted capillary discharge},
journal = {Physics of Plasmas},
volume = {28},
number = {5},
pages = {053104},
year = {2021},
doi = {10.1063/5.0046428},
URL = {https://doi.org/10.1063/5.0046428},
eprint = {https://doi.org/10.1063/5.0046428}
}

@article{Xu_2016,
	Author = {Xu, X. L. and Hua, J. F. and Wu, Y. P. and Zhang, C. J. and Li, F. and Wan, Y. and Pai, C.-H. and Lu, W. and An, W. and Yu, P. and Hogan, M. J. and Joshi, C. and Mori, W. B.},
	Doi = {10.1103/PhysRevLett.116.124801},
	Issue = {12},
	Journal = {Phys. Rev. Lett.},
	Month = {3},
	Numpages = {5},
	Pages = {124801},
	Publisher = {American Physical Society},
	Title = {Physics of Phase Space Matching for Staging Plasma and Traditional Accelerator Components Using Longitudinally Tailored Plasma Profiles},
	Url = {https://link.aps.org/doi/10.1103/PhysRevLett.116.124801},
	Volume = {116},
	Year = {2016}}

@article{Lindstrom_2018,
	Author = {Lindstr\o{}m, C. A. and Adli, E. and Boyle, G. and Corsini, R. and Dyson, A. E. and Farabolini, W. and Hooker, S. M. and Meisel, M. and Osterhoff, J. and R\"ockemann, J.-H. and Schaper, L. and Sjobak, K. N.},
	Doi = {10.1103/PhysRevLett.121.194801},
	Issue = {19},
	Journal = {Phys. Rev. Lett.},
	Month = {11},
	Numpages = {6},
	Pages = {194801},
	Publisher = {American Physical Society},
	Title = {Emittance Preservation in an Aberration-Free Active Plasma Lens},
	Url = {https://link.aps.org/doi/10.1103/PhysRevLett.121.194801},
	Volume = {121},
	Year = {2018}}

@article{vanTilborg_2017,
	Author = {van Tilborg, J. and Barber, S. K. and Tsai, H.-E. and Swanson, K. K. and Steinke, S. and Geddes, C. G. R. and Gonsalves, A. J. and Schroeder, C. B. and Esarey, E. and Bulanov, S. S. and Bobrova, N. A. and Sasorov, P. V. and Leemans, W. P.},
	Doi = {10.1103/PhysRevAccelBeams.20.032803},
	Issue = {3},
	Journal = {Phys. Rev. Accel. Beams},
	Month = {3},
	Numpages = {6},
	Pages = {032803},
	Publisher = {American Physical Society},
	Title = {Nonuniform discharge currents in active plasma lenses},
	Url = {https://link.aps.org/doi/10.1103/PhysRevAccelBeams.20.032803},
	Volume = {20},
	Year = {2017}}

@article{Deng_2019,
	Author = {Deng, A. and Karger, O. S. and Heinemann, T. and Knetsch, A. and Scherkl, P. and Manahan, G. G. and Beaton, A. and Ullmann, D. and Wittig, G. and Habib, A. F. and Xi, Y. and Litos, M. D. and O'Shea, B. D. and Gessner, S. and Clarke, C. I. and Green, S. Z. and Lindstr{\o}m, C. A. and Adli, E. and Zgadzaj, R. and Downer, M. C. and Andonian, G. and Murokh, A. and Bruhwiler, D. L. and Cary, J. R. and Hogan, M. J. and Yakimenko, V. and Rosenzweig, J. B. and Hidding, B.},
	Da = {2019/11/01},
	Doi = {10.1038/s41567-019-0610-9},
	Id = {Deng2019},
	Issn = {1745-2481},
	Journal = {Nature Physics},
	Number = {11},
	Pages = {1156--1160},
	Title = {Generation and acceleration of electron bunches from a plasma photocathode},
	Ty = {JOUR},
	Url = {https://doi.org/10.1038/s41567-019-0610-9},
	Volume = {15},
	Year = {2019}}

@article{Manahan_2019,
	Author = {Manahan, G. G. and Habib, A. F. and Scherkl, P. and Ullmann, D. and Beaton, A. and Sutherland, A. and Kirwan, G. and Delinikolas, P. and Heinemann, T. and Altuijri, R. and Knetsch, A. and Karger, O. and Cook, N. M. and Bruhwiler, D. L. and Sheng, Z. -M. and Rosenzweig, J. B. and Hidding, B.},
	Booktitle = {Philosophical Transactions of the Royal Society A: Mathematical, Physical and Engineering Sciences},
	Date = {2019-08-12},
	Doi = {10.1098/rsta.2018.0182},
	Journal = {Philosophical Transactions of the Royal Society A: Mathematical, Physical and Engineering Sciences},
	Journal1 = {Philosophical Transactions of the Royal Society A: Mathematical, Physical and Engineering Sciences},
	M3 = {doi: 10.1098/rsta.2018.0182},
	Month = {8},
	Number = {2151},
	Pages = {20180182},
	Publisher = {Royal Society},
	Title = {Advanced schemes for underdense plasma photocathode wakefield accelerators: pathways towards ultrahigh brightness electron beams},
	Ty = {JOUR},
	Url = {https://doi.org/10.1098/rsta.2018.0182},
	Volume = {377},
	Year = {2019}}

@article{Doss_2019,
	Author = {Doss, C. E. and Adli, E. and Ariniello, R. and Cary, J. and Corde, S. and Hidding, B. and Hogan, M. J. and Hunt-Stone, K. and Joshi, C. and Marsh, K. A. and Rosenzweig, J. B. and Vafaei-Najafabadi, N. and Yakimenko, V. and Litos, M.},
	Doi = {10.1103/PhysRevAccelBeams.22.111001},
	Issue = {11},
	Journal = {Phys. Rev. Accel. Beams},
	Month = {11},
	Numpages = {9},
	Pages = {111001},
	Publisher = {American Physical Society},
	Title = {Laser-ionized, beam-driven, underdense, passive thin plasma lens},
	Url = {https://link.aps.org/doi/10.1103/PhysRevAccelBeams.22.111001},
	Volume = {22},
	Year = {2019}}

@article{Xi_2013,
	Author = {Xi, Y. and Hidding, B. and Bruhwiler, D. and Pretzler, G. and Rosenzweig, J. B.},
	Doi = {10.1103/PhysRevSTAB.16.031303},
	Issue = {3},
	Journal = {Phys. Rev. ST Accel. Beams},
	Month = {3},
	Numpages = {9},
	Pages = {031303},
	Publisher = {American Physical Society},
	Title = {Hybrid modeling of relativistic underdense plasma photocathode injectors},
	Url = {https://link.aps.org/doi/10.1103/PhysRevSTAB.16.031303},
	Volume = {16},
	Year = {2013}}

@misc{Gessner_2020,
	Archiveprefix = {arXiv},
	Author = {S. Gessner and the AWAKE Collaboration},
	Eprint = {2006.09991},
	Primaryclass = {physics.acc-ph},
	Title = {Evolution of a plasma column measured through modulation of a high-energy proton beam},
	Year = {2020}}

@misc{IBMQC,
title = {{IBM} Quantum Computing},
year = {2020 (accessed March 08, 2022)}, 
howpublished = "\url{https://www.ibm.com/quantum-computing/}"
}

@article{wang2019quantum,
  title={Quantum Fast {P}oisson Solver: the algorithm and modular circuit design},
  author={Wang, Shengbin and Wang, Zhimin and Li, Wendong and Fan, Lixin and Wei, Zhiqiang and Gu, Yongjian},
  journal={arXiv preprint arXiv:1910.09756},
  year={2019}
}

@misc{MicrosoftQDev,
title = {Q\# and the {Q}uantum {D}evelopment {K}it},
year = {2020 (accessed March 08, 2022)}, 
howpublished = "\url{https://www.microsoft.com/en-us/quantum/development-kit}"
}

@article{rebentrost2014quantum,
  title={Quantum support vector machine for big data classification},
  author={Rebentrost, Patrick and Mohseni, Masoud and Lloyd, Seth},
  journal={Physical review letters},
  volume={113},
  number={13},
  pages={130503},
  year={2014},
  publisher={APS}
}

@article{chatterjee2016generalized,
  title={Generalized coherent states, reproducing kernels, and quantum support vector machines},
  author={Chatterjee, Rupak and Yu, Ting},
  journal={arXiv preprint arXiv:1612.03713},
  year={2016}
}

@article{low2014quantum,
  title={Quantum inference on {B}ayesian networks},
  author={Low, Guang Hao and Yoder, Theodore J and Chuang, Isaac L},
  journal={Physical Review A},
  volume={89},
  number={6},
  pages={062315},
  year={2014},
  publisher={APS}
}

@article{cong2019quantum,
  title={Quantum convolutional neural networks},
  author={Cong, Iris and Choi, Soonwon and Lukin, Mikhail D},
  journal={Nature Physics},
  volume={15},
  number={12},
  pages={1273--1278},
  year={2019},
  publisher={Nature Publishing Group}
}

@article{zhao2019bayesian,
  title={Bayesian deep learning on a quantum computer},
  author={Zhao, Zhikuan and Pozas-Kerstjens, Alejandro and Rebentrost, Patrick and Wittek, Peter},
  journal={Quantum Machine Intelligence},
  volume={1},
  number={1},
  pages={41--51},
  year={2019},
  publisher={Springer}
}

@article{dunjko2016quantum,
  title={Quantum-enhanced machine learning},
  author={Dunjko, Vedran and Taylor, Jacob M and Briegel, Hans J},
  journal={Physical review letters},
  volume={117},
  number={13},
  pages={130501},
  year={2016},
  publisher={APS}
}

@inproceedings{dunjko2017advances,
  title={Advances in quantum reinforcement learning},
  author={Dunjko, Vedran and Taylor, Jacob M and Briegel, Hans J},
  booktitle={2017 IEEE International Conference on Systems, Man, and Cybernetics (SMC)},
  pages={282--287},
  year={2017},
  organization={IEEE}
}

@article{preskill2018quantum,
  title={Quantum computing in the {NISQ} era and beyond},
  author={Preskill, John},
  journal={Quantum},
  volume={2},
  pages={79},
  year={2018},
  publisher={Verein zur F{\"o}rderung des Open Access Publizierens in den Quantenwissenschaften}
}

@article{Xiao2019,
	Author = {Xiao, L. and others},
	Doi = {10.1109/JMMCT.2019.2954946},
	Journal = {IEEE J. Multiscale and Multiphysics Comp. Tech.},
	Pages = {298--306},
	Title = {Advances in Multiphysics Modeling for Parallel Finite-Element Code Suite {ACE3P}},
	Volume = {4},
	Year = {2019}
}

@INPROCEEDINGS{akcelik2008,
  author = {V. Akcelik and L.-Q. Lee and Z. Li and C.-K. Ng and L. Xiao and K. Ko},
  title = {{SRF} Cavity Imperfection Studies Using Advanced Shape Uncertainty Quantification Tools},
  booktitle = {Proceedings of {LINAC08}},
  year = {2008},
  date = {2008-09-29/2008-10-03},
  location = {Victoria, BC, Canada}
}

@INPROCEEDINGS{lunin2018,
  author = {A. Lunin and T. Khabiboulline and N. Solyak and A. Sukhanov and V. Yakovlev},
  title = {Statistical analysis of the eigenmode spectrum in the SRF cavities with mechanical imperfections},
  booktitle = {Proceedings of 13th Int. Computational Accelerator Physics Conf.},
  year = {2018},
  date = {2018-09-20/2018-09-24},
  location = {Key West, FL, USA}
}

@INPROCEEDINGS{Qiang2000b,
  author = {J. Qiang, and R. D. Ryne, and S. Habib},
  title = {Self-consistent Langevin simulation of Coulomb collisions in charged-particle beams},
  booktitle = {Proceedings of the IEEE/ACM SC2000 Conference},
  year = {2000},
  date = {2000-11-04/2000-11-10},
  location = {Dallas Texas, USA}
}

@article{Yu2009,
  title={Lattice and beam dynamics for the pulse mode of the laser-electron storage ring
for a Compton x-ray source},
  author={P. Yu, and Y. Wang, and W. Huang},
  journal={Phys. Rev. ST Accel. Beams},
  volume={12},
  pages={061301},
  year={2009}
}

@article{Mayes2021,
  title={Computational approaches to Coherent Synchrotron Radiation in two and three dimensions},
  author={C. Mayes},
  journal={Journal of Instrumentation},
  volume={16},
  pages={P10010},
  year={2021}
}

@article{Marzouk2021,
  title={Efficient algorithm for high fidelity collisional charged particle beam dynamics},
  author={A. Al Marzouk and H. D. Schaumburg and S. Abeyratne and B. Erdelyi},
  journal={Phys. Rev. ST Accel. Beams},
  volume={24},
  pages={074601},
  year={2021}
}

@article{Ivanov2020,
author = {Ivanov, Andrei and Agapov, Ilya},
doi = {10.1103/PhysRevAccelBeams.23.074601},
issn = {2469-9888},
journal = {Physical Review Accelerators and Beams},
month = {7},
number = {7},
pages = {074601},
publisher = {American Physical Society},
title = {{Physics-based deep neural networks for beam dynamics in charged particle accelerators}},
url = {https://doi.org/10.1103/PhysRevAccelBeams.23.074601 https://link.aps.org/doi/10.1103/PhysRevAccelBeams.23.074601},
volume = {23},
year = {2020}
}

@misc{HEPAP,
  title = {{Accelerating Discovery: A Strategic Plan for Accelerator R\&D in the U.S., Report of the  High Energy Physics Advisory Panel (HEPAP) Accelerator Research and Development Subpanel, 2015. \url{https://science.osti.gov/-/media/hep/hepap/pdf/Reports/Accelerator_RD_Subpanel_Report.pdf}}},
}

@article{Qiangcpc2001,
  title={Parallel 3D Poisson Solver for a Charged
  Beam in a Conducting Pipe},
  author={J. Qiang and R. D. Ryne},
  journal={Comp. Phys. Comm.},
  volume={138},
  pages={18},
  year={2001}
}

@article{Qiangcpc2006,
  title={A 3D model for ion beam formation and transport simulation},
  author={J. Qiang, and D. Todd, and D. Leitner},
  journal={Comp. Phys. Comm.},
  volume={175},
  pages={416},
  year={2006}
}

@article{Qiangcpc2016,
  title={Efficient Three-Dimensional Poisson Solvers in Open Rectangular Conducting Pipe},
  author={J. Qiang},
  journal={Comp. Phys. Comm.},
  volume={203},
  pages={122},
  year={2016}
}

@article{Qiangnim2017,
  title={A Fast Numerical Integrator for Relativistic Charged Particle Tracking},
  author={J. Qiang},
  journal={Instruments \& Methods in Physics Research A},
  volume={885},
  pages={55},
  year={2017}
}

@article{QiangPRAB2018,
  title={Symplectic particle-in-cell model for space-charge beam dynamics simulation},
  author={J. Qiang},
  journal={Phys. Rev. ST Accel. Beams},
  volume={21},
  pages={054201},
  year={2018}
}

@article{QiangPRAB2019,
  title={Fast 3D Poisson solvers in elliptical conducting pipe for space-charge simulation},
  author={J. Qiang},
  journal={Phys. Rev. ST Accel. Beams},
  volume={22},
  pages={104601},
  year={2019}
}

@article{Qiang2022,
  title={Simulation of space-charge effects using a quantum Schrodinger
    approach},
  author={J. Qiang},
  journal={Phys. Rev. ST Accel. Beams},
  volume={25},
  pages={034602},
  year={2022}
}

@article{Qiangnim2022,
  title={X-ray FEL linear accelerator design via start-to-end global optimization},
  author={Qiang, Ji},
  journal={Nuclear Instruments and Methods in Physics Research Section A: Accelerators, Spectrometers, Detectors and Associated Equipment},
  pages={166294},
  year={2022},
  publisher={Elsevier}
}

@inproceedings{Qiangnapac2016,
  title={Start-to-End Beam Dynamics Optimization of X-Ray FEL Light Source Accelerators},
  author={Qiang, Ji},
  booktitle={Proceedings of North American PAC2016},
  pages={838-842},
  year={2016},
  month = {October 9-14},
  location = {Chicago, IL, USA},
}

@article{Qiangprab2014,
  title={Start-to-end simulation of x-ray radiation of a next generation light source using the real number of electrons},
  author={Qiang, J and Corlett, J and Mitchell, CE and Papadopoulos, CF and Penn, G and Placidi, M and Reinsch, M and Ryne, RD and Sannibale, F and Sun, C and others},
  journal={Physical Review Special Topics-Accelerators and Beams},
  volume={17},
  number={3},
  pages={030701},
  year={2014},
  publisher={APS}
}

@article{bizzozero2021,
  author = {D.A. Bizzozero and J. Qiang and L. Ge and Z. Li and C.-K. Ng and L. Xiao},
  title = {Multi-objective optimization with an integrated electromagnetics and beam dynamics workflow},
  journal = {Nucl. Instrum. Methods Phys. Res. A},
  year = {2021},
  volume = {1020},
  pages = {165844}
}

@misc{Diederichs2022,
  author = {Severin Diederichs and Carlo Benedetti and Axel Huebl and Rémi Lehe and Andrew Myers and Alexander Sinn and Jean-Luc Vay and Weiqun Zhang and Maxence Thévenet},
  title = {HiPACE++: a portable, 3D quasi-static Particle-in-Cell code},
  year = {2022},
  eprint = {arXiv:2109.10277},
  url = {https://arxiv.org/abs/2109.10277}
}

@article{PlasmonicsIEEE2021,
    title={Nanomaterials Based Nanoplasmonic Accelerators and Light-Sources Driven by Particle- Beams},
    author={Sahai, A. A.}, 
    journal={IEEE Access}, 
    volume={9}, 
    pages={54831-54839},
    year={2021},
    publisher={IEEE},
    Url = {https://ieeexplore.ieee.org/document/9395099},
    doi = {10.1109/ACCESS.2021.3070798}}

@inproceedings{PlasmonicsSPIE2021,
    title={Emergence of TeraVolts per meter plasmonics using relativistic surface plasmonic modes},
    author={Sahai, A. A.}, 
    booktitle={Proc. Vol. 11797, Plasmonics: Design, Materials, Fabrication, Characterization, and Applications XIX},
    volume={117972A}, 
    year={2021},
    publisher={SPIE},
    doi = {10.1117/12.2596637},
    Url = {https://www.spiedigitallibrary.org/conference-proceedings-of-spie/11797/2596637/Emergence-of-TeraVolts-per-meter-plasmonics-using-relativistic-surface-plasmonic/10.1117/12.2596637.short?SSO=1}}

@misc{CUDenverPCT,
  author = {Sahai, A. A.},
  title = {anostructure nanoplasmonic accelerator, high-energy photon source, and related methods},
  volume = {PCT WO2021216424A1},
  year = {2021},
  Url = {https://patentimages.storage.googleapis.com/55/0a/6c/7b757efb9e9547/WO2021216424A1.pdf}
}

@inproceedings{NanofocusingSPIE2022,
    title={Plasmonic nano-focusing of particle beams by surface crunch-in plasmons excited in tapered tubes},
    author={Sahai, A. A.}, 
    booktitle={Proc. SPIE 11999, Ultrafast Phenomena and Nanophotonics XXVI},
    volume={1199903}, 
    year={2022},
    publisher={SPIE},
    doi = {10.1117/12.2605722},
    Url = {https://www.spiedigitallibrary.org/conference-proceedings-of-spie/11999/2605722/Plasmonic-nano-focusing-of-particle-beams-by-surface-crunch-in/10.1117/12.2605722.short}}

@article{Plasmon1953,
    title={A Collective Description of-Electron Interactions: III. Coulomb Interactions in a Degenerate Electron Gas},
    author={Bohm, D. and Pines, D.}, 
    journal={Phys. Rev.}, 
    volume={92}, 
    pages={609},
    year={1953},
    publisher={APS},
    Url = {https://ieeexplore.ieee.org/document/9395099},
    doi = {10.1109/ACCESS.2021.3070798}}

@article{Wavebreaking1959,
    title={Nonlinear Electron Oscillations in a Cold Plasma},
    author={Dawson, J. M.}, 
    journal={Phys. Rev. (1959)}, 
    volume={113}, 
    pages={383},
    year={1959},
    publisher={APS},
    Url = {https://ieeexplore.ieee.org/document/9395099},
    doi = {10.1109/ACCESS.2021.3070798}}

@misc{PVm-plasmonics-snowmass,
    title="{PetaVolts per meter Plasmonics: Snowmass21 White Paper}",
    author={A. Sahai and collaborators},
    year={2022},
    archivePrefix={arXiv},
    Url = {https://arxiv.org/abs/2203.11623}}

@misc{Snowmass_WP_EOD,
      title="{Snowmass AF1 White Paper: Strategies in Education, Outreach, and Inclusion to Enhance the US Workforce in Accelerator Science and Engineering}", 
      author={Bai, M. and others},
      year={2022},
      archivePrefix={arXiv}
}

@article{karniadakis2021physics,
  title={Physics-informed machine learning},
  author={Karniadakis, George Em and Kevrekidis, Ioannis G and Lu, Lu and Perdikaris, Paris and Wang, Sifan and Yang, Liu},
  journal={Nature Reviews Physics},
  volume={3},
  number={6},
  pages={422--440},
  year={2021},
  publisher={Nature Publishing Group}
}

@article{zhu2021high,
  title = {High-Fidelity Prediction of Megapixel Longitudinal Phase-Space Images of Electron Beams Using Encoder-Decoder Neural Networks},
  author = {Zhu, J. and Chen, Y. and Brinker, F. and Decking, W. and Tomin, S. and Schlarb, H.},
  journal = {Phys. Rev. Applied},
  volume = {16},
  issue = {2},
  pages = {024005},
  numpages = {9},
  year = {2021},
  month = {Aug},
  publisher = {American Physical Society},
  doi = {10.1103/PhysRevApplied.16.024005},
  url = {https://link.aps.org/doi/10.1103/PhysRevApplied.16.024005}
}

@article{Berz:1988aj,
    author = "Berz, M.",
    title = "{Differential Algebraic Description of Beam Dynamics to Very High Orders}",
    reportNumber = "SSC-152",
    journal = "Part. Accel.",
    volume = "24",
    pages = "109--124",
    year = "1989"
}

@inproceedings{Wu2019Practical,
author = {Wu, Jian and Toscano-Palmerin, Saul and Frazier, Peter and Wilson, Andrew},
year = {2019},
title = {Practical Multi-fidelity Bayesian Optimization for Hyperparameter Tuning},
booktitle ={Uncertainty Quantification in Artificial Intelligence}
}

@inproceedings{eriksson2019scalable,
	author = {Eriksson, David and Pearce, Michael and Gardner, Jacob and Turner, Ryan D and Poloczek, Matthias},
	booktitle = {Advances in Neural Information Processing Systems},
	editor = {H. Wallach and H. Larochelle and A. Beygelzimer and F. d\textquotesingle Alch\'{e}-Buc and E. Fox and R. Garnett},
	publisher = {Curran Associates, Inc.},
	title = {Scalable Global Optimization via Local Bayesian Optimization},
	url = {https://proceedings.neurips.cc/paper/2019/file/6c990b7aca7bc7058f5e98ea909e924b-Paper.pdf},
	volume = {32},
	year = {2019}}

@article{Sanchez-Gonzalez:2019gis,
    author = "Sanchez-Gonzalez, Alvaro and Bapst, Victor and Cranmer, Kyle and Battaglia, Peter",
    title = "{Hamiltonian Graph Networks with ODE Integrators}",
    eprint = "1909.12790",
    archivePrefix = "arXiv",
    primaryClass = "cs.LG",
    month = "9",
    year = "2019"
}

@InProceedings{neiswanger2021bayesian,
  title = 	 {Bayesian Algorithm Execution: Estimating Computable Properties of Black-box Functions Using Mutual Information},
  author =       {Neiswanger, Willie and Wang, Ke Alexander and Ermon, Stefano},
  booktitle = 	 {Proceedings of the 38th International Conference on Machine Learning},
  pages = 	 {8005--8015},
  year = 	 {2021},
  editor = 	 {Meila, Marina and Zhang, Tong},
  volume = 	 {139},
  series = 	 {Proceedings of Machine Learning Research},
  month = 	 {18--24 Jul},
  publisher =    {PMLR},
  url = 	 {https://proceedings.mlr.press/v139/neiswanger21a.html}
}

@article{Gupta_2021,
	author = {Lipi Gupta and Auralee Edelen and Nicole Neveu and Aashwin Mishra and Christopher Mayes and Young-Kee Kim},
	doi = {10.1088/2632-2153/ac27ff},
	journal = {Machine Learning: Science and Technology},
	month = {oct},
	number = {4},
	pages = {045025},
	publisher = {{IOP} Publishing},
	title = {Improving surrogate model accuracy for the {LCLS}-{II} injector frontend using convolutional neural networks and transfer learning},
	url = {https://doi.org/10.1088/2632-2153/ac27ff},
	volume = {2},
	year = 2021}

@ARTICLE{deb2002afast,

  author={Deb, K. and Pratap, A. and Agarwal, S. and Meyarivan, T.},

  journal={IEEE Transactions on Evolutionary Computation}, 

  title={A fast and elitist multiobjective genetic algorithm: NSGA-II}, 

  year={2002},

  volume={6},

  number={2},

  pages={182-197},

  doi={10.1109/4235.996017}}

@conference{EdelenNeurIPS2017,
author = {Edelen, A. and  Edelen, J. and  Milton, S.  and Biedron, S.  and  Van der Slot,P.J.M.},
title = {Using Neural Network Control Policies for Rapid Switching Between Beam Parameters in a free electron laser},
 booktitle        = "NeurIPS 2017",
 pages          = "",
 year           = "2017",
 address        = "Long Beach, CA", 
 note = {\url{https://dl4physicalsciences.github.io/files/nips_dlps_2017_16.pdf}}
}

@conference{EdelenNeurIPS2019,
author = {Edelen, A. and  Neveu, N. and Emma, C. and Ratner, D. and Mayes, C.},
title = {Machine Learning Models for Optimization and Control of X-ray Free Electron Lasers},
 booktitle        = "NeurIPS 2019",
 pages          = "",
 year           = "2019",
 address        = "Vancouver, Canada", 
  note = {\url{https://ml4physicalsciences.github.io/2019/files/NeurIPS_ML4PS_2019_90.pdf}}
}

@article{scheinker2021adaptiveML,
  title={Adaptive machine learning for time-varying systems: low dimensional latent space tuning},
  author={Scheinker, Alexander},
  journal={Journal of Instrumentation},
  volume={16},
  number={10},
  pages={P10008},
  year={2021},
  publisher={IOP Publishing},
  url={https://doi.org/10.1088/1748-0221/16/10/P10008}
}

@article{pan2009survey,
  title={A survey on transfer learning},
  author={Pan, Sinno Jialin and Yang, Qiang},
  journal={IEEE Transactions on knowledge and data engineering},
  volume={22},
  number={10},
  pages={1345--1359},
  year={2009},
  publisher={IEEE}
}

@article{nagabandi2018deep,
  title={Deep online learning via meta-learning: Continual adaptation for model-based RL},
  author={Nagabandi, Anusha and Finn, Chelsea and Levine, Sergey},
  journal={arXiv preprint arXiv:1812.07671},
  year={2018}
}

@article{mishra2021uncertainty,
  title={Uncertainty quantification for deep learning in particle accelerator applications},
  author={Mishra, Aashwin Ananda and Edelen, Auralee and Hanuka, Adi and Mayes, Christopher},
  journal={Physical Review Accelerators and Beams},
  volume={24},
  number={11},
  pages={114601},
  year={2021},
  publisher={APS}
}

@article{duris2020bayesian,
  title={Bayesian optimization of a free-electron laser},
  author={Duris, Joseph and Kennedy, Dylan and Hanuka, Adi and Shtalenkova, Jane and Edelen, Auralee and Baxevanis, P and Egger, Adam and Cope, T and McIntire, M and Ermon, S and others},
  journal={Physical review letters},
  volume={124},
  number={12},
  pages={124801},
  year={2020},
  publisher={APS}
}

@article{edelen2016first,
  title={First steps toward incorporating image based diagnostics into particle accelerator control systems using convolutional neural networks},
  author={Edelen, AL and Biedron, SG and Milton, SV and Edelen, JP},
  journal={arXiv preprint arXiv:1612.05662},
  year={2016}
}

@article{raissi2019physics,
  title={Physics-informed neural networks: A deep learning framework for solving forward and inverse problems involving nonlinear partial differential equations},
  author={Raissi, Maziar and Perdikaris, Paris and Karniadakis, George E},
  journal={Journal of Computational physics},
  volume={378},
  pages={686--707},
  year={2019},
  publisher={Elsevier}
}

@article{hanuka2021physics,
  title={Physics model-informed Gaussian process for online optimization of particle accelerators},
  author={Hanuka, Adi and Huang, Xiaobiao and Shtalenkova, Jane and Kennedy, Dylan and Edelen, Auralee and Zhang, Z and Lalchand, VR and Ratner, D and Duris, J},
  journal={Physical Review Accelerators and Beams},
  volume={24},
  number={7},
  pages={072802},
  year={2021},
  publisher={APS}
}

@article{scheinker2018demonstration,
  title={Demonstration of model-independent control of the longitudinal phase space of electron beams in the linac-coherent light source with femtosecond resolution},
  author={Scheinker, Alexander and Edelen, Auralee and Bohler, Dorian and Emma, Claudio and Lutman, Alberto},
  journal={Physical review letters},
  volume={121},
  number={4},
  pages={044801},
  year={2018},
  publisher={APS},
  url={https://doi.org/10.1103/PhysRevLett.121.044801}
}

@article{roussel2022differentiable,
  title={Differentiable Preisach Modeling for Characterization and Optimization of Accelerator Systems with Hysteresis},
  author={Roussel, R and Edelen, A and Ratner, D and Dubey, K and Gonzalez-Aguilera, JP and Kim, YK and Kuklev, N},
  journal={arXiv preprint arXiv:2202.07747},
  year={2022}
}

@article{adelmann:opal,
	title = {{OPAL} a {Versatile} {Tool} for {Charged} {Particle} {Accelerator} {Simulations}},
	url = {http://arxiv.org/abs/1905.06654},
	urldate = {2019-05-17},
	journal = {arXiv:1905.06654 [physics]},
	author = {Adelmann, Andreas and Calvo, Pedro and Frey, Matthias and Gsell, Achim and Locans, Uldis and Metzger-Kraus, Christof and Neveu, Nicole and Rogers, Chris and Russell, Steve and Sheehy, Suzanne and Snuverink, Jochem and Winklehner, Daniel},
	month = may,
	year = {2019},
	note = {arXiv: 1905.06654}
}

@article{winklehner:spiral,
	title = {Realistic simulations of a cyclotron spiral inflector within a particle-in-cell framework},
	volume = {20},
	url = {https://link.aps.org/doi/10.1103/PhysRevAccelBeams.20.124201},
	doi = {10.1103/PhysRevAccelBeams.20.124201},
	number = {12},
	urldate = {2018-10-15},
	journal = {Physical Review Accelerators and Beams},
	author = {Winklehner, Daniel and Adelmann, Andreas and Gsell, Achim and Kaman, Tulin and Campo, Daniela},
	month = dec,
	year = {2017},
	pages = {124201},
}

@article{emma2018machine,
  title = {Machine learning-based longitudinal phase space prediction of particle accelerators},
  author = {Emma, C. and Edelen, A. and Hogan, M. J. and O'Shea, B. and White, G. and Yakimenko, V.},
  journal = {Phys. Rev. Accel. Beams},
  volume = {21},
  issue = {11},
  pages = {112802},
  numpages = {6},
  year = {2018},
  month = {Nov},
  publisher = {American Physical Society},
  doi = {10.1103/PhysRevAccelBeams.21.112802},
  url = {https://link.aps.org/doi/10.1103/PhysRevAccelBeams.21.112802}
}

@article{edelen2020machine,
  title = {Machine learning for orders of magnitude speedup in multiobjective optimization of particle accelerator systems},
  author = {Edelen, Auralee and Neveu, Nicole and Frey, Matthias and Huber, Yannick and Mayes, Christopher and Adelmann, Andreas},
  journal = {Phys. Rev. Accel. Beams},
  volume = {23},
  issue = {4},
  pages = {044601},
  numpages = {23},
  year = {2020},
  month = {Apr},
  publisher = {American Physical Society},
  doi = {10.1103/PhysRevAccelBeams.23.044601},
  url = {https://link.aps.org/doi/10.1103/PhysRevAccelBeams.23.044601}
}

@article{edelen2016neural,

  author={Edelen, A. L. and Biedron, S. G. and Chase, B. E. and Edstrom, D. and Milton, S. V. and Stabile, P.},

  journal={IEEE Transactions on Nuclear Science}, 

  title={Neural Networks for Modeling and Control of Particle Accelerators}, 

  year={2016},

  volume={63},

  number={2},

  pages={878-897},

  doi={10.1109/TNS.2016.2543203}}

@article{roussel2021turnkey,
  title = {Turn-key constrained parameter space exploration for particle accelerators using Bayesian active learning.},
  author = {Roussel, Ryan, and Gonzalez-Aguilera, Juan Pablo, and Kim, Young-Kee, and Wisniewski, Eric, and Liu, Wanming, and Piot, Philippe, and Power, John, and Hanuka, Adi, and Edelen, Auralee},
  journal = {Nature Communications},
  volume = {12},
  issue = {1},
  year = {2021},
  url = {https://doi.org/10.1038/s41467-021-25757-3}
}

@article{roussel2021multiobj,
  title = {Multiobjective Bayesian optimization for online accelerator tuning},
  author = {Roussel, Ryan and Hanuka, Adi and Edelen, Auralee},
  journal = {Phys. Rev. Accel. Beams},
  volume = {24},
  issue = {6},
  pages = {062801},
  numpages = {14},
  year = {2021},
  month = {Jun},
  publisher = {American Physical Society},
  doi = {10.1103/PhysRevAccelBeams.24.062801},
  url = {https://link.aps.org/doi/10.1103/PhysRevAccelBeams.24.062801}
}

@article{convery2021uncertainty,
  title = {Uncertainty quantification for virtual diagnostic of particle accelerators},
  author = {Convery, Owen and Smith, Lewis and Gal, Yarin and Hanuka, Adi},
  journal = {Phys. Rev. Accel. Beams},
  volume = {24},
  issue = {7},
  pages = {074602},
  numpages = {8},
  year = {2021},
  month = {Jul},
  publisher = {American Physical Society},
  doi = {10.1103/PhysRevAccelBeams.24.074602},
  url = {https://link.aps.org/doi/10.1103/PhysRevAccelBeams.24.074602}
}

@article{ogren2021surrogate,
	author = {J. {\"O}gren and C. Gohil and D. Schulte},
	doi = {10.1088/1748-0221/16/05/p05012},
	journal = {Journal of Instrumentation},
	month = {may},
	number = {05},
	pages = {P05012},
	publisher = {{IOP} Publishing},
	title = {Surrogate modeling of the {CLIC} final-focus system using artificial neural networks},
	url = {https://doi.org/10.1088/1748-0221/16/05/p05012},
	volume = {16},
	year = 2021}

\end{document}